\documentclass[twocolumn,showpacs,preprintnumbers,amsmath,amssymb]{revtex4}

\usepackage{graphicx}
\usepackage{dcolumn}
\usepackage{bm}

\begin{document}



\title{Angular momentum nonconservation and conservation in quasiclassical Positronium}


\author{David C. Lush } 
\affiliation{%
 d.lush@comcast.net \\}%


\date{\today}

\begin{abstract} 

It is shown that due to Thomas precession, angular momentum is not generally a constant of the motion in a quasiclassical model of the Positronium atom consisting of circular-orbiting point charges with intrinsic spin and associated magnetic moment.  Despite absence of externally-applied torque, angular momentum is a constant of the motion only if the electron and positron intrinsic angular momentum vector components perpendicular to the orbital angular momentum are antiparallel and of equal magnitude.
   
\end{abstract} 




\maketitle






\section{Introduction}

The timing and sequence of events in the development of quantum theory seem to be such that little attempt was made historically to incorporate electrodynamical effects of the electron intrinsic spin into the classical atom model as pioneered by Rutherford \cite{Rutherford1911,Rutherford1914}.  In the absence of knowledge of the intrinsic spin and its magnitude involving the Planck constant (\(h\)), there was apparently no means to relate its occurrence as a proportionality constant in energy levels observed in atomic spectra, other than by postulate, as was done by Bohr \cite{Bohr1914} to significant success.  Subsequently, both Sommerfeld's \cite{Sommerfeld1916} refinement and extension of the Bohr model, and de Broglie's \cite{deBroglie1925} postulate of an electron wave character, also occurred prior to general recognition that the electron must possess an intrinsic angular momentum with magnitude proportional to \(h\).  The published inference of the spin existence by Goudsmit and Uhlenbeck \cite{GoudsmitUhlenbeck1926} in February, 1926, was nearly simultaneous with the inventions of Heisenberg's matrix mechanics \cite{Heisenberg1925} and Schr\"odinger's wave mechanics \cite{Schroedinger1926}, in 1925 and 1926.  Furthermore, although these first modern quantum theories were created without much if any knowledge of an intrinsic spin involving the Planck constant, it was soon incorporated in the new quantum theories by Pauli \cite{Pauli1927}, and its fundamental necessity in quantum theory was recognized through the work of Dirac \cite{Dirac1928} published in 1928. 

Yet one must wonder what attempts might have been made to incorporate the intrinsic spin into the classical atom model of Rutherford, had not the modern quantum theory emerged so abruptly and successfully.  The Bohr-Sommerfeld model was itself remarkably successful, in its heyday, but had only a postulatory basis.  Surely the discovery of intrinsic angular momentum and involving the Planck constant would have suggested the possibility of a dynamical basis for the Bohr postulate that orbital angular momentum is quantized in whole multiples of the reduced Planck constant, \(\hbar\equiv h/(2\pi)\). It would have been natural to suppose that a basis for this postulate (as well as the two additional postulates involving \(\hbar\) proposed by Sommerfeld soon after) might emerge from the classical electrodynamics of elementary particles carrying magnetic dipoles as well as electric charge.

Instead, with the rapid emergence of modern quantum theory nearly simultanteously with the discovery that subatomic particles possess an intrinsic angular momentum, interest in the old quantum theory based on modifications of the classical, point-charge Rutherford atom model rapidly waned.  However, at least prior to Dirac's work, a strong impetus existed to look closely at the classical atom model with spin, due to the so-called spin-orbit coupling anomaly. The proposal of the spin by Goudsmit and Uhlenbeck had been based on atomic emission spectrum measurements made by Zeeman for discharging gas under the influence of an externally-applied magnetic field.  But, at the time of its introduction, the intrinsic spin idea had a serious problem. Electrodynamically, it did not account consistently and simultaneously for the magnitudes of the Zeeman effect and of the spin-orbit coupling that manifested as spectral line splitting even in the absence of externally-applied magnetic field.  The spin magnitude inferred from the Zeeman effect is \(h/4\pi\) (\textit{i.e.}, \( \hbar/2\)), while the calculation of the emission line splitting due to spin-orbit coupling, assuming this spin magnitude, obtained a result too large by a factor of two.  

It was L. H. Thomas who proposed \cite{Thomas1927} that the spin-orbit coupling anomaly could be accounted for by the relativistic effect now known as the Thomas precession.  The Thomas precession reduces the rate of precession of the electron spin axis by a factor of about one-half, and assuming the classical-physics relationship between rate of precession and applied torque continues to apply, the spin-orbit coupling magnitude must also be approximately halved from its value expected in the absence of the Thomas precession.  Thomas observed that this explanation required that ``secular'', or orbit-averaged, total angular momentum be a constant of the motion, which he found to be the case.  In recent decades however it has been recognized that there must be included in the electrodynamics of magnetic dipoles an effect due to the presence of ``hidden'' momentum.  When hidden momentum is included in the analysis, Thomas' result of secular angular momentum conservation can no longer be obtained for  hydrogenic atoms with heavy nuclei that lack a substantial magnetic moment \cite{LushY9}.  

Hidden momentum is mechanical momentum imparted to the charged current carriers of a current-loop magnetic dipole by an electric field \cite{ShockleyJames1967,Coleman1968,Griffiths}.  It is equal and opposite the field momentum of the dipole magnetic field crossed with the electric field, so that the total of the field plus mechanical momentum of the stationary magnetic dipole in an electric field may vanish. Under any applied force, present an electric field, the equation of motion of a magnetic dipole must account for the change in the hidden momentum that accompanies translaional and orientational motion of the dipole, and so its omission leads to erroneous and nonsensical results.  In particular, omitting the hidden momentum leads to violation of Newton's law of action and reaction in the interaction of magnetic dipoles with electric monopoles. Therefore, the quasiclasical model of hydrogenic atoms considered by Thomas, where the electron intrinsic spin and magnetic moment is accounted for but the nuclear spin and intrinsic magnetic moment is negligible, and the hidden momentum is omitted, is inconsistent. The attractive force on the electron toward the nucleus does not equal the attractive force on the nucleus toward the electron.

Incorporating the hidden momentum does not entirely remedy the problems with the quasiclassical hydrogen model.  Incorporating the hidden momentum in the electrodynamics obtains consistency between the two alternative calculations of the binding force \cite{KholmetskiiY10a,LushY10a,KholmetskiiY10b,LushY10b}, but results in nonconservation of the total angular momentum \cite{LushY9oc1}. The rate of electron spin precession, which must include the effect of the Thomas precession, in agreement with Thomas, is approximately halved compared to its value absent Thomas precession, but the orbit precession frequency is doubled compared to the value calculated by Thomas.  The orbit precssion frequency magnitude is doubled due to the presence of the hidden momentum, and is unaffected by the Thomas precession that halves the spin precession frequency.  Simple dynamical considerations then lead to an inescapable conclusion that the total angular momentum is not a constant of the motion.  The total angular momentum magnitude is constant but the vector total angular momentum precesses around a fixed direction, even absent an externally applied magnetic field. This situation is not remedied either by accounting for the field momentum, nor by a careful relativistic treatment that takes proper account of the difference between the center of energy and center of mass.  The field angular momentum can be shown to vanish identically for the hydrogenic atom \cite{LushY9oc2}, and since the observation of angular momentum nonconservation can be made even for the minimally relativistic situation of quasiclassical Hydrogen with a circular-orbiting electron at the ground-state Bohr radius, relativistic effects other than the Thomas precession are negligible and so cannot change this outcome.

Beyond angular momentum nonconservation due to the Thomas precession, there is a problem of consistency of the spin-orbit interaction energy calculation, when the hidden momentum is included in the calculation, if it is assumed that the Thomas precssion causes a reduction of the spin-orbit coupling by the Thomas factor about one-half.  The problem is that the orbital magnetic moment in the quasiclassical picture also precesses but isn't reduced by the Thomas precession, and so the energy associated with the orbit precession is not reduced by the Thomas factor, and so stands in disagreement with the calculation focusing on the spin and including the Thomas factor.  This problem of inconsistency of description seems even more serious than the problem of angular momentum nonconservation and should lead to reconsideration of whether the Thomas precession can be accepted as plausibly reducing the spin-orbit coupling energy in the quasiclassical description.  
                
That the system consisting of a spinning magnetic electron bound to a non-magnetic proton does not conserve angular momentum should perhaps not be considered surprising, in light of result obtained by Kiessling \cite{Kiessling1999}, that classical electron theory does not conserve angular momentum in the absence of intrinsic spin. It therefore seems na\"ive to expect it in the quasiclassical hydrogen atom model with a non-magnetic proton. Even though the proton possesses spin, without a magnetic moment it cannot couple into the electrodynamics as necessary to recover angular momentum conservation.  The same consideration applies to accounting for the known intrinsic magnetic moment of the proton.  As can be easily verified, it is far too weak to correct the angular momentum nonconservation.

The situation of angular momentum nonconservation in classical spin theory presents serious problems.  One immediate consequence is that the picture that results is not even self-consistent, as already observed.  There are two different spin-orbit coupling magnitudes that result, depending on the focus of the calculation.  However there is another possibilty that might plausibly lead to a more satisfying result.  In the hydrogenic atom model where the angular momentum nonconservation is observed, no accounting has been made of the compositeness of the nucleus or proton.  Considering even Hydrogen as the simplest case, and unknown to Thomas, it is now known that the proton consists of three particles with spin equal to the electron and presumably also possessing intrinsic magnetic moment. It seems plausible that the difficulties already described could be remedied if these could be taken into account, but the difficulties of doing such in the classical picture are considerable.  This provides motivation to explore the quasiclassical electrodynamical model of the Positronium atom.  In Positronium, the problem of one particle lacking a useful intrinsic magnetic moment is well remedied, yet the simplicity of the minimally-relativistic two-body problem is retained.  However, one simplifying advantage of the Hydrogen problem is lost in that there is no longer any advantage in working in the electron rest frame.  

In the quasiclassical model of Hydrogen originated by Thomas and retained in many textbooks to the present day, there is significant advantage to working in a reference frame centered on the spinning particle, when the other particle can be considerd to be non-magnetic.  Then, the more-complicated electrodynamics due to a moving magnetic dipole do not arise.  This benefit is mooted in the analysis of Positronium, but nonetheless it is not too difficult to work directly in the laboratory frame.  It can easily be verified that the techniques employed including the motion of the magnetic dipoles yield the same result in a laboratory-frame analysis of Hydrogen as is obtained in the electron rest frame.  

Based on the above motivating considerations, the objective of the present analysis is to determine the angular momentum conservation properties, including any effects due to Thomas precession, of the quasiclassical Positronium model consisting of two point-charge particles, the electron and positron, also endowed with intrinsic spin and associated intrinsic magnetic dipole moment.  This will be determined by developing an equation of motion of the total angular momentum that includes all terms to the order of the Thomas precession contribution.  

In the quasiclassical model, the particle positions and velocities, as well as their spin magnitudes and orientations, are perfectly well defined at all times.  To make the analysis as simple as possible, the particles are assumed to orbit each other circularly under Coulomb attraction.  The orbital radius will be taken to be no smaller than the ground-state radius of the Bohr model of Positronium, which will ensure that relativistic effects other than the Thomas precession are negligible. Radiative effects and radiation reaction due to motions of the charges are also negligible over the course of a single orbit, as is the effect of delay. (These facts are demonstrated in an appendix.)  It will then be readily seen, the total angular momentum is not generally a constant of the motion, for arbitrary orientations of the particle spins relative to each other and the orbital plane.  Rather, angular momentum is a constant of the motion only if the particle intrinsic spin components in the orbital plane are aligned antiparallel and of equal magnitude.  Thus, an expectation of angular momentum constancy gives rise to relative orientational constraints on the spins and orbit that may be akin to some behaviors of electron spins in atomic eigenstates in quantum theory.  

A sequel is planned to assess the system radiativity and radiation reaction forces related to the dipolar nature of the particles, and to determine if there exist forces that will make the special relative orientations preferred that obtain angular momentum constancy.  It seems plausible that in this way a new classical-physics mechanism of atomic radiative decay may be discovered, that may be compared with observation. However it does not seem likely that they will lead to general angular momentum conservation, because radiative fields are expected to be accompanied by radiation damping forces that themselves balance the radiated angular momentum.  Also, if these radiation fields due to intrinsic dipole motion are no larger than those due to the motion of the electric dipole due to charge separation, which are evaluated and shown to be inconsequential herein (see Appendices B and E), they can have no substantial effect. 

It should be stressed that the angular momentum conservation violation according to the present analysis exists despite accounting for angular momentum of the non-radiative electromagnetic field, which is not insignificant.  Both the field and hidden angular momentum contributions are essential in obtaining that the motion of the total angular momentum vanishes except for the contribution due to Thomas precession.
          
Counter to the results presented herein, the expectation that an analysis fully incorporating the Thomas precession and all electromagnetic forces and fields should obtain conservation of angular momentum is bolstered by the work of Schild and Schlosser \cite{SchildSchlosser1965,SchildSchlosser1968}, who showed that angular momentum is generally conserved in the electromagnetic two-body problem of charged particles with spin. Their manifestly-covariant general analysis accounts for both the Thomas precession and all ({\em i.e.} both nonradiative and radiative) electromagnetic fields.  However, the illustrative special case they considered, similar to the present system but restricted to aligned spins and orbit, was too restricted to uncover the interesting phenomonology to be described in the present work.  This suggests a course of further investigation, to apply the theory of Schild and Schlosser to the situation investigated here.

\section{Overview of the Analysis}

The analysis objective is to develop an equation of motion for the total angular momentum of the quasiclassical Positronium atom model for the circular orbit and including the effect of the Thomas precession, for arbitrary relative orientations of the spins and orbit.  The model is quasiclassical in the sense of being based on classical electrodynamics of point-charge particles with an {\em ad hoc} incorporation of intrinsic spin and magnetic moment. By assuming the quasiclassical Positronium atom is in the Bohr model ground state, the orbital velocity is small enough so that the primary significant relativistic effect is the Thomas precession.  (There is also the relativistic effect that the moving magnetic dipoles acquire electric dipole moment as well.) It is shown in the appendix that any dynamical effects of radiation reaction and delay, as well as  other relativistic effects, are  negligible to the order of the present analysis.  Therefore the central result of the present work, that according to the quasiclassical analysis, total angular momentum is not generally conserved by electromagnetically-interacting point particles with intrinsic spin, cannot be negated by a future analysis that includes radiation reaction and relativistic corrections to all orders.  

The total angular momentum to be considered has sources of four distinct types.  These are the intrinsic spins, the kinetic orbital angular momentum due to the motion of the particle masses, the so-called ``hidden'' orbital angular momentum, and the field angular momentum.  Equations of motion are developed for each source individually, and then summed to obtain an equation of motion of the total angular momentum.  

The equations of motions of the electron and positron intrinsic spins are developed in Section III using the equation of motion of the spin axis developed by Thomas \cite{Thomas1927oc1}, which can be shown to be derivable from relativistically-covariant Bargmann-Michel-Telegdi (BMT) equation \cite{BMT1959}.  This equation of motion is straightforwardly specialized to the case of each particle moving in the electromagnetic field of the other.      

The equation of motion of the kinetic orbital angular momentum, that is, the orbital angular momentum of the particle masses, is developed in Section IV based on the various electrodynmic forces acting on the particles due to their attributes of electric charge and magnetic dipole moment.  Importantly, due to the translational motion of the magnetic dipole carrying particles, they will acquire electric dipole moment that is not negligible to the analysis.  Translational (Stern-Gerlach) forces are then present due to the field gradients acting on the dipoles, as well as Coulomb and Biot-Savart forces acting on the charges.  Additional non-negligible forces arise on the charges due to translational motions of the dipoles.  The translational forces act to change the particles' mechanical momentum in accordance with Newton's law of motion, but the mechanical momentum must include the hidden momentum.  The time variation of the hidden momentum that must be accounted for in the equations of translational motion of the particles appears correspondingly in the equation of motion of the kinetic orbital angular momentum.  

When the general equation of motion of the particles' kinetic orbital angular momenta is available in terms of the total torque acting on the orbits, the next step is to identify all of the relevant translational forces and the torques they generate.  After taking inventory of the forces and torques in Section V, they are evaluated explicitly in Section VI.

In Section VII the equation of motion of the total angular momentum is derived in two steps.  The first step is to evaluate separately the rates of change of the kinetic, hidden, and field orbital angular momenta.  These quantities are then summed to obtain an equation of motion of the total orbital angular momentum.  The second and final step is then to sum the motion of the total orbital angular momentum with the motion of the electron and positron spin angular momenta to obtain the motion of the total angular momentum.  The resulting equation of motion shows that the total angular momentum is not generally a constant of the motion, due to Thomas precession, for arbitrary relative orientations of the particle spins and orbital angular momenta.  Angular momentum is a constant of the motion only for certain specific relative configurations of the spins and orbit.   

A set of appendices provide some standard results for reference and help to justify the approximations used. 

The analysis is performed in the laboratory frame throughout.  

\subsection{Notation}

Particle positions relative to the center of mass are represented by \(\mbox{\boldmath$r$}_e\) and \(\mbox{\boldmath$r$}_p\) for the electron and proton respectively.  In generic relations where the particle identity is unimportant the subscript may be dropped for brevity, but in result equations \(\mbox{\boldmath$r$}\) represents the vector from the positron to the electron, and \(r = |\mbox{\boldmath$r$}|\) to represent the electron-positron separation.   Thus  \(\mbox{\boldmath$r$}=\mbox{\boldmath$r$}_e-\mbox{\boldmath$r$}_p\) and for the equal-mass particles \(\mbox{\boldmath$r$}_e=-\mbox{\boldmath$r$}_p\).  Similarly, \(\mbox{\boldmath$v$} \equiv  d\mbox{\boldmath$r$}/dt\) will represent the electron velocity relative to the positron, while \(\mbox{\boldmath$v$}_e\) and \(\mbox{\boldmath$v$}_p\) will represent the particle velocities in the center-of-mass frame.

Since there are orbital angular momenta from three contributing sources (due to motion of the masses, ``hidden'', and field) and due to each particle separately, it is important distinguish them clearly in the notation.  The orbital angular momentum due to the motion of the electron and positron masses, that is, the kinetic orbital angular momentum, is represented as \(\mbox{\boldmath$L$}_e\) and \(\mbox{\boldmath$L$}_p\), and their sum \(\mbox{\boldmath$L$} \equiv \mbox{\boldmath$L$}_e + \mbox{\boldmath$L$}_p\), is the total kinetic orbital angular momentum.  The total hidden orbital angular momentum is represented as \(\mbox{\boldmath$L$}_{\text{hidden}} \equiv \mbox{\boldmath$L$}_{\text{hidden,}e} + \mbox{\boldmath$L$}_{\text{hidden,}p}\).  The total field angular momentum is represented as \(\mbox{\boldmath$L$}_{\text{field}} \equiv \mbox{\boldmath$L$}_{\text{field,}e} + \mbox{\boldmath$L$}_{\text{field,}p}\).  

\subsection{Representation of Intrinsic Magnetic Moments}

The representation of the intrinsic spin and associated magnetic dipole assumed for the present analysis is an ideal classical current-loop magnetic dipole.  The dipole-carrying particles are considered sufficiently separated so that only the magnetic far-field of the ideal dipole is significant.  

The intrinsic spin vector that follows from this representation has a well-defined orientation at all times.

The magnitude of the spin angular momentum is assumed in the present analysis to be \(\hbar/2\). This is taken as the total magnitude, as opposed to being the magnitude along a particular axis as in quantum theory.  This magnitude assumption is primarily for convenience and the central result is insensitive to whether the spin magnitude is \(\hbar/2\) or (as in quantum theory) \(\sqrt{3}/2\). 

It has also been suggested by various authors (see, for example \cite{Hestenes1990} and references therein) that the electron intrinsic spin may be due to a classical but relativisitic circulatory motion of a charged point-like particle, and that evidence for this identification may be found in the so-called \(\mbox{\boldmath$\alpha$} \cdot \mbox{\boldmath$E$}\) term of the Dirac equation for a free electron in an electric field.  The existence of hidden momentum and its assoctiation with intrinsic magnetic moments of elementary particles, is not inconsistent with this classical ``zitterbewegung'' model for the spin.  (This use of the term ``zitterbewegung'' in this way should not be equated with the usage by Schr\"odinger in his analysis of the Dirac quantum mechanical relativistic electron theory.) The model used herein is explicitly assumed to have one property not possessed by the classical zitterbewegung model for the spin; that is, no time variation of the intrinsic fields that is faster than the classical orbital frequency of the particles.  The effect of relaxing this assumption and considering a possibly more realistic model for the intrinsic fields (such as provided by Rivas \cite{Rivas}) is beyond the scope of the present work.  However, as Rivas shows that the time-averaged fields of the relativistic classical zitterbewegung are asymptotic to the fields of an ideal current loop, it seems plausible that the results found herein may apply to the zitterbewegung model as well.

\subsection{Level of Approximation}

The level of approximation taken herein may be thought of as minimally relativistic in the sense that the relativistic \(\gamma\) factor is usually approximated as unity. Two relativistic effects that are however central to the analysis and not neglected are the Thomas precession and the necessary electric dipole moment that accompanies a translating magnetic dipole.   Attempts are made through analysis to justify neglecting other relativistic effects such as propagation delay and the exact relativisitic description of the electromagnetic fields due to moving point particles, as the need arises.

It is also assumed that the electron ``g-factor,'' that is, the ratio of the electron magnetic dipole strength to that of a classical magnetic dipole constant current-carrier charge-to-mass ratio and equal angular momentum, is exactly two.  That this assumption is only valid to a little better than the first order in \(v/c\) in the present application will be shown to be not significant to the result. 

The analysis is also restricted to circular orbit.  This restriction is arbitrary but provides significant simplification.  

\subsection{Explanatation of the Particular Sequence of the Analysis}

As stated above, the general analysis approach is to obtain separate equations of motion for each of the four identified angular momentum contributors, and then sum them to obtain the equation of motion of the total angular momentum.  It is to be expected however that these separate equations of motion of the angular momentum contributors are not independent of each other.  For example and most obviously, the electromagnetic field angular momentum will depend entirely (and neglecting delay, which is defended in the appendix) on the instantaneous kinematical description of the particles including their spin orientations and orientational motions as well as the particles' positions, velocities, and accelerations.  The field configuration also affects the dynamical behavior through radiation damping forces, but fortunately expressions for radiation damping forces are available from textbooks in terms of the charge and magnetic dipole accelerations, and in the current application these can be neglected, which is also shown in the appendix.  Based on these considerations, the field angular momentum and its time rate of change will be evaluated as the last contribution to the total, after the others are sufficiently defined so that the motion of the field angular momentum may be easily evaluated.

The hidden angular momentum is determined by the kinematics and the instantaneous field configuration.  The hidden angular momentum will be seen to have a description that is formally identical to the field angular momentum.  Therefore it will also be evaluated after the kinematics have been specified by the equations of motion of the spins and kinetic orbital angular momentum.

Although the formula for the hidden angular momentum is formally similar to that for the field angular momentum, hidden momentum differs from field momentum in an important respect.  Unlike field momentum, hidden momentum enters directly into the equations of motion of intrinsically-magnetic particles, because any applied force must act on both kinetic and hidden momentum.  In determining the equation of motion of the kinetic angular momentum, this will be handled by accounting for hidden momentum in the equations of translational motion of the particles.   

An equation of motion of the spin orientations that depends only on the instantaneous fields is available from the literature.  The orientation and motion of the spins will be seen to enter the dynamics directly and non-negligibly.  Also, the hidden angular momentum and its motion will be seen to follow in a straightforward fashion from the spins orientations and motion, as well as the particle kinematics.  Therefore the equations of motion of the spins will be evaluated as the first significant analysis step.  Although the standard spin equation of motion is provided in terms of the fields, it will be convenient to rewrite them in terms of the other-particle relative position, velocity, and spin axis orientation. The particle accelerations will not need to be included due to the negligibility here of the acceleration field, as is shown in the appendix. 

Evaluation of the motion of the kinetic orbital angular momentum follows sequentially after the motions of the electron and positron spins have been determined in terms of the kinematic description.  When all the angular momentum motion contributions are finally available, the motion of the total angular momentum is evaluated as their sum.

\section{Equations of Motion of the Spins}

The following equation of motion of the spin vector \(\mbox{\boldmath$s$}\), for a particle  with intrinsic spin and magnetic moment, is derived by Jackson \cite{jcksn:classelec3} from the BMT equation, and who also notes it to be equivalent to that developed earlier by L. H. Thomas \cite{Thomas1927oc1}:

\begin{widetext}

\begin{equation}
\frac{d\mbox{\boldmath$s$}}{dt} = \frac{q}{mc}\mbox{\boldmath$s$} \times \left[\left(\frac{g}{2}-1+\frac{1}{\gamma} \right)\mbox{\boldmath$B$} - \left(\frac{g}{2}-1\right)\frac{\gamma}{\gamma+1}(\mbox{\boldmath$\beta$} \cdot \mbox{\boldmath$B$})\mbox{\boldmath$\beta$} -  \left(\frac{g}{2}-\frac{\gamma}{\gamma+1}\right)\mbox{\boldmath$\beta$} \times \mbox{\boldmath$E$}\right],
\label{ThomasEq}
\end{equation}

\end{widetext}

where \(q\) is the particle charge, \(\mbox{\boldmath$B$}\) is the magnetic field, \(\mbox{\boldmath$E$}\) is the electric field, \(\mbox{\boldmath$\beta$} = \mbox{\boldmath$v$}/c\) where \(\mbox{\boldmath$v$}\) is the particle velocity, and \(\gamma = 1/\sqrt{1-\beta^2}\), \(m\) is the particle mass, and \(t\) is time in the laboratory frame.  This is Jackson's form except that the (signed) charge value of \(e\) has been replaced with \(q\), so that \(e\) can be reserved for the fundamental charge magnitude.   Although this equation is developed assuming an absence of field gradients, and there are clearly nonvanishing field gradients in the present application, the error of omitting the field gradient terms can be evaluated based on the work of Good \cite{Good1961}, and is not significant to the present analysis. In general, as will be shown explicitly, higher-order corrections to the motion of the spins are not significant to the present analysis. It can be noted as well, Thomas applied Equation (\ref{ThomasEq}) in his analysis reconciling spin-orbit coupling with the anomalous Zeeman effect, thereby disregarding the field gradients in the atomic domain.

The equation of motion of the spin angular momentum vector of Equation (\ref{ThomasEq}) will be applied to determine the motion of the electron and positron spins in the quasiclassical model of positronium.  Pursuant to this objective it will be useful to first reduce the complexity of this equation in accordance with the approximations that are valid in the minimally-relativistic domain with respect to the quasiclassical Positronium atom in its Bohr model ground state.  For example (see Appendix A), approximating the relativistic \(\gamma\) factor as unity introduces an error of one part in \(1/\beta^2\), with \(\beta\) here approximately a factor of \(1/\alpha^2\) where \(\alpha\) is the fine-structure constant, or about \(5 \times 10^{-5}\).  However, this level accuracy (that is, one part in \(\alpha^2\)) is more than is needed here.  Since the stated objective of the analysis is to obtain an equation of motion of the total angular momentum that includes all contributions significant to the order of the contribution of the Thomas precession, it's worth examining what is the contribution of the Thomas precession to Eq. (\ref{ThomasEq}). 

The Thomas precession contribution to the motion of the spin vector in Eq. (\ref{ThomasEq}) is [cite Jackson?]

\begin{equation}
\frac{q}{mc}\mbox{\boldmath$s$} \times \left[ \left(\frac{\gamma}{\gamma+1}\right)\mbox{\boldmath$\beta$} \times \mbox{\boldmath$E$}\right] \approx \frac{q}{mc}\mbox{\boldmath$s$} \times \left[ \left(\frac{1}{2}\right)\mbox{\boldmath$\beta$} \times \mbox{\boldmath$E$}\right].
\nonumber
\label{TPContribToThomasEq}
\end{equation}

It will be seen below that in the present application \(\mbox{\boldmath$\beta$} \times \mbox{\boldmath$E$}\) is of the same order of magnitude as \(\mbox{\boldmath$B$}\). Returning to consideration of Eq. (\ref{ThomasEq}), it can then be observed that the magnitude of the center term of the bracketed terms on the right hand side has a factor of \(\beta^2\) relative to the other two terms, and also relative to the contribution of the Thomas precession.  Therefore this term is negligible in the present analysis, apart from that it will vanish under an approximation of the electron and positron g-factors as exactly 2.

Approximating \(g\) as \(2\) in Eq. (\ref{ThomasEq}) is appealing simply on account of the algebraic simplifications it will provide, but the approximation that \(g=2\) is only accurate to slightly better (about one part in \(10^3\) here) than the first order in \(\beta\).  An examination of the potential effects of taking the approximation \(g=2\) is therefore warranted.  Now, in the equation of motion of the total angular momentum, it is to be expected that the motion of the spins will enter directly in accordance with Eq. (\ref{ThomasEq}), since the total angular momentum itself will include the electron and positron spins directly. In this contribution the approximation that \(g=2\) will have no untoward effect beyond a slight error of the net motion of the spins, that is negligible compared to the effect of the Thomas precession in Eq. (\ref{ThomasEq}).  Additionally, the rate of change of orientation of the spin may enter also into the part of the motion of the total angular momentum that is due to the motion of the orbital angular momentum.  However it will be seen in Section VII that the motion of the spins enters into the equation of motion of the orbital angular momentum only at order \(\beta^2\) compared to the instantaneous orientations of the spins.  Therefore, in the equation of motion of the total angular momentum that contains all terms to the order of the Thomas precession contribution, the contribution of the rate of change of the spins in the motion of the orbital angular momentum is negligible.  (This is shown explicitly in Section VIIB, Eq. (\ref{ProofInsignificant}).)  Therefore there will be no further error introduced by taking the further approximation of the electron and positron \(g\)-factors to be exactly two, as opposed to their exact values of about one part in \(10^{3}\) larger, other than changing the rate of motion of the spins slightly.  This error will have no effect on the central result of the present analysis, that the total angular momentum is not generally a constant of the motion, as a consequence of the Thomas precession.  Later it will be straightforward if desired to repeat the analysis with more precise \(g\)-factor values, to confirm there is no effect on the central conclusion, but for the initial analysis it is a significant simplification to let \(g = 2\).  

As justified by the considerations above, neglecting terms of relative order of \(\beta^2\), and approximating \(\gamma\) as unity and the \(g\)-factor as two results in the following equation of motion for the electron spin \(\mbox{\boldmath$s$}_e\) in the quasiclassical Positronium atom, where \(\mbox{\boldmath$E$}\) and \(\mbox{\boldmath$B$}\) are the electric field and magnetic field of the positron:

\begin{equation}
\frac{d\mbox{\boldmath$s$}_e}{dt} = \frac{q_e}{mc}\mbox{\boldmath$s$}_e \times \left[\mbox{\boldmath$B$}-  \left(\frac{1}{2}\right)\mbox{\boldmath$\beta$}_e \times \mbox{\boldmath$E$}\right],
\label{ThomasEqRed}
\end{equation}

where the generic \(q\) for the charge has been replaced with an electron-specific quantity \(q_e\) so that the polarities of the electron and positron may be more easily accounted for.  That is, \(q_e \equiv -e\) where \(e\) is the fundamental charge magnitude, and  \(q_p \equiv e\), where \(q_p\) is the positron charge.  Here and henceforth, \(m=m_e=m_p\) is the electron or positron mass.

The magnetic field at the electron is due to the intrinsic magnetic moment of the positron and to the motion of the positron charge.  This can be expressed as \(\mbox{\boldmath$B$} = \mbox{\boldmath$B$}_\mu + \mbox{\boldmath$B$}_v\).  With the intrinsic magnetic moments modeled as classical current loop dipoles as described above, the magnetic field component due to the positron intrinsic magnetic moment is

\begin{equation}
\mbox{\boldmath$B$}_\mu = \frac{3\mbox{\boldmath$n$} \left(
\mbox{\boldmath$n$} \cdot \mbox{\boldmath$\mu$}_p \right)  - \mbox{\boldmath$\mu$}_p}{r^3},
\label{Bdue2mup}
\end{equation}

where \(\mbox{\boldmath$\mu$}_p = q_p\mbox{\boldmath$s$}_p/mc \) is the positron intrinsic magnetic moment, and \(\mbox{\boldmath$n$} = \mbox{\boldmath$r$}/r\) is a unit vector in the direction from the positron to the electron.  

The magnetic field at the electron due to the orbital motion of the positron charge may be approximated as

\begin{equation}
\mbox{\boldmath$B$}_v = \frac{q_p}{c r^3}\mbox{\boldmath$v$}_p \times \mbox{\boldmath$r$} = -\frac{2q_p}{c r^3}\mbox{\boldmath$v$}_p \times \mbox{\boldmath$r$}_p = \frac{2q_p}{m c r^3}\mbox{\boldmath$L$}_p, 
\label{Bdue2poscharge}
\end{equation}

with \(\mbox{\boldmath$r$} \equiv \mbox{\boldmath$r$}_e - \mbox{\boldmath$r$}_p  = -2 \mbox{\boldmath$r$}_p = 2 \mbox{\boldmath$r$}_e \), and \( \mbox{\boldmath$L$}_p  = \mbox{\boldmath$r$}_p \times m \mbox{\boldmath$v$}_p \) is the positron kinetic orbital angular momentum. (This expression for the magnetic field due to a translating point charge may be obtained from the exact Li\'enard-Wiechert magnetic field of a moving point charge by neglecting the acceleration (radiation) field, neglecting delay, and approximating \(\gamma\) as unity.  As shown in Appendix B2, this expression represents the exact magnetic field of the moving positron charge accurately to the first order in \(\beta\).  Therefore by the same reasoning as above, this will introduce only an order \(\beta^3\) error in the motion of the total angular momentum.)  Noting \( \mbox{\boldmath$L$} \equiv \mbox{\boldmath$L$}_e + \mbox{\boldmath$L$}_p  = 2 \mbox{\boldmath$L$}_p \) obtains that

\begin{equation}
\mbox{\boldmath$B$}_v = \frac{q_p}{m c r^3}\mbox{\boldmath$L$}.
\label{Bdue2poscharge}
\end{equation}

The electric field at the electron consists of the fieled due to the positron's electric charge plus an additional contribution due to the motion of the positron intrinsic magnetic moment.  The electric field due to translational and orientational motions of the intrinsic magnetic dipoles is considered in detail below.  The latter total magnitude is of order \(\beta^2\) compared to the Coulomb field.  In general, the term in Eq. (\ref{ThomasEq}) involving the electric field is of the same order as the term involving the magnetic field directly, and therefore by the considerations stated above, that relative order \(\beta^2\) terms in the motion of the spin are irrelevant to the present analysis, it must be concluded that the contribution to the motion of the spin axes of the electric field due to motion of the intrinsic magnetic dipoles is negligible to the present analysis of the motion of the total angular momentum.     

The Coulomb field of the positron at the electron is

\begin{equation}
\mbox{\boldmath$E$} = \frac{q_p\mbox{\boldmath$r$}}{r^3}.
\label{PositronCoulombField}
\end{equation}

It is shown in the appendix that the description of the electric field as the instantaneous Coulomb field deviatiates only by an order \(\beta^4\) error from the exact delayed Li\'enard-Wiechert velocity electric field of the electron or positron in the Bohr model ground state.  It is also shown, the Li\'enard-Wiechert acceleration electric field term is of order \(\beta^3\) here and so is also negligible.  Therefore the electric field will be represented in the equations of motion of the spins as simply the undelayed Coulomb field as represented by Eq. (\ref{PositronCoulombField}), and the equivalent expression for the electron Coulomb field in the equation of motion of the positron spin.

Based on the above considerations the equation of motion of the electron spin vector of Eq. (\ref{ThomasEqRed}) becomes

\begin{widetext}

\begin{equation}
\frac{d\mbox{\boldmath$s$}_e}{dt} = \frac{q_e}{m c}\mbox{\boldmath$s$}_e \times \left[\left(\frac{3\mbox{\boldmath$n$} \left(
\mbox{\boldmath$n$} \cdot \mbox{\boldmath$\mu$}_p   \right)  -
\mbox{\boldmath$\mu$}_p}{r^3} + \frac{q_p}{m c r^3}\mbox{\boldmath$L$}\right) -  \left(\frac{1}{2}\right)\mbox{\boldmath$\beta$}_e \times \left(\frac{q_p\mbox{\boldmath$r$}}{r^3}\right)\right],
\label{ThomasEqRed1}
\end{equation}

where \(\mbox{\boldmath$\beta$}_e = \mbox{\boldmath$v$}_e/c \) is the electron velocity in the laboratory frame, and \(\mbox{\boldmath$r$} = 2\mbox{\boldmath$r$}_e \) is the vector from the positron to the electron.

The total kinetic orbital angular momentum \(\mbox{\boldmath$L$}\) is twice the electron kinetic orbital angular momentum \(\mbox{\boldmath$L$}_e = \mbox{\boldmath$r$}_e \times m c \mbox{\boldmath$\beta$}_e = -m c \mbox{\boldmath$\beta$}_e \times  \mbox{\boldmath$r$}_e \), and using again that \(\mbox{\boldmath$r$} = 2 \mbox{\boldmath$r$}_e \), so that \(m c\mbox{\boldmath$\beta$}_e \times  \mbox{\boldmath$r$} = 2 m c \mbox{\boldmath$\beta$}_e \times \mbox{\boldmath$r$}_e = -2\mbox{\boldmath$L$}_e = -\mbox{\boldmath$L$}\) obtains

\begin{equation}
\frac{d\mbox{\boldmath$s$}_e}{dt} \equiv \dot{\mbox{\boldmath$s$}}_e = \frac{q_e}{m c}\mbox{\boldmath$s$}_e \times \left[\left( \frac{q_p}{m c}\left( \frac{3\mbox{\boldmath$n$} \left(
\mbox{\boldmath$n$} \cdot \mbox{\boldmath$s$}_p   \right)  -
\mbox{\boldmath$s$}_p}{r^3} \right) + \frac{q_p}{m c r^3}\mbox{\boldmath$L$} \right) + \left(\frac{1}{2}\right) \left(\frac{q_p}{m c r^3}\right)\mbox{\boldmath$L$} \right].
\label{ThomasEqRed1}
\end{equation}

\end{widetext}

Replacing \(q_p\) with its value of the fundamental charge \(+e\), and \(q_e\) by \(-e\), and collecting similar factors obtains finally 

\begin{equation}
\dot{\mbox{\boldmath$s$}}_e = -\frac{e^2}{m^2 c^2 r^3}\mbox{\boldmath$s$}_e \times \left[ 3\mbox{\boldmath$n$} \left(
\mbox{\boldmath$n$} \cdot \mbox{\boldmath$s$}_p\right)   -
\mbox{\boldmath$s$}_p + \left(\frac{3}{2}\right) \mbox{\boldmath$L$} \right].
\label{ThomasEqRedElectron}
\end{equation}

It is obtained similarly for the positron that 

\begin{equation}
\dot{\mbox{\boldmath$s$}}_p = -\frac{e^2}{m^2 c^2 r^3}\mbox{\boldmath$s$}_p \times \left[ 3\mbox{\boldmath$n$} \left(
\mbox{\boldmath$n$} \cdot \mbox{\boldmath$s$}_e\right)   -
\mbox{\boldmath$s$}_e + \left(\frac{3}{2}\right) \mbox{\boldmath$L$} \right].
\label{ThomasEqRedPositron}
\end{equation}

It is also of interest to make note of what is the contribution of the Thomas precession to Equations (\ref{ThomasEqRedElectron}) and (\ref{ThomasEqRedPositron}).  The contribution of the Thomas precession to Equation (\ref{ThomasEqRedElectron}) is

\begin{equation}
\dot{\mbox{\boldmath$s$}}_{\text{TP},e} = \frac{e^2}{m^2 c^2 r^3}\mbox{\boldmath$s$}_e \times \left[\left(\frac{1}{2}\right) \mbox{\boldmath$L$} \right].
\label{ThomasEqRedElectronTPContrib}
\end{equation}

The contribution of the Thomas precession to Equation (\ref{ThomasEqRedPositron}) is

\begin{equation}
\dot{\mbox{\boldmath$s$}}_{\text{TP},p} = \frac{e^2}{m^2 c^2 r^3}\mbox{\boldmath$s$}_p \times \left[\left(\frac{1}{2}\right) \mbox{\boldmath$L$} \right].
\label{ThomasEqRedPositronTPContrib}
\end{equation}

In Section VIII, Equations (\ref{ThomasEqRedElectron}) and (\ref{ThomasEqRedPositron}) are used  to determine the contributions of the electron and positron spin motions to the total angular momentum.  An upper limit on the magnitude of the time rate of change of the spin vectors in the Bohr model ground state of Positronium is determined in Appendix D.  This limit value is used to show the neglibility of certain effects and so help to simplify the analysis.

Before proceeding, and although it won't be pursued further herein, it is perhaps worthwhile to make an observation about the nature of the motion of the spins in the quasiclassical Positronium atom model, as embodied in Eqs. (\ref{ThomasEqRedElectron}) and (\ref{ThomasEqRedPositron}).  It is interesting that there is an additional component in the motion of the spin vectors that is a much more rapid motion than the regular precessional motions, that are essentially Larmor precession when viewed in the particle rest frame. Specifically, the first term in the square brackets of either  equation varies at the orbital frequency, which, in the present application, is about eight decimal orders of magnitude larger than the Larmor precession frequency of the spin-orbit coupling, represented by the third term in the square brackets (and as modified by the Thoms precession when viewed in the laboratory frame).  Since radiation intensity is proportional to the square of the magnitude of dipole acceleration, and for harmonic motion the acceleration is proportional to the square of the frequency, this represents a very great increase in radiativity compared to the radiativity associated with the Larmor precession of the spin-orbit coupling.  Also, as is well known and will be treated explicitly in Section VI and elsewhere herein, the magnetic-dipole-carrying particles will acquire electric dipole moments due to their motion, and so acceleration of the spin vectors will potentially cause electric dipole radiation as well as magnetic dipole radiation.  Such radiative mechanisms don't seem to have been considered previously and warrant further examination to determine the degree of correspondence, if any, with observation.  Dipole radiation due to motion of point charges is after all a good match for observation in the limit of large quantum numbers, as expected in accordance with the Bohr correspondence principle. The analysis here hints that there may be additional classical phenomena that can extend the correspondence further into the quantum domain.

\section{Equation of Motion of the Kinetic Orbital Angular Momentum}

In this section the equations governing the time rates of change of the electron and positron kinetic orbital angular momentum vectors are developed. As previously outlined in Section IID, the magnitude of the spin angular momenta and the contribution of the Thomas precession for the individual particles determines the accuracy needed for the hidden and field angular momentum contributions to the total angular momentum.  In determining the effect of the Thomas precession, there is no need to be concerned with terms that are orders of magnitude in \(\beta\) smaller than its own contribution. Also, although the kinetic orbital angular momentum is distinct from the hidden angular momentum, the presence of hidden momentum enters directly into the equation of motion of the kinetic orbital angular momentum, in a significant fashion. 

In the quasiclassical Positronium model viewed in the laboratory frame, the electron and positron are orbiting each other influenced by electric and magnetic fields due to the other-particle electric charge and dipole moment.  The predominant influence is the Coulomb attractive force between the particles, but non-Coulomb forces are present that cause deviation from Keplerian motion.  Herein as noted it is assumed that the orbits are circular, but it cannot be expected that an exact circular orbit will result dynamically if the forces are not purely radial and conservative. However, assuming an initial condition of perfect circularity, the deviations from circularity on the time scale of an orbit can be shown to be small enough so as to be insignificant to the present analysis.  

The approach to determining the motion of the kinetic orbital angular momentum is as follows.  First, the basic equation of motion of the kinetic orbital angular momentum is developed in terms of the kinetic linear momentum that in the nonrelativistic regime is simply the mass times the particle velocity, separately for each particle.  This reduces the problem of determining the motion of the kinetic orbital angular momentum to one of determining the change of particle velocity in response to applied forces.  Then, the equations of translational motion of the particles are developed based on the applied forces acting in accordance with Newton's law of motion, where the momentum changing under the applied force is the total mechanical momentum that includes hidden momentum.  Finally, the equation of motion of the total kinetic orbital angular momentum is developed in terms of the total torques acting on the individual particle orbits.

\subsection{Motion of the Kinetic Orbital Angular Momentum of Each Particle Separately}

The positron kinetic orbital angular momentum is defined as

\begin{equation}
\mbox{\boldmath$L$}_p \equiv \mbox{\boldmath$r$}_p \times \mbox{\boldmath$P$}_p = \mbox{\boldmath$r$}_p \times m\mbox{\boldmath$v$}_p, 
\label{KineticOrbitalAngularMomentum}
\end{equation}

where \(\mbox{\boldmath$P$}_p\) is the positron kinetic momentum given in the nonrelativistic limit as \(\mbox{\boldmath$P$}_p = m\mbox{\boldmath$v$}_p\). The motion of the positron kinetic orbital angular momentum is then given by

\begin{equation}
\dot{\mbox{\boldmath$L$}}_p = \dot{\mbox{\boldmath$r$}}_p \times m\mbox{\boldmath$v$}_p +  \mbox{\boldmath$r$}_p \times m\dot{\mbox{\boldmath$v$}}_p = \mbox{\boldmath$r$}_p \times m\dot{\mbox{\boldmath$v$}}_p,
\label{OrbitalAngularMomentum}
\end{equation}

with \(\mbox{\boldmath$v$}_p \equiv \dot{\mbox{\boldmath$r$}}_p\). 

Similarly, for the electron, \(\dot{\mbox{\boldmath$L$}}_e = \mbox{\boldmath$r$}_e \times m\dot{\mbox{\boldmath$v$}}_e\).

The equations for the motion of the kinetic orbital angular momentum of the individual particles show the need for evaluating the particles' acceleration.

\subsection{Hidden Linear Momentum and the Equation of Translational Motion of a Magnetic Dipole}

In general a force  \(\mbox{\boldmath$F$}\) on a classical rigid body is equal to its time rate of change of mechanical momentum, \(d\mbox{\boldmath$P$}_{\text{mech}}/dt\), where \(\mbox{\boldmath$P$}_{\text{mech}}\) is the mechanical momentum of the body.  In the nonrelativistic limit, and for a classical body without a magnetic moment, the mechanical momentum is the kinetic momentum \(\mbox{\boldmath$P$} = 
m\mbox{\boldmath$v$}\).  However, as described in the introduction, it is now generally recognized that if the body carries a net current in its rest frame, and if an electric field is present, then there is mechanical momentum in addition to the kinetic momentum. This momentum is nonvanishing in the rest frame of the current-carrying body and is often termed ``hidden'' momentum.  For a classical current loop magnetic dipole of moment \(\mbox{\boldmath$m$}\), in an electric field \(\mbox{\boldmath$E$}\), the hidden momentum is \(\mbox{\boldmath$P$}_{\text{hidden}} = \mbox{\boldmath$m$} \times \mbox{\boldmath$E$}/c\) \cite{Hnizdo1992,Jackson3rdEdoc2}.  The existence of hidden momentum implies that the equation of translational motion of a classical current-loop magnetic dipole must take into account the time rate of change of the hidden as well as the kinetic momentum of the body.  Omission of the hidden momentum from the equation of motion will lead to a nonphysical dynamical description of the interaction of electric monopoles and magnetic dipoles. For example, if the hidden momentum is not taken into account, it will appear that the force on a stationary magnetic dipole due to a moving electric monopole is not equal and opposite the force on the electric monopole due to the magnetic dipole.  Newton's law of action and reaction will thus be violated, if hidden momentum is disregarded.

There is also support in the literature \cite{Shockley68, MunozY1} for there being a hidden momentum associated with the intrinsically-magnetic electron, in the presence of an electric field external to the electron.  In any case, a quasiclassical analysis of particles with intrinsic magnetic moment will find Newton's law of action and reaction violated if hidden momentum is not included.  The problem of violation of the law of action and reaction led to the discovery of hidden momentum, and applies in the case of particles with intrinsic magnetic moments, when treated classically, just as it does in the case of classical current-carrying rigid bodies.  Violation of the law of action and reaction would seem to lead inevitably to impossibility of constructing a consistent dynamical description, and so cannot be considered physical.  Therefore for the present work it will be assumed that for a particle with intrinsic magnetic dipole moment \(\mbox{\boldmath$\mu$}\), the total mechanical momentum must include a hidden momentum \(\mbox{\boldmath$\mu$} \times \mbox{\boldmath$E$}/c\). The electron mechanical momentum is then

\begin{equation}
\mbox{\boldmath$P$}_{\text{mech},e} = \mbox{\boldmath$P$}_e + \mbox{\boldmath$P$}_{\text{hidden},e} = 
\mbox{\boldmath$P$}_e + \mbox{\boldmath$\mu$}_e \times \mbox{\boldmath$E$}(\mbox{\boldmath$r$}_e)/c. 
\label{MechanicalMomentum}
\end{equation}

The rate of change of the electron mechanical momentum is accordingly

\begin{equation}
\dot{\mbox{\boldmath$P$}}_{\text{mech},e} = 
\dot{\mbox{\boldmath$P$}}_e + \mbox{\boldmath$\mu$}_e \times \dot{\mbox{\boldmath$E$}}(\mbox{\boldmath$r$}_e)/c + \dot{\mbox{\boldmath$\mu$}}_e \times \mbox{\boldmath$E$}(\mbox{\boldmath$r$}_e)/c. 
\label{MechMomentumDot}
\end{equation}

Equating the rate of change of the mechanical momentum to the applied force obtains the rate of change of the electron kinetic momentum as

\begin{equation}
\dot{\mbox{\boldmath$P$}}_e  =  \mbox{\boldmath$F$}_e -
\mbox{\boldmath$\mu$}_e \times \dot{\mbox{\boldmath$E$}}(\mbox{\boldmath$r$}_e)/c - \dot{\mbox{\boldmath$\mu$}}_e \times \mbox{\boldmath$E$}(\mbox{\boldmath$r$}_e)/c, 
\label{MassTimesAccel}
\end{equation}

where \(\mbox{\boldmath$F$}_e\) is the force applied to the electron.  The latter will consist of the Lorentz force on the electron charge as well as forces due to field gradients acting on the electron intrinsic magnetic dipole moment, and on the electric dipole moment it aquires due to translational motion.

\subsection{Approximate Equation of Translational Motion}

The objective of the present analysis, as stated above, is to obtain an equation of motion of the total angular momentum of quasiclassical Positronium atom model that is accurate to order \(\beta^2\).  The angular momentum contribution of the time rate of change of the kinetic momentum as given by Eq. (\ref{MassTimesAccel}) is simply of the form \(\mbox{\boldmath$r$}_e \times \dot{\mbox{\boldmath$P$}}_e\); that is, each term in Eq. (\ref{MassTimesAccel}) will contribute proportionately to the orbital angular momentum.  Therefore, if any terms on the right hand side of Eq. (\ref{MassTimesAccel}) are smaller than \(\beta^2\) times the largest term, their corresponding terms in the equation of motion of the orbital angular momentum may be neglected.    

In order to assess the relative magnitudes of the three  terms on the right hand side of Eq. (\ref{MassTimesAccel}) contributing to the time rate of change of the kinetic momentum, for the electron and positron in the quasiclassical Positronium atom, it will be useful to approximate the electric field as simply the instantaneous Coulomb field of the charge.  The degree to which this is a valid approximation is considered in detail in Appendix B. In particular, the electric field contributions due to the translational motion of the electron intrinsic magnetic dipole, and its relativistically-acquired electric dipole moment are not negligible to the present analysis and will be taken into account.  However a small influence on a term that is already small may be negligble.

It is shown in Appendix B that the electric field at the electron due to the positron, in the quasiclassical Positronium model with circular orbit, accurate to order of \(\beta^2\), is

\begin{eqnarray}
\mbox{\boldmath$E$}(\mbox{\boldmath$r$}_e) = \frac{e\mbox{\boldmath$n$}}{r^2} +  \frac{1}{c r^3}\left[3\mbox{\boldmath$n$}(\mbox{\boldmath$v$}_p \times \mbox{\boldmath$\mu$}_p) \cdot \mbox{\boldmath$n$} - 2(\mbox{\boldmath$v$}_p \times \mbox{\boldmath$\mu$}_p)\right].
\label{ElectricFieldToOrderBetaSqd}
\end{eqnarray}

(This neglects the acceleration field, which is an order \(\beta^3\) correction, and deviation between the instantaneous Coulomb field and the exact delayed Lienard-Wiechert velocity electric field, which is an order \(\beta^4\) correction.  These errors are evaluated in the appendix.)

Preparatory to considering what terms are relevant in the equation of motion of the kinetic momentum with an approximation for the field, it must also be considered what errors may be introduced into the mechanical momentum itself (as given by Eq. (\ref{MechanicalMomentum})) by any field approximation.  This is essential because even though these errors are small in magnitude, they may be rapidly varying and so may contribute disproportionately to the equation of motion of the kinetic momentum.  However, since some of the terms are already varying at the orbital frequency, any terms of smaller magnitude will not become larger after time differentiation if they do not vary faster than the orbital frequency.  It is therefore assumed that there is no time variation faster than at the orbital rate in smaller-magnitude electric field terms in Eq. (\ref{MechanicalMomentum}).

In order to determine what terms of the electric field  and electric field time derivative should be retained when calculating the time variation of the kinetic momentum according to Eq. (\ref{MassTimesAccel}), it is useful to assess the relative magnitudes of the terms under the assumption that the electric field is simply the Coulomb field.  Since it is established that corrections to the Coulomb field and its first time derivative arise only at the order \(\beta^2\), these corrections may be negligible in an equation of motion of the momentum where contributions of terms involving the hidden momentum are small compared to the kinetic momentum.  Therefore Eq. (\ref{MassTimesAccel}) will be evaluated initially under the assumption of a purely Coulomb electric field, although later a more exact expression for the electric field will be used to determine the motion of the electron and positron orbits.

The first term on the right hand side of Eq. (\ref{MassTimesAccel}) is the translational force on the electron.  This force consists of the Lorentz force on the electron charge, plus field-gradient forces acting on the intrinsically-magnetic electron.  All non-Coulomb forces are evaluated in Section VI and shown to be of order \(\beta^2\) or smaller than the Coulomb force, in the Bohr model of Positronium ground state.  Therefore similar considerations as support approximating the electric field as the Coulomb field support approximating the translational force as the Coulomb electrostatic force, in determining the approximate relative magnitude of the terms on the right hand side of Eq. (\ref{MassTimesAccel}).

The Coulomb field of a particle of charge, \(q\), is

\begin{equation}
\mbox{\boldmath$E$}_{\text{Coulomb}} = 
\frac{q \mbox{\boldmath$r$}}{r^3}.  
\label{CoulombField}
\end{equation}

With the electric field approximated as the Coulomb field, and the translational force approximated as the electrostatic Coulomb force, the first term on the right hand side of Eq. (\ref{MassTimesAccel}) is then approximated as 

\begin{equation}
\mbox{\boldmath$F$}_e \approx e\mbox{\boldmath$E$} \approx
-\frac{e^2 \mbox{\boldmath$r$}}{r^3}.  
\label{T1mag1}
\end{equation}

The magnitude of the force on the electron is then given to the same order of approximation as

\begin{equation}
\left|\mbox{\boldmath$F$}_e\right| \approx 
\frac{e^2}{r^2}.  
\label{T1mag1}
\end{equation}

At the Positronium Bohr radius \(r_{\text{B}} = 2\hbar^2/me^2\) this becomes

\begin{equation}
\left|\mbox{\boldmath$F$}_e\right| \approx 
e^2 \frac{e^4 m^2}{4\hbar^4}  = \frac{e^6 m^2}{4\hbar^4}.
\label{T1mag1}
\end{equation}

The magnitude upper limit of the second term on the right hand side of Eq. (\ref{MassTimesAccel}), assuming a Coulomb electric field and a circular orbit, is

\begin{equation}
\left| \mbox{\boldmath$\mu$}_e \times \dot{\mbox{\boldmath$E$}}/c \right| \le \mu \frac{ev}{cr^3} = \frac{e^3\hbar}{\sqrt{2} m^{3/2} c^2 r^{7/2}}.
\label{T2mag1}
\end{equation}

where the Bohr magneton formula \(\mu_{\text{B}} = e \hbar/2 m\) and the result from Appendix A (Equation (\ref{eprelvel})) that the electron-positron relative velocity magnitude in the circular-orbit Rutherford model is \(v = e\sqrt{2/mr}\) have been used. At the Bohr radius this becomes

\begin{equation}
\left| \mbox{\boldmath$\mu$}_e \times \dot{\mbox{\boldmath$E$}}/c \right| \le \frac{1}{16}\left(\frac{e^{10} m^2}{c^2 \hbar^6}\right).
\label{T2mag1}
\end{equation}

The ratio of the second term on the right hand side of Eq. (\ref{MassTimesAccel}) magnitude to the translational force magnitude is thus bounded for the Bohr model ground state as

\begin{equation}
\frac{\left| \mbox{\boldmath$\mu$}_e \times \dot{\mbox{\boldmath$E$}}/c \right|} {\left|\mbox{\boldmath$F$}_e\right|} \le \frac{\alpha^2}{4},
\label{T2mag1}
\end{equation}

where \(\alpha = e^2/\hbar c\) is the fine structure constant. This shows that the second term is nominally an order \(\beta^2\) order correction to the change in mechanical momentum due to the Coulomb force alone.  Therefore corrections to this term that are of order \(\beta\) or smaller may be neglected in the equation of motion of the mechanical momentum that is accurate to order \(\beta^2\).

The magnitude upper limit of the third term on the right hand side of Eq. (\ref{MassTimesAccel}), assuming a Coulomb electric field and a circular orbit, is

\begin{equation}
\left| \dot{\mbox{\boldmath$\mu$}}_e \times \mbox{\boldmath$E$}/c \right| \le \frac{e\dot{\mu}}{cr^2} = \frac{e}{cr^2}\frac{e \dot{s}}{c m} = \frac{e^2}{c^2 r^2 m}\dot{s}.
\label{T3mag1}
\end{equation}

At the Bohr radius this becomes

\begin{equation}
\left| \dot{\mbox{\boldmath$\mu$}}_e \times \mbox{\boldmath$E$}/c \right| \le  \frac{e^2}{c^2  m}\left(\frac{2\hbar^2}{e^2 m} \right)^{-2} \dot{s} = \left(\frac{e^6 m}{4 c^2 \hbar^4} \right) \dot{s}.
\label{T3mag1}
\end{equation}

An upper bound for the magnitude of the time rate of change of the spin vector, for the quasiclassical positronium atom in the Bohr ground state, is developed in Appendix D (see Eq. (\ref{sdot_bound_appendix})), as

\begin{equation}
\left|\dot{\mbox{\boldmath$s$}}_e\right| \le \frac{7}{32} \frac{e^8 m}{\hbar^4 c^2}, 
\label{sdot_bound_copy1}
\end{equation}

so

\begin{equation}
\left| \dot{\mbox{\boldmath$\mu$}}_e \times \mbox{\boldmath$E$}/c \right| \le  \left(\frac{e^6 m}{4 c^2 \hbar^4} \right) \frac{7}{32} \frac{e^8 m}{\hbar^4 c^2} = \frac{7}{128} \frac{e^{14} m^2}{\hbar^8 c^4}. 
\label{T3mag1}
\end{equation}

The ratio of this bound to the first term magnitude is 

\begin{equation}
\frac{\left| \dot{\mbox{\boldmath$\mu$}}_e \times \mbox{\boldmath$E$}/c \right|} {\left|\mbox{\boldmath$F$}_e\right|} \le \frac{7}{128} \frac{e^{14} m^2}{\hbar^8 c^4}\left(\frac{e^6 m^2}{\hbar^4} \right)^{-1},
\label{T3mag1}
\end{equation}

or

\begin{equation}
\frac{\left| \dot{\mbox{\boldmath$\mu$}}_e \times \mbox{\boldmath$E$}/c \right|} {\left|\mbox{\boldmath$F$}_e\right|} \le \frac{7}{128} \frac{e^{8}}{\hbar^4 c^4} = \frac{7}{128} \alpha^4.
\label{T3mag1}
\end{equation}

The third term on the right hand side of Eq. (\ref{MassTimesAccel}), assuming a Coulomb electric field and a circular orbit, is thus at most a factor of \(\beta^4\) as large as the first term that is the force on the electron.  Therefore it may be considered negligible in the present analysis.

In Appendix C it is shown that non-Coulomb terms in the electric field contribute to the rate of change of the electric field only at the order of \(\beta^2\) or below compared to the time rate of change of the Coulomb electric field, under the assumptions of the electron intrinsic magnetic moment equivalent to that of a classical current loop.  Since the electric field time-derivative terms assuming the Coulomb field contributed to Eq. (\ref{MassTimesAccel}) only at order \(\beta^2\) or below to the time rate of change of the kinetic momentum, non-Coulomb field contributions will not be needed in constructing an approximation for Eq. (\ref{MassTimesAccel}) that is accurate to order \(\beta^2\).     

Neglecting the third term on the right hand side of Eq. (\ref{MassTimesAccel}), and approximating the electric field in the second term as the Coulomb field, based on the considerations as described above, results in the following equation of translational motion for the dipole-carrying electron in the field of the co-circular-orbiting positron:

\begin{equation}
\dot{\mbox{\boldmath$P$}}_e \equiv m\dot{\mbox{\boldmath$v$}}_e = \mbox{\boldmath$F$}_e - \frac{q_p}{c r^3} \mbox{\boldmath$\mu$}_e \times   \dot{\mbox{\boldmath$r$}}, 
\label{ElecTranslationalEq}
\end{equation}

where \(\mbox{\boldmath$F$}_e\) is the total translational force on the electron. 

It is important to recognize that although it was assumed that the electric field was the Coulomb field in calculating the relative magnitude of the three terms on the right hand side of Eq. (\ref{MassTimesAccel}), when the dynamical behavior is investigated the force on the electron charge will be based on the electric field approximation given by Eq. (\ref{ElectricFieldToOrderBetaSqd}), which includes non-Coulomb terms due to the positron intrinsic dipole moment, and is accurate to order \(\beta^2\).  Forces on the electron due to its dipolar nature and the presence of field gradients are significant to the analysis and will be included as well.   

\subsection{Motion of the Kinetic Orbital Angular Momentum}

The equation of motion of the electron kinetic orbital angular momentum around the center of mass based on Equation (\ref{ElecTranslationalEq}) for the electron translational motion is

\begin{equation}
\dot{\mbox{\boldmath$L$}}_e \equiv \mbox{\boldmath$r$}_e \times  m\dot{\mbox{\boldmath$v$}}_e = \mbox{\boldmath$r$}_e \times  \mbox{\boldmath$F$}_e - \frac{q_p}{c r^3} \mbox{\boldmath$r$}_e \times (\mbox{\boldmath$\mu$}_e \times   \dot{\mbox{\boldmath$r$}}). 
\label{EoTN}
\end{equation}

For the circular orbit and using the vector identity \(\mbox{\boldmath$a$} \times (\mbox{\boldmath$b$} \times \mbox{\boldmath$c$}) = (\mbox{\boldmath$c$} \cdot\mbox{\boldmath$a$})\mbox{\boldmath$b$} - (\mbox{\boldmath$b$} \cdot \mbox{\boldmath$a$})\mbox{\boldmath$c$} 
\), and with \(\mbox{\boldmath$v$} \equiv \dot{\mbox{\boldmath$r$}} = 2 \dot{\mbox{\boldmath$r$}}_e \), and noting that for the circular orbit \(\mbox{\boldmath$v$} \cdot \mbox{\boldmath$r$} \equiv 0 \), this can be rewritten in what will turn out to be a more convenient form as

\begin{equation}
\dot{\mbox{\boldmath$L$}}_e =  \mbox{\boldmath$\tau$}_e + \frac{e}{2 c r^3} (\mbox{\boldmath$r$} \cdot  \mbox{\boldmath$\mu$}_e) \mbox{\boldmath$v$}, 
\label{EoTN}
\end{equation}

where \(\mbox{\boldmath$\tau$}_e \equiv \mbox{\boldmath$r$}_e \times  \mbox{\boldmath$F$}_e\).  For the positron it is similarly obtained that

\begin{equation}
\dot{\mbox{\boldmath$L$}}_p =  \mbox{\boldmath$\tau$}_p - \frac{e}{2 c r^3} (\mbox{\boldmath$r$} \cdot  \mbox{\boldmath$\mu$}_p) \mbox{\boldmath$v$}. 
\label{EoTN}
\end{equation}

The equation of motion of the total kinetic orbital angular momentum, \(\mbox{\boldmath$L$} \equiv \mbox{\boldmath$L$}_e + \mbox{\boldmath$L$}_p \), is then

\begin{equation}
\dot{\mbox{\boldmath$L$}} = \dot{\mbox{\boldmath$L$}}_e + \dot{\mbox{\boldmath$L$}}_p = \mbox{\boldmath$\tau$}_e +  \mbox{\boldmath$\tau$}_p + \frac{e}{2 c r^3} (\mbox{\boldmath$r$} \cdot  (\mbox{\boldmath$\mu$}_e - \mbox{\boldmath$\mu$}_p))  \mbox{\boldmath$v$}. 
\label{EoTN}
\end{equation}

In terms of spin vectors rather than intrinsic magnetic moments, with electron and positron g-factors \(g_e = g_p = 2\), and in terms of the total torque,

\begin{equation}
\dot{\mbox{\boldmath$L$}} = \mbox{\boldmath$\tau$} - \frac{e^2}{2 m c^2 r^3} (\mbox{\boldmath$r$} \cdot  (\mbox{\boldmath$s$}_e + \mbox{\boldmath$s$}_p))  \mbox{\boldmath$v$}, 
\label{Ldot_final}
\end{equation}

where \(\mbox{\boldmath$\tau$} \equiv \mbox{\boldmath$\tau$}_e + \mbox{\boldmath$\tau$}_p\) is the total of the separate torques on the electron and positron orbits. 

With the general equation of motion of the kinetic orbital angular momentum in hand in Eq. (\ref{Ldot_final}), it remains to determine the total torque \(\mbox{\boldmath$\tau$}\), in order to evaluate the contribution of the kinetic orbital angular momentum to the total orbital angular moment, and in turn to the total angular momentum. 

\section{Inventory of Translational Forces on the Particles and Resultant Torques on the Orbit}

Preparatory to evaluating the rate of change of the kinetic orbital angular momentum using Eq. (\ref{Ldot_final}), it will be useful to take inventory of all of the translational forces acting on the particles, as well as which of them will yield torques on the orbit.  This will provide a road map for evaluating the total torque as needed according to Eq. (\ref{Ldot_final}).  

\subsection{Inventory of Translational Forces on the Particles}

The laboratory-frame translational forces on the particles may be enumerated as follows:

1) Coulomb force.

2) Biot-Savart force on the charges transiting the magnetic field (consisting of the intrinsic magnetic field of the other particle and due to the motion of the other particle's charge).  The former force will be referred to as \(\mbox{\boldmath$F$}_{\text{Biot-Savart 1}}\).
The latter force will be referred to as \(\mbox{\boldmath$F$}_{\text{Biot-Savart 2}}\).

3) Force on the intrinsic magnetic moments due to anisotropy of the magnetic field (the Stern-Gerlach force).  The Stern-Gerlach force due to anisotropy of the magnetic field due to the other charge motion will be referred to as \(\mbox{\boldmath$F$}_{\text{Stern-Gerlach 1}}\). The Stern-Gerlach force due to anisotropy of the other particle's intrinsic magnetic field will be referred to as \(\mbox{\boldmath$F$}_{\text{Stern-Gerlach 2}}\).

4) Force on the charges due to electric field induced by motion of the other particle magnetic moment. The force due to the electric field induced by translational motion of the other particle intrinsic magnetic moment will be referred to as \(\mbox{\boldmath$F$}_{\mbox{\boldmath$v$} \times \mbox{\boldmath$\mu$}}\).  This force will also include the influence of the electric field due to the motion-acquired  electric dipole moment of the other particle.  (The influence of the electric field induced by orientational motion of the other particle intrinsic magnetic moment will be shown to be insignificant to the present analysis).

5) Force on the motion-acquired electric dipole moments due to anisotropy of the electric field of the other charge and the other particle's motion-acquired electric dipole moment (the latter is shown negligible).  The non-negligible former force will be referred to as \(\mbox{\boldmath$F$}_{\text{Stern-Gerlach like}}\). 

It is important to recognize that the forces act to change the total mechanical momentum of the particle that includes both the momentum of the moving mass and the ``hidden'' mechanical momentum of the intrinsic magnetic moment in the electric field present at each particle's location in the Positronium atom.

\subsection{Summary of Torques on the Orbit}

The forces enumerated immediately above will lead to torques on the particle orbits if they have non-radial components with respect to the center of mass.

The Coulomb force is exactly radial in the approximation of the present analysis (which is shown in Appendix B to be valid to order \(\beta^3\)).  Therefore it does not contribute any torque on either the electron or positron orbits.  All of the other forces enumerated above may generate torques separately on the electron and positron orbits.  It is desirable to identify them separately by particle furthermore because they are added on the one hand in calculating the motion of the orbital angular momentum, and differenced on the other when calculating the motion of the orbital magnetic moment, as will be needed in a planned sequel.

\section{Calculation of the Torques on the Orbits}

In this section the individual torques on the electron and positron orbits as identified above are explicitly calculated. (The reader not immediately interested in the details of the calculation of the torques may skip to Section VII, where the results of the torque calculations are summarized prior to their application in determining the equation of motion of the total orbital angular momentum.)

\subsection{Torque on Orbit Due to Biot-Savart Force in Magnetic Field due to Motion of the Other Charge}

The laboratory-frame magnetic field at the positron due to the motion of the electron may be written as

\begin{equation}
\mbox{\boldmath$B$} =  \frac{q_e}{cr^3} \left(\dot{\mbox{\boldmath$r$}}_e \times (-\mbox{\boldmath$r$}) \right). 
\label{tottorqueorb3}
\end{equation}

where the minus sign of \((-\mbox{\boldmath$r$})\) accounts for that \(\mbox{\boldmath$r$}\) is defined as the displacement from the positron to the electron, rather than towards the positron as the field point as needed here.

The Biot-Savart force on the positron traversing the magnetic field is 

\begin{equation}
\mbox{\boldmath$F$}_{\text{Biot-Savart 1,}p} \equiv   \frac{q_p\dot{\mbox{\boldmath$r$}}_p}{c} \times \left(\frac{q_e}{cr^3} \left(\dot{\mbox{\boldmath$r$}}_e \times (-\mbox{\boldmath$r$}) \right)\right). 
\label{tottorqueorb3}
\end{equation}

The torque on the positron orbit due to this force is 

\begin{equation}
\mbox{\boldmath$\tau$}_{\text{Biot-Savart 1,}p}  =  -\mbox{\boldmath$r$}_p \times \left[ \frac{q_p\dot{\mbox{\boldmath$r$}}_p}{c} \times \left(\frac{q_e}{cr^3} \left(\dot{\mbox{\boldmath$r$}}_e \times \mbox{\boldmath$r$} \right)\right)\right], 
\label{tottorqueorb3}
\end{equation}

or


\begin{equation}
\mbox{\boldmath$\tau$}_{\text{Biot-Savart 1,}p}  =  -\mbox{\boldmath$r$}_p \times \left[ \dot{\mbox{\boldmath$r$}}_e  \left(\dot{\mbox{\boldmath$r$}}_p \cdot \mbox{\boldmath$r$} \right) -\mbox{\boldmath$r$} \left(\dot{\mbox{\boldmath$r$}}_p \cdot \dot{\mbox{\boldmath$r$}}_e \right)   \right] \left( \frac{q_p q_e}{c^2 r^3} \right).
 \label{tottorqueorb3}
\end{equation}


Since \(\mbox{\boldmath$r$}_p = -\mbox{\boldmath$r$}/2\), the cross product of \(\mbox{\boldmath$r$}_p\) with \(\mbox{\boldmath$r$}\) vanishes, leaving

\begin{equation}
\mbox{\boldmath$\tau$}_{\text{Biot-Savart 1,}p}  =  -\mbox{\boldmath$r$}_p \times \left[ \dot{\mbox{\boldmath$r$}}_e \left(\dot{\mbox{\boldmath$r$}}_p \cdot \mbox{\boldmath$r$} \right)  \right]\left( \frac{q_p q_e}{c^2 r^3} \right).
\label{tottorqueorb3}
\end{equation}

This torque vanishes for the circular orbit, due to perpendicularity of the radius and the velocity vectors, as does the similar torque on the electron orbit, and so need not be considered further herein.  

\subsection{Torque on Orbit due to Biot-Savart Force on Charge Motion through Intrinsic Magnetic Fields}

The force on the positron charge due to traversing the  electron intrinsic magnetic field is 

\begin{equation}
\mbox{\boldmath$F$}_{\text{Biot-Savart 2,}p} = 
\frac{e}{c}\mbox{\boldmath$v$}_p \times \frac{3\mbox{\boldmath$n$}
\left( \mbox{\boldmath$n$} \cdot \mbox{\boldmath$\mu$}_e   \right) -
\mbox{\boldmath$\mu$}_e}{r^3},
\label{bs1fonp}
\end{equation}

with \( \mbox{\boldmath$n$} = -\mbox{\boldmath$r$}/ r \) here, and \( \mbox{\boldmath$v$}_p\) the proton velocity as measured in the electron rest frame.

The torque on the positron orbit due to the electron intrinsic magnetic moment is then

\begin{equation}
\mbox{\boldmath$\tau$} = \mbox{\boldmath$r$}_p \times
\mbox{\boldmath$F$}_p = \mbox{\boldmath$r$}_p \times \left(
\frac{e}{c}\mbox{\boldmath$v$}_p \times \frac{3\mbox{\boldmath$n$}
\left( \mbox{\boldmath$n$} \cdot \mbox{\boldmath$\mu$}_e   \right) -
\mbox{\boldmath$\mu$}_e}{r^3} \right).
\label{bs1top1}
\end{equation}

The vector triple product involving \(\mbox{\boldmath$\mu$}_e\) can be expanded as

\begin{equation}
\mbox{\boldmath$r$}_p \times (\mbox{\boldmath$v$}_p \times \mbox{\boldmath$\mu$}_e) = (\mbox{\boldmath$r$}_p \cdot\mbox{\boldmath$\mu$}_e)\mbox{\boldmath$v$}_p - (\mbox{\boldmath$r$}_p \cdot \mbox{\boldmath$v$}_p)\mbox{\boldmath$\mu$}_e 
= (\mbox{\boldmath$r$}_p \cdot\mbox{\boldmath$\mu$}_e)\mbox{\boldmath$v$}_p,
\label{vtpident2}
\end{equation}

since the velocity and position vectors are orthogonal for the circular orbit so that the second term in the center vanishes.  Similarly the vector triple product of Eq. (\ref{bs1top1}) involving \(  \mbox{\boldmath$n$} \)  yields that \( \mbox{\boldmath$r$}_p \times (\mbox{\boldmath$v$}_p \times \mbox{\boldmath$n$}) = r_p \mbox{\boldmath$v$}_p \) and so Eq. (\ref{bs1top1}) becomes

\begin{equation}
\mbox{\boldmath$\tau$} =  \frac{e}{cr^3} \left[3\left( \mbox{\boldmath$n$} \cdot \mbox{\boldmath$\mu$}_e   \right) r_p \mbox{\boldmath$v$}_p
-(\mbox{\boldmath$r$}_p \cdot \mbox{\boldmath$\mu$}_e) \mbox{\boldmath$v$}_p \right], 
\label{torqueporbdtemm3}
\end{equation}

or, 

\begin{equation}
\mbox{\boldmath$\tau$} \equiv \mbox{\boldmath$\tau$}_{\text{Biot-Savart 2},p} = \frac{e}{cr^3} \left[2(\mbox{\boldmath$r$}_p \cdot \mbox{\boldmath$\mu$}_e) \mbox{\boldmath$v$}_p \right]. 
\label{torqueporbdtemm3}
\end{equation}

Recognizing that \(\mbox{\boldmath$r$} = -2\mbox{\boldmath$r$}_p \) and \(\mbox{\boldmath$v$} = -2\mbox{\boldmath$v$}_p\), and replacing the electron intrinsic magnetic moment with its equivalent in terms of intrinsic spin, obtains the final form for the torque on the positron orbit due to the Biot-Savart force in the electron intrinsic magnetic field as 

\begin{equation}
\mbox{\boldmath$\tau$}_{\text{Biot-Savart 2},p} = -\frac{e^2}{2 m c^2 r^3} \mbox{\boldmath$v$} \left[ \mbox{\boldmath$r$} \cdot \mbox{\boldmath$s$}_e\right]. 
\label{BS1torquepfinal}
\end{equation}

The torque on the electron orbit due to the Biot-Savart force in the positron intrinsic magnetic field is found similarly to be 

\begin{equation}
\mbox{\boldmath$\tau$}_{\text{Biot-Savart 2},e} = -\frac{e^2}{2 m c^2 r^3} \mbox{\boldmath$v$} \left[ \mbox{\boldmath$r$} \cdot \mbox{\boldmath$s$}_p\right]. 
\label{HzizdoY91_Eq1a}
\end{equation}

Since the other type of Biot-Savart force-derived torque vanishes, the total torque on the orbit due to Biot-Savart forces is thus

\begin{equation}
\mbox{\boldmath$\tau$}_{\text{Biot-Savart}} = -\frac{e^2}{2 m c^2 r^3} \mbox{\boldmath$v$} \left[ \mbox{\boldmath$r$} \cdot (\mbox{\boldmath$s$}_e +\mbox{\boldmath$s$}_p)\right]. 
\label{TorqueBiotSavart}
\end{equation}

\subsection{Torque on Orbit Due to Stern-Gerlach Force in Magnetic Field due to Motion of Positron Charge}

The force on the electron intrinsic magnetic moment due to the motion of the positron charge is given (in the laboratory frame) by

\begin{equation}
\nabla(\mbox{\boldmath$\mu$}_e\cdot\mbox{\boldmath$B$}) \equiv \mbox{\boldmath$F$}_{\text{Stern-Gerlach 1,}e} = \nabla\left(\frac{q_p}{cr^3}\left(\mbox{\boldmath$\mu$}_e\cdot \left(
\mbox{\boldmath$v$}_p \times \mbox{\boldmath$r$}  \right)  \right) \right),
\label{Expandt1}
\end{equation}

which evaluates to

\begin{equation}
\mbox{\boldmath$F$}_{\text{Stern-Gerlach 1,}e} =  \frac{q_p}{c r^3}\left[\mbox{\boldmath$\mu$}_e\times 
\mbox{\boldmath$v$}_p  - 3\mbox{\boldmath$\mu$}_e\cdot \left(
\mbox{\boldmath$v$}_p \times \mbox{\boldmath$n$}  \right)\mbox{\boldmath$n$} \right].
\label{Expandt1a}
\end{equation}

In terms of the electron-positron relative velocity, \(\mbox{\boldmath$v$}= -2\mbox{\boldmath$v$}_p\), this becomes

\begin{equation}
\mbox{\boldmath$F$}_{\text{Stern-Gerlach 1,}e} =  -\frac{q_p}{2c r^3}\left[\mbox{\boldmath$\mu$}_e\times 
\mbox{\boldmath$v$}  - 3\mbox{\boldmath$\mu$}_e\cdot \left(
\mbox{\boldmath$v$} \times \mbox{\boldmath$n$}  \right)\mbox{\boldmath$n$} \right].
\label{Expandt1a}
\end{equation}

The torque on the electron orbit due to the Stern-Gerlach force on the electron intrinsic magnetic moment, in the magnetic field due to the motion of positron charge is then 

\begin{equation}
\mbox{\boldmath$\tau$}_{\text{Stern-Gerlach 1,}e}  \equiv \mbox{\boldmath$r$}_e\times\left[\nabla(\mbox{\boldmath$\mu$}_e\cdot\mbox{\boldmath$B$})\right] = \mbox{\boldmath$r$}_e\times\left[  -\frac{q_p}{2c r^3}\left[\mbox{\boldmath$\mu$}_e\times 
\mbox{\boldmath$v$}  \right] \right]
\label{Expandt1}
\end{equation}

The vector triple product can be expanded as \(\mbox{\boldmath$r$}_e \times (\mbox{\boldmath$v$} \times \mbox{\boldmath$\mu$}_e) = (\mbox{\boldmath$r$}_e \cdot\mbox{\boldmath$\mu$}_e)\mbox{\boldmath$v$} - (\mbox{\boldmath$r$}_e \cdot \mbox{\boldmath$v$})\mbox{\boldmath$\mu$}_e 
= r(\mbox{\boldmath$n$} \cdot\mbox{\boldmath$\mu$}_e)\mbox{\boldmath$v$}/2\),  
since \(  \mbox{\boldmath$n$} = \mbox{\boldmath$r$} / r = 2\mbox{\boldmath$r$}_e / r\), and recognizing that the proton velocity and position vectors are orthogonal for the circular orbit so that the second term in the center vanishes.  It is thus obtained that 

\begin{equation}
\mbox{\boldmath$\tau$}_{\text{Stern-Gerlach 1,}e} = 
\frac{q_p}{4 c r^2}
(\mbox{\boldmath$n$} \cdot \mbox{\boldmath$\mu$}_e) \mbox{\boldmath$v$}. 
\label{torqueporbdtemm3}
\end{equation}

The similar torque on the positron orbit is 

\begin{equation}
\mbox{\boldmath$\tau$}_{\text{Stern-Gerlach 1,}p} \equiv \mbox{\boldmath$r$}_p\times\left[\nabla(\mbox{\boldmath$\mu$}_p\cdot\mbox{\boldmath$B$})\right] = 
\frac{q_e}{4 c r^2}
(\mbox{\boldmath$n$} \cdot \mbox{\boldmath$\mu$}_p) \mbox{\boldmath$v$} .
\label{torqueporbdtemm3}
\end{equation}

With \(\mbox{\boldmath$r$}_p = -\mbox{\boldmath$r$}_e \), the total torque on the orbit due to the Stern-Gerlach 1 force is thus

\begin{equation}
\mbox{\boldmath$\tau$}_{\text{Stern-Gerlach 1}} = 
\frac{e}{4 c r^2}
\left( \mbox{\boldmath$n$} \cdot \left(\mbox{\boldmath$\mu$}_e - \mbox{\boldmath$\mu$}_p\right) \right) \mbox{\boldmath$v$}. 
\label{torqueporbdtemm3}
\end{equation}

This is rewritten  in terms of intrinsic spins rather than intrinsic magnetic moments, and replacing  \(\mbox{\boldmath$n$}\) with \(\mbox{\boldmath$r$}/r\), as

\begin{equation}
\mbox{\boldmath$\tau$}_{\text{Stern-Gerlach 1}} = 
-\frac{e^2}{4 m c^2 r^3}
\left( \mbox{\boldmath$r$}  \cdot \left(\mbox{\boldmath$s$}_e + \mbox{\boldmath$s$}_p\right) \right) \mbox{\boldmath$v$}. 
\label{TorqueSternGerlach1}
\end{equation}

\subsection{Torque on Orbit Due to Stern-Gerlach Force due to Intrinsic Magnetic Field of Other Particle}

The force on the electron due to anisotropy of the magnetic field due to the positron intrinsic magnetic moment is given by

\begin{equation}
\mbox{\boldmath$F$}_{\text{Stern-Gerlach 2,}e} \equiv \nabla(\mbox{\boldmath$\mu$}_e\cdot\mbox{\boldmath$B$}_{\mu_p}), 
\label{Expandt1}
\end{equation}

where

\begin{equation}
\mbox{\boldmath$B$}_{\mu_p}  = \frac{3\mbox{\boldmath$n$} \left(
\mbox{\boldmath$n$} \cdot \mbox{\boldmath$\mu$}_p   \right)  -
\mbox{\boldmath$\mu$}_p}{r^3}. 
\label{Expandt1}
\end{equation}

The force on the electron is then

\begin{equation}
\mbox{\boldmath$F$}_{\text{Stern-Gerlach 2,}e}  = \nabla\left(\mbox{\boldmath$\mu$}_e\cdot\left(\frac{3\mbox{\boldmath$n$} \left(
\mbox{\boldmath$n$} \cdot \mbox{\boldmath$\mu$}_p   \right)  -
\mbox{\boldmath$\mu$}_p}{r^3}\right)\right), 
\label{FSE2e1}
\end{equation}

which evaluates to 

\begin{widetext}

\begin{eqnarray}
\mbox{\boldmath$F$}_{\text{Stern-Gerlach 2,}e} = - \left(\frac{3\mbox{\boldmath$n$}}{r^4}\right)\left(\mbox{\boldmath$\mu$}_e \cdot \mbox{\boldmath$\mu$}_p\right)  -  3\left(\mbox{\boldmath$\mu$}_e\cdot \mbox{\boldmath$n$} \right)\left(\mbox{\boldmath$n$} \cdot \mbox{\boldmath$\mu$}_p\right)\left(\frac{3\mbox{\boldmath$n$}}{r^4}\right) + \nonumber \\ \left(\frac{3}{r^3}\right)\left(\mbox{\boldmath$\mu$}_e 
\cdot \mbox{\boldmath$n$} \right)\frac{1}{r}\left[\mbox{\boldmath$\mu$}_p - \mbox{\boldmath$n$} (\mbox{\boldmath$\mu$}_p\cdot\mbox{\boldmath$n$})\right] +    \left(\frac{3}{r^3}\right)\left(\mbox{\boldmath$n$} \cdot \mbox{\boldmath$\mu$}_p\right)\frac{1}{r}\left[\mbox{\boldmath$\mu$}_e - \mbox{\boldmath$n$} (\mbox{\boldmath$\mu$}_e\cdot\mbox{\boldmath$n$})\right].
\label{Expandt1}
\end{eqnarray}

The torque on the electron orbit in the laboratory frame is then given by

\begin{equation}
\mbox{\boldmath$\tau$}_{\text{Stern-Gerlach 2,}e} = \mbox{\boldmath$r$}_e\times\left[\nabla(\mbox{\boldmath$\mu$}_e\cdot\mbox{\boldmath$B$})\right]=\mbox{\boldmath$r$}_e\times\left[\mbox{\boldmath$\mu$}_p\left(\mbox{\boldmath$\mu$}_e 
\cdot \mbox{\boldmath$n$} \right) +    \mbox{\boldmath$\mu$}_e\left(\mbox{\boldmath$n$} \cdot \mbox{\boldmath$\mu$}_p\right)\right]\left(\frac{3}{r^4}\right).
\label{Expandt1}
\end{equation}

Similarly, for \(\mbox{\boldmath$B$}\) at the positron due to the electron intrinsic field,

\begin{equation}
\mbox{\boldmath$\tau$}_{\text{Stern-Gerlach 2,}p} =\mbox{\boldmath$r$}_p\times\left[\nabla(\mbox{\boldmath$\mu$}_p\cdot\mbox{\boldmath$B$})\right] = \mbox{\boldmath$r$}_p\times\left[ 
\mbox{\boldmath$\mu$}_e \left(\mbox{\boldmath$\mu$}_p\cdot(-\mbox{\boldmath$n$})\right) + \left(
(-\mbox{\boldmath$n$}) \cdot \mbox{\boldmath$\mu$}_e  \right)\mbox{\boldmath$\mu$}_p \right]\left(\frac{3}{r^4}\right),
\label{Expandt1}
\end{equation}

or, since for equal-mass particles \(\mbox{\boldmath$r$}_p = -\mbox{\boldmath$r$}_e\),

\begin{equation}
\mbox{\boldmath$\tau$}_{\text{Stern-Gerlach 2,}p} = \mbox{\boldmath$r$}_p\times\left[\nabla(\mbox{\boldmath$\mu$}_p\cdot\mbox{\boldmath$B$})\right] = \mbox{\boldmath$r$}_e\times\left[\mbox{\boldmath$\mu$}_e \left(\mbox{\boldmath$\mu$}_p\cdot\mbox{\boldmath$n$}\right) + \left(
\mbox{\boldmath$n$} \cdot \mbox{\boldmath$\mu$}_e\right)\mbox{\boldmath$\mu$}_p \right]\left(\frac{3}{r^4}\right).
\label{Expandt1}
\end{equation}

The total torqe on the orbit due to this category of force is then (with \(\mbox{\boldmath$r$} \equiv \mbox{\boldmath$r$}_e - \mbox{\boldmath$r$}_p =  2\mbox{\boldmath$r$}_e\))

\begin{equation}
\mbox{\boldmath$\tau$}_{\text{Stern-Gerlach 2}} = \mbox{\boldmath$r$}\times\left[ 
\mbox{\boldmath$\mu$}_p\left(\mbox{\boldmath$\mu$}_e\cdot\mbox{\boldmath$n$}\right) + \left(\mbox{\boldmath$n$} \cdot \mbox{\boldmath$\mu$}_p \right)\mbox{\boldmath$\mu$}_e \right]\left(\frac{3}{r^4}\right),
\label{Expandt1}
\end{equation}

or, in terms of the intrinsic spins, 

\begin{equation}
\mbox{\boldmath$\tau$}_{\text{Stern-Gerlach 2}} = - \mbox{\boldmath$r$}\times\left[ 
\mbox{\boldmath$s$}_p \left(\mbox{\boldmath$s$}_e\cdot\mbox{\boldmath$n$}\right) + \left(\mbox{\boldmath$n$} \cdot \mbox{\boldmath$s$}_p\right) \mbox{\boldmath$s$}_e \right] \frac{3e^2}{ m^2 c^2 r^4}.
\label{Expandt1}
\end{equation}

This can be put into a form that can be more directly used later as

\begin{equation}
\mbox{\boldmath$\tau$}_{\text{Stern-Gerlach 2}} = \frac{e^2}{ m^2 c^2 r^3} \left[ \mbox{\boldmath$s$}_p\times
3 \mbox{\boldmath$n$} \left(\mbox{\boldmath$s$}_e\cdot\mbox{\boldmath$n$}\right) + \mbox{\boldmath$s$}_e\times3\mbox{\boldmath$n$} \left(\mbox{\boldmath$n$} \cdot \mbox{\boldmath$s$}_p\right)  \right].
\label{TorqueSternGerlach2}
\end{equation}

\end{widetext}

\subsection{Torque on Orbit Due to Electric Force due to Motions of Intrinsic Magnetic Dipoles}

A moving magnetic dipole gives rise to electric field due to the time-varying magnetic field it causes.  The electric field due to the motion of each magnetic dipole-carrying particle acts on the other particle's charge.  Also, a translating magnetic dipole must acquire an electric dipole moment \cite{Fisher71}.  This also gives rise to electrical forces acting on the other-particle charge. These forces give rise to orbital torques if they have components that are nonradial around the center of mass.

The total electric field due to translational motion of the magnetic dipole, for the Positronium atom model Bohr ground state, and accurate to within a factor of \(\beta^2\) times the Coulomb electric field strength, is found in Appendix B4 to be

\begin{equation}
\mbox{\boldmath $E$}_{\mbox{\boldmath$v$} \times \mbox{\boldmath$\mu$}} =  \frac{1}{c r^3}\left(3((\mbox{\boldmath$v$} \times \mbox{\boldmath$\mu$}) \cdot \mbox{\boldmath$n$})\mbox{\boldmath$n$} - 2(\mbox{\boldmath$v$} \times \mbox{\boldmath$\mu$})\right).
\label{elecfldmmorb}
\end{equation}

The force accurate to order \((v/c)^2\) acting on the electron due to the circular orbiting intrinsically-magnetic positron is then

\begin{equation}
\mbox{\boldmath $F$}_{\mbox{\boldmath$v$} \times \mbox{\boldmath$\mu$}} =  -\frac{e}{cr^3}\left(3((\mbox{\boldmath$v$}_p \times \mbox{\boldmath$\mu$}_p) \cdot \mbox{\boldmath$n$})\mbox{\boldmath$n$} - 2(\mbox{\boldmath$v$}_p \times \mbox{\boldmath$\mu$}_p)\right).
\label{elecfldmmorb}
\end{equation}

The torque on the electron orbit is thus

\begin{equation}
\mbox{\boldmath$\tau$}_{\mbox{\boldmath$v$} \times \mbox{\boldmath$\mu$},e} = \mbox{\boldmath$r$}_e \times \left[ \frac{e}{cr^3}2(\mbox{\boldmath$v$}_p \times \mbox{\boldmath$\mu$}_p) \right],
\label{elecfldmmorb}
\end{equation}

or, returning to the convention that \(\mbox{\boldmath$v$}\) is the electron velocity relative to the positron, and in terms of the positron spin rather than the intrinsic moment,

\begin{equation}
\mbox{\boldmath$\tau$}_{\mbox{\boldmath$v$} \times \mbox{\boldmath$\mu$},e} = -\mbox{\boldmath$r$} \times \left[ \frac{e^2}{2mc^2r^3}(\mbox{\boldmath$v$} \times \mbox{\boldmath$s$}_p) \right].
\label{elecfldmmorb}
\end{equation}

The torque on the positron orbit is found similarly to be

\begin{equation}
\mbox{\boldmath$\tau$}_{\mbox{\boldmath$v$} \times \mbox{\boldmath$\mu$},e} = -\mbox{\boldmath$r$} \times \left[ \frac{e^2}{2mcr^3}(\mbox{\boldmath$v$} \times \mbox{\boldmath$s$}_e) \right],
\label{elecfldmmorb}
\end{equation}

and the total torque,  

\begin{equation}
\mbox{\boldmath$\tau$}_{\mbox{\boldmath$v$} \times \mbox{\boldmath$\mu$}} = -\mbox{\boldmath$r$} \times \frac{e^2}{2mcr^3} \left[ \mbox{\boldmath$v$} \times \left(\mbox{\boldmath$s$}_e + \mbox{\boldmath$s$}_p \right) \right].
\label{elecfldmmorb}
\end{equation}

For the circular orbit this can be simplified as 

\begin{equation}
\mbox{\boldmath$\tau$}_{\mbox{\boldmath$v$} \times \mbox{\boldmath$\mu$}} = - \frac{e^2}{2mcr^3} \left[  (\mbox{\boldmath$r$} \cdot\left(\mbox{\boldmath$s$}_e + \mbox{\boldmath$s$}_p \right))\mbox{\boldmath$v$} \right].
\label{TorqueVCrossMu}
\end{equation}

\subsection{Torque on Orbit due to Stern-Gerlach-like Force on Electric Dipole}

A magnetic dipole \(\mbox{\boldmath$m$}\) translating  with velocity \(\mbox{\boldmath$v$}\) acquires an electric dipole moment,  \(\mbox{\boldmath$d$}\),  as  \cite{Fisher71}

\begin{equation}
\mbox{\boldmath$d$} = \mbox{\boldmath$v$} \times \mbox{\boldmath$m$}/c.
\label{Edipole_dtv}
\end{equation}

The intrinsically-magnetic electron thus acquires an electric dipole moment \(\mbox{\boldmath$d$}_e\) due to its translational motion given by \(\mbox{\boldmath$d$}_e = \mbox{\boldmath$v$}_e \times \mbox{\boldmath$\mu$}_e/c\).

The force on an electric dipole \(\mbox{\boldmath$d$}\) in a electric field \(\mbox{\boldmath$E$}\) is 

\begin{equation}
\mbox{\boldmath$F$} = \nabla(\mbox{\boldmath$d$}\cdot\mbox{\boldmath$E$}) = (\mbox{\boldmath$d$}\cdot\nabla)\mbox{\boldmath$E$}, 
\label{ForceOnDipole}
\end{equation}

where the second equality applies provided that \(\nabla \times \mbox{\boldmath$E$}\) vanishes or is negligible as in the present application.

For the Coulomb electric field due to the positron charge, acting on the electron, this becomes

\begin{equation}
\mbox{\boldmath$F$}_e = (\mbox{\boldmath$d$}_e\cdot\nabla)\frac{q_p \mbox{\boldmath$r$}}{r^3}. 
\label{HzizdoY91_Eq1a}
\end{equation}

With

\begin{equation}
(\mbox{\boldmath$d$}\cdot\nabla) \left(\frac{\mbox{\boldmath$r$}}{r^3} \right) = \frac{\mbox{\boldmath$d$}}{r^3} - \frac{3(\mbox{\boldmath$d$} \cdot \mbox{\boldmath$n$}) \mbox{\boldmath$n$}}{r^3}, 
\label{HzizdoY91_Eq1a}
\end{equation}

the force on the electron becomes

\begin{equation}
\mbox{\boldmath$F$}_e = \frac{q_p}{c r^3} \left[ (\mbox{\boldmath$v$}_e \times \mbox{\boldmath$\mu$}_e) - 3((\mbox{\boldmath$v$}_e \times \mbox{\boldmath$\mu$}_e) \cdot \mbox{\boldmath$r$}) \mbox{\boldmath$r$}\right].
\label{HzizdoY91_Eq1a}
\end{equation}

The torque on the orbit due to this force is thus

\begin{equation}
\mbox{\boldmath$\tau$}_e = \mbox{\boldmath$r$} \times \mbox{\boldmath$F$}_e = \frac{q_p}{c r^3} \mbox{\boldmath$r$}_e \times \left[ (\mbox{\boldmath$v$}_e \times \mbox{\boldmath$\mu$}_e) \right], 
\label{HzizdoY91_Eq1a}
\end{equation}

or, in terms of the spin instead of the magnetic moment, and with \(q_p = - q_e = e\), and \( \mbox{\boldmath$r$} = 2\mbox{\boldmath$r$}_e\) and \(\mbox{\boldmath$v$}=2\mbox{\boldmath$v$}_e \),

\begin{equation}
\mbox{\boldmath$\tau$}_e = -\frac{e^2}{4 m c^2 r^3} \mbox{\boldmath$r$} \times \left[ \mbox{\boldmath$v$} \times \mbox{\boldmath$s$}_e \right]. 
\label{HzizdoY91_Eq1a}
\end{equation}

The torque on the positron orbit is similarly

\begin{equation}
\mbox{\boldmath$\tau$}_p = -\frac{e^2}{4 m c^2 r^3} \mbox{\boldmath$r$} \times \left[ (\mbox{\boldmath$v$} \times \mbox{\boldmath$s$}_p) \right]. 
\label{HzizdoY91_Eq1a}
\end{equation}

The total torque on the orbit due to this type of force is thus

\begin{equation}
\mbox{\boldmath$\tau$}_e + \mbox{\boldmath$\tau$}_p = -\frac{e^2}{4 m c^2 r^3} \mbox{\boldmath$r$} \times \left[ \mbox{\boldmath$v$} \times (\mbox{\boldmath$s$}_e +\mbox{\boldmath$s$}_p )\right]. 
\label{HzizdoY91_Eq1a}
\end{equation}

For the circular orbit this torque can be rewritten as

\begin{equation}
\mbox{\boldmath$\tau$}_{\text{Stern-Gerlach-like}} = -\frac{e^2}{4 m c^2 r^3} \mbox{\boldmath$v$} \left[ \mbox{\boldmath$r$} \cdot (\mbox{\boldmath$s$}_e +\mbox{\boldmath$s$}_p )\right]. 
\label{TorqueSternGerlachLike}
\end{equation}

\subsection{Negligibility of Electric Dipole-to-Dipole Force and Torque}

A Stern-Gerlach-like force and corresponding torque exists due to anisotropy of the electric field of the motion-acquired electric dipole moment of one particle, acting on the motion-acquired electric dipole moment of the other.  This torque is shown to be negligible to the present analysis.

Using the first equality of Eq. (\ref{ForceOnDipole}) obtains for the electric field due to the positron electric dipole moment acting on the electron the force

\begin{equation}
\mbox{\boldmath$F$} = \nabla\left(\mbox{\boldmath$d$}_e\cdot\frac{3(\mbox{\boldmath$n$}\cdot\mbox{\boldmath$d$}_p)\mbox{\boldmath$n$} - \mbox{\boldmath$d$}_p}{r^3}\right). 
\label{HzizdoY91_Eq1a}
\end{equation}

This force is formally similar to the Stern-Gerlach force on the electron intrinsic magnetic moment in the intrinsic magnetic field of the positron (as given by Eq. (\ref{FSE2e1})).  It is weaker by a \(\beta^2\) factor however due to the \(v/c\) in Eq. (\ref{Edipole_dtv}) for the acquired electric dipole moment and is therefore negligible in the present analysis.

\section{Motion of the total orbital angular momentum}

The total orbital angular momentum, referred to here as \(\mbox{\boldmath$L$}_{\text{total}}\), consists of the total kinetic orbital angular momentum, that is here represented as \(\mbox{\boldmath$L$} = \mbox{\boldmath$L$}_e + \mbox{\boldmath$L$}_p\),  the total hidden mechanical orbital angular  momentum, represented as \(\mbox{\boldmath$L$}_{\text{hidden}}\), and the total electromagnetic field orbital angular momentum.  The latter is due to the electron and positron intrinsic magnetic fields crossed onto the electric field of the other particle, and will be represented as \(\mbox{\boldmath$L$}_{\text{field}}\).   The equation of motion of the total orbital angular momentum is then expressible as   

\begin{equation}
\dot{\mbox{\boldmath$L$}}_{\text{total}} =\dot{\mbox{\boldmath$L$}} + \dot{\mbox{\boldmath$L$}}_{\text{hidden}} + \dot{\mbox{\boldmath$L$}}_{\text{field}}.
\end{equation}

\begin{widetext}

\subsection{Rate of Change of the Kinetic Orbital Angular Momentum}

In this section the rate of change of the total kinetic orbital angular momentum, that is, the total orbital angular momentum of the translational motion of the electron and positron masses, is calculated.

It was determined previously (Equation (\ref{Ldot_final})) that 

\begin{equation}
\dot{\mbox{\boldmath$L$}} \equiv \dot{\mbox{\boldmath$L$}}_e + \dot{\mbox{\boldmath$L$}}_p = \mbox{\boldmath$\tau$} - \frac{e^2}{2 m c^2 r^3} (\mbox{\boldmath$r$} \cdot  (\mbox{\boldmath$s$}_e + \mbox{\boldmath$s$}_p))  \mbox{\boldmath$v$} 
\nonumber
\end{equation}

where \(\mbox{\boldmath$\tau$} \equiv \mbox{\boldmath$\tau$}_e + \mbox{\boldmath$\tau$}_p\) is the sum of the torques on the electron and positron orbits.  From the results of the previous section, the sum of the non-negligible torques is

\begin{equation}
\mbox{\boldmath$\tau$} = \mbox{\boldmath$\tau$}_{\text{Biot-Savart}} + \mbox{\boldmath$\tau$}_{\text{Stern-Gerlach 1}} + \mbox{\boldmath$\tau$}_{\text{Stern-Gerlach 2}} + \mbox{\boldmath$\tau$}_{\mbox{\boldmath$v$} \times \mbox{\boldmath$\mu$}} + \mbox{\boldmath$\tau$}_{\text{Stern-Gerlach like}}
\label{TorqueTotal}
\end{equation}

where, from Equations (\ref{TorqueBiotSavart}), (\ref{TorqueSternGerlach1}), (\ref{TorqueSternGerlach2}), (\ref{TorqueVCrossMu}), (\ref{TorqueSternGerlachLike}), and with \(\mbox{\boldmath$n$} \equiv \mbox{\boldmath$r$}/r\),  

\begin{equation}
\mbox{\boldmath$\tau$}_{\text{Biot-Savart}} = -\frac{e^2}{2 m c^2 r^2} \mbox{\boldmath$v$} \left[ \mbox{\boldmath$n$} \cdot (\mbox{\boldmath$s$}_e +\mbox{\boldmath$s$}_p)\right] 
\nonumber
\end{equation}

\begin{equation}
\mbox{\boldmath$\tau$}_{\text{Stern-Gerlach 1}} = 
-\frac{e^2}{4 m c^2 r^2}
\left( \mbox{\boldmath$n$} \cdot \left(\mbox{\boldmath$s$}_e + \mbox{\boldmath$s$}_p\right) \right) \mbox{\boldmath$v$} 
\nonumber
\end{equation}

\begin{equation}
\mbox{\boldmath$\tau$}_{\text{Stern-Gerlach 2}} = \frac{e^2}{m^2 c^2 r^3} \left[ \mbox{\boldmath$s$}_p\times
3 \mbox{\boldmath$n$} \left(\mbox{\boldmath$s$}_e\cdot\mbox{\boldmath$n$}\right) + \mbox{\boldmath$s$}_e\times3\mbox{\boldmath$n$} \left(\mbox{\boldmath$n$} \cdot \mbox{\boldmath$s$}_p\right)  \right]
\nonumber
\end{equation}

\begin{equation}
\mbox{\boldmath$\tau$}_{\mbox{\boldmath$v$} \times \mbox{\boldmath$\mu$}} = - \frac{e^2}{2mcr^3} \left[  (\mbox{\boldmath$r$} \cdot\left(\mbox{\boldmath$s$}_e + \mbox{\boldmath$s$}_p \right))\mbox{\boldmath$v$} \right]
\nonumber
\end{equation}

\begin{equation}
\mbox{\boldmath$\tau$}_{\text{Stern-Gerlach like}} = -\frac{e^2}{4 m c^2 r^2} \mbox{\boldmath$v$} \left[ \mbox{\boldmath$n$} \cdot (\mbox{\boldmath$s$}_e +\mbox{\boldmath$s$}_p )\right] 
\nonumber
\end{equation}

Evaluating the sum of the torques according to Eq. (\ref{TorqueTotal}) obtains the total torque as 

\begin{equation}
\mbox{\boldmath$\tau$} = -\frac{3e^2}{2 m c^2 r^2}
\left( \mbox{\boldmath$n$} \cdot \left(\mbox{\boldmath$s$}_e + \mbox{\boldmath$s$}_p\right) \right) \mbox{\boldmath$v$}  + \frac{e^2}{r^3 m^2 c^2} \left[ \mbox{\boldmath$s$}_p\times
3 \mbox{\boldmath$n$} \left(\mbox{\boldmath$s$}_e\cdot\mbox{\boldmath$n$}\right) + \mbox{\boldmath$s$}_e\times3\mbox{\boldmath$n$} \left(\mbox{\boldmath$n$} \cdot \mbox{\boldmath$s$}_p\right)  \right].
\label{Expandt1}
\end{equation}

The rate of change of the total kinetic orbital angular momentum evaluated according to Eq. (\ref{Ldot_final}) is thus

\begin{equation}
\dot{\mbox{\boldmath$L$}} = -\frac{2e^2}{m c^2 r^2}
\left( \mbox{\boldmath$n$} \cdot \left(\mbox{\boldmath$s$}_e + \mbox{\boldmath$s$}_p\right) \right) \mbox{\boldmath$v$}  + \frac{e^2}{r^3 m^2 c^2} \left[ \mbox{\boldmath$s$}_p\times
3 \mbox{\boldmath$n$} \left(\mbox{\boldmath$s$}_e\cdot\mbox{\boldmath$n$}\right) + \mbox{\boldmath$s$}_e\times3\mbox{\boldmath$n$} \left(\mbox{\boldmath$n$} \cdot \mbox{\boldmath$s$}_p\right)  \right].
\label{TotalLdot}
\end{equation}

It will turn out to be useful to understand the relative magnitude of the two terms summed on the right hand side of Eq. (\ref{TotalLdot}), as well as the overall magnitude of \(\dot{\mbox{\boldmath$L$}}\).  This will aid in demonstrating the negligibility of some contributions to the motion of the total angular momentum, and so simplify the analysis.  The first term on the right hand side of Eq. (\ref{TotalLdot}) is bounded as

\begin{equation}
\left| \frac{2e^2}{m c^2 r^2}
\left( \mbox{\boldmath$n$} \cdot \left(\mbox{\boldmath$s$}_e + \mbox{\boldmath$s$}_p\right) \right) \mbox{\boldmath$v$} \right| \le \frac{2e^2}{m c^2 r^2} \hbar \frac{e\sqrt{2}}{m^{1/2}r^{1/2}} =  \frac{ 2\sqrt{2}e^3\hbar}{c^2 m^{3/2}r^{5/2}}.
\label{LdotKineticB1}
\end{equation}

The second term on the right hand side of Eq. (\ref{TotalLdot}) is bounded as

\begin{equation}
\left| \frac{e^2}{r^3 m^2 c^2} \left[ \mbox{\boldmath$s$}_p\times
3 \mbox{\boldmath$n$} \left(\mbox{\boldmath$s$}_e\cdot\mbox{\boldmath$n$}\right) + \mbox{\boldmath$s$}_e\times3\mbox{\boldmath$n$} \left(\mbox{\boldmath$n$} \cdot \mbox{\boldmath$s$}_p\right)  \right] \right| \le \frac{6 e^2}{r^3 m^2 c^2} \frac{\hbar^2}{4}.
\label{LdotKineticB2}
\end{equation}

The ratio of the second term bound to the first is then

\begin{equation}
\frac{6 e^2}{r^3 m^2 c^2} \frac{\hbar^2}{4} \left(\frac{ 2\sqrt{2}e^3\hbar}{c^2 m^{3/2}r^{5/2}} \right)^{-1} = \frac{3}{4\sqrt{2}} \left(\frac{\hbar}{ e m^{1/2} r^{1/2}} \right).
\label{LdotKineticB2}
\end{equation}

At the Positronium Bohr radius \(r_{\text{B}}= 2\hbar^2/(e^2m)\), this becomes

\begin{equation}
\frac{3}{4\sqrt{2}} \left(\frac{\hbar}{ e m^{1/2}} \right)\left(\frac{2\hbar^2}{e^2 m} \right)^{-1/2} = \frac{3}{4\sqrt{2}} \left(\frac{\hbar}{ e m^{1/2}} \right)\left(\frac{e m^{1/2}}{\sqrt{2}\hbar} \right) = \frac{3}{8}.
\label{LdotKineticB2}
\end{equation}

The two terms on the right hand side of Eq. (\ref{TotalLdot}) being of similar magnitude are thus both significant to the rate of change of the kinetic orbital angular momentum, for the quasiclassical Positronium atom in the Bohr ground state.

\subsection{Rate of Change of Hidden Orbital Angular Momentum}

With the hidden momentum of a magnetic dipole of moment \(\mbox{\boldmath$m$}\) as given previously as \(\mbox{\boldmath$m$} \times \mbox{\boldmath$E$}/c\), the hidden angular momentum around the center of mass, of the electron intrinsic magnetic moment in the Coulomb field of the positron, is defined as

\begin{equation}
\mbox{\boldmath$L$}_{\text{hidden},e}  =  \frac{1}{c}\mbox{\boldmath$r$}_e \times \left(\mbox{\boldmath$\mu$}_e \times \mbox{\boldmath$E$}\right)  =  \frac{e}{c r^3}\mbox{\boldmath$r$}_e \times \left(\mbox{\boldmath$\mu$}_e \times \mbox{\boldmath$r$}\right). 
\label{L_hidden_e}
\end{equation}

(Since the hidden orbital angular momentum magnitude is about a factor of \(\beta^2\) smaller than the kinetic orbital angular momentum magnitude, there is no need to include higher order terms in the description of the electric field here.)  Expanding the vector triple product and with \(\mbox{\boldmath$r$} = 2\mbox{\boldmath$r$}_e\) obtains that

\begin{equation}
\mbox{\boldmath$L$}_{\text{hidden},e}  =  \frac{e}{2 c r^3}\left[ r^2\mbox{\boldmath$\mu$}_e - \mbox{\boldmath$r$} \left(\mbox{\boldmath$\mu$}_e \cdot \mbox{\boldmath$r$}\right) \right]. 
\label{L_hidden_e_reduced}
\end{equation}

For the circular orbit (where \(\dot{r}\equiv 0\)),

\begin{equation}
\frac{d\mbox{\boldmath$L$}_{\text{hidden}e},e}{dt} \equiv \dot{\mbox{\boldmath$L$}}_{\text{hidden},e}  =  \frac{e}{2 c r^3}\left[ r^2\dot{\mbox{\boldmath$\mu$}}_e - \mbox{\boldmath$v$} \left(\mbox{\boldmath$\mu$}_e \cdot \mbox{\boldmath$r$}\right) - \mbox{\boldmath$r$} \left(\left(\dot{\mbox{\boldmath$\mu$}}_e \cdot \mbox{\boldmath$r$}\right) + \left(\mbox{\boldmath$\mu$}_e \cdot \mbox{\boldmath$v$}\right)\right) \right].
\label{L_hidden}
\end{equation}

Rearranging and in terms of the electron spin instead of the intrinsic magnetic moment,

\begin{equation}
\dot{\mbox{\boldmath$L$}}_{\text{hidden},e}  =  \frac{e^2}{2 m c^2 r^3}\left[\mbox{\boldmath$v$} \left(\mbox{\boldmath$s$}_e \cdot \mbox{\boldmath$r$}\right) + \mbox{\boldmath$r$}\left(\mbox{\boldmath$s$}_e \cdot \mbox{\boldmath$v$}\right)\right] - \frac{e^2}{2 m c^2 r}\left[ \dot{\mbox{\boldmath$s$}}_e - \mbox{\boldmath$n$} \left(\dot{\mbox{\boldmath$s$}}_e \cdot \mbox{\boldmath$n$}\right) \right]. 
\label{L_hidden_dot_e_all}
\end{equation}

Now, the second term in the brackets on the right hand side can be further evaluated by substitution for \(\dot{\mbox{\boldmath$s$}}_e\) as given by Eq. (\ref{ThomasEqRedElectron}), but it is worthwhile instead to consider the relative magnitude of the two terms being differenced in Eq. (\ref{L_hidden_dot_e_all}).  As has already been used in Section IVC, an upper bound for the magnitude of the time rate of change of the spin vector, for the quasiclassical positronium atom in the Bohr ground state, is developed in Eq. (\ref{sdot_bound_appendix}) in Appendix D. Taking \(r\) to be the Bohr radius then obtains that 

\begin{equation}
\left|\frac{e^2}{2 m c^2 r}\left[ \dot{\mbox{\boldmath$s$}}_e - \mbox{\boldmath$n$} \left(\dot{\mbox{\boldmath$s$}}_e \cdot \mbox{\boldmath$n$}\right) \right] \right| \le \frac{e^2}{2 m c^2} \frac{e^2 m}{\hbar^2}\frac{7}{32} \frac{e^8 m}{\hbar^4 c^2} = \frac{7}{64} \frac{e^{12} m}{\hbar^6 c^4}.
\label{sdot_bound_copy2}
\end{equation}

An upper bound on the first term on the right hand side of Eq. (\ref{L_hidden_dot_e_all}) can be developed as

\begin{equation}
\left|\frac{e^2}{2 m c^2 r^3}\left[\mbox{\boldmath$v$} \left(\mbox{\boldmath$s$}_e \cdot \mbox{\boldmath$r$}\right) + \mbox{\boldmath$r$}\left(\mbox{\boldmath$s$}_e \cdot \mbox{\boldmath$v$}\right)\right] \right| \le \frac{e^2vs}{2 m c^2 r^2} = \frac{e^2}{2 m c^2 r^2} \frac{e\sqrt{2}}{m^{1/2}r^{1/2}} \frac{\hbar}{2} = \frac{e^3\sqrt{2}}{c^2m^{3/2}r^{5/2}} \frac{\hbar}{4},
\label{Ldot_hidden_t1_bound_gen}
\end{equation}

which at the Bohr radius becomes

\begin{equation}
\left|\frac{e^2}{2 m c^2 r^3}\left[\mbox{\boldmath$v$} \left(\mbox{\boldmath$s$}_e \cdot \mbox{\boldmath$r$}\right) + \mbox{\boldmath$r$}\left(\mbox{\boldmath$s$}_e \cdot \mbox{\boldmath$v$}\right)\right] \right| \le  \frac{e^3\sqrt{2}}{c^2m^{3/2}} \frac{\hbar}{4} \left(\frac{2 \hbar^2}{e^2 m} \right)^{-5/2}  = \frac{e^8 m}{16c^2\hbar^4}. 
\label{Ldot_hidden_t1_bound_Bohr}
\end{equation}

The relative magnitude of the two bounds can now be evaluated as

\begin{equation}
\frac{7}{64} \frac{e^{12} m}{\hbar^6 c^4} \left( \frac{e^8 m}{16c^2\hbar^4} \right)^{-1} = \frac{7}{64} \frac{e^{12} m}{\hbar^6 c^4} \left( \frac{16c^2\hbar^4}{e^8 m} \right) = \frac{7}{4} \frac{e^{4}}{\hbar^2 c^2} = \frac{7}{4} \alpha^2.
\label{sdot_bound_copy2}
\end{equation}

The time rate of change of the electron hidden orbital angular momentum can now be written accurate to one part in \(1/\beta^2\) as

\begin{equation}
\dot{\mbox{\boldmath$L$}}_{\text{hidden},e}  =    \frac{e^2}{2 m c^2 r^3}\left[\mbox{\boldmath$r$}\left(\mbox{\boldmath$s$}_e \cdot \mbox{\boldmath$v$}\right) + \mbox{\boldmath$v$} \left(\mbox{\boldmath$s$}_e \cdot \mbox{\boldmath$r$}\right) \right]. 
\label{Ldot_hidden_e}
\end{equation}

The rate of change of the hidden orbital angular momentum of the positron intrinsic magnetic moment in the electric field of the electron is evaluated similarly to obtain

\begin{equation}
\dot{\mbox{\boldmath$L$}}_{\text{hidden},p} =  \frac{e^2}{2 m c^2 r^3}\left[\mbox{\boldmath$r$}\left(\mbox{\boldmath$s$}_p \cdot \mbox{\boldmath$v$}\right) + \mbox{\boldmath$v$} \left(\mbox{\boldmath$s$}_p \cdot \mbox{\boldmath$r$}\right) \right]. 
\label{Ldot_hidden_p}
\end{equation}

The rate of change of the total hidden orbital angular momentum is then (accurate to one part in \(1/\beta^2\))

\begin{equation}
\dot{\mbox{\boldmath$L$}}_{\text{hidden}} =  \frac{e^2}{2 m c^2 r^3}\left[\mbox{\boldmath$r$}\left(\left(\mbox{\boldmath$s$}_e + \mbox{\boldmath$s$}_p \right)\cdot \mbox{\boldmath$v$}\right) + \mbox{\boldmath$v$} \left( \left(\mbox{\boldmath$s$}_e + \mbox{\boldmath$s$}_p\right)\cdot \mbox{\boldmath$r$}\right) \right].
\label{Ldot_hidden_total}
\end{equation}

In order to confirm that Eq. (\ref{Ldot_hidden_total}) is sufficiently accurate for the present purpose, the relative magnitude of the rate of change of the hidden orbital angular momentum to that of the kinetic orbital angular momentum as given by Eq. (\ref{TotalLdot}) can be evaluated.  Since the two terms on the right hand side of  Eq. (\ref{TotalLdot}) were found to be of the same order of magnitude (at the Bohr ground state), either can be taken as representative of the general magnitude of the rate of change of the kinetic orbital angular momentum for the Bohr ground state.  The ratio  of the bound on the larger term of the hidden orbital angular momentum motion, from Eq. (\ref{Ldot_hidden_t1_bound_gen}), to the bound on the first term on the right hand side of Eq. (\ref{TotalLdot}), as given by Eq. (\ref{LdotKineticB1}), is

\begin{equation}
\frac{e^3\sqrt{2}}{c^2m^{3/2}r^{5/2}} \frac{\hbar}{2} \left(\frac{ 2\sqrt{2}e^3\hbar}{c^2 m^{3/2}r^{5/2}} \right)^{-1} = \frac{1}{4}
\label{LdotKineticB2}
\end{equation}

This result establishes that the magnitude of the rate of change of the hidden orbital angular momentum is similar to that of the kinetic orbital angular momentum.  

It is also of interest to compare the magnitude of the term involving \(\dot{\mbox{\boldmath$s$}}_e\), in the motion of the electron hidden orbital angular momentum of Eq. (\ref{L_hidden_dot_e_all}), to the magnitude of the contribution to the motion of the total angular momentum attributable to the Thomas precession.  The contribution of the Thomas precession to the motion of the electron spin vector in the quasiclassical Positronium atom as given by Eq. (\ref{ThomasEqRedElectron}), evaluated at the Bohr radius, is bounded as

\begin{equation}
\left|\frac{e^2}{m^2 c^2 r^3}\mbox{\boldmath$s$}_e \times \left[  \left(\frac{1}{2}\right) \mbox{\boldmath$L$} \right]\right| \le \frac{e^2}{m^2 c^2 r^3}\frac{\hbar^2}{4} = \frac{e^2}{m^2 c^2} \frac{\hbar^2}{4}\left(\frac{2\hbar^2}{e^2m}\right)^{-3} = \frac{e^8m}{32 c^2\hbar^4}. 
\label{ThomasElectron}
\end{equation}

The ratio of the magnitude of the term involving \(\dot{\mbox{\boldmath$s$}}_e\) in Eq. (\ref{L_hidden_dot_e_all}) to the magnitude of the contribution to the motion of the total angular momentum attributable to the Thomas precession is then

\begin{equation}
\frac{7}{64} \frac{e^{12} m}{\hbar^6 c^4} \left(\frac{e^8m}{32 c^2\hbar^4}\right)^{-1}  = \frac{7}{2} \frac{e^{4}}{\hbar^2 c^2} = \frac{7}{2} \alpha^2. 
\label{ProofInsignificant}
\end{equation}

Since \(\beta\) is the same order as \(\alpha\) here, neglecting the term involving \(\dot{\mbox{\boldmath$s$}}_e\) in Eq. (\ref{L_hidden_dot_e_all}) for the hidden orbital angular momentum rate of change will introduce only order \(\beta^2\) errors compared to the effect of the Thomas precession, in the rate of change of the total orbital angular momentum.

\subsection{Rate of Change of Field Angular Momentum}

The total angular momentum can be expected to be a conserved quantity only if the field angular momentum is included \cite{Hnizdo1992oc1}.  For a system consisting of a stationary current distribution and a particle with electric charge \(q\), and with \(\mbox{\boldmath$r$}\) the displacement of the charged particle from the curent distribution, there is field angular momentum around a fixed point \(\mbox{\boldmath$r$}_1\) evaluated as \cite{Trammel1964}

\begin{equation}
\mbox{\boldmath$L$}_{\text{field}} = \frac{1}{4\pi c} \int (\mbox{\boldmath$r$}-\mbox{\boldmath$r$}_1) \times \left( \mbox{\boldmath$E$} \times  \mbox{\boldmath$B$} \right) d^3\mbox{\boldmath$r$} = (q/c)(\mbox{\boldmath$r$}-\mbox{\boldmath$r$}_1) \times \mbox{\boldmath$A$} (\mbox{\boldmath$r$}),
\label{L_field_Trammel}
\end{equation}

where \(\mbox{\boldmath$A$} (\mbox{\boldmath$r$})\) is the vector potential of the current distribution evaluated at the location of the charged particle.

If the vector potential in Eq. (\ref{L_field_Trammel}) is taken to be that of the positron magnetic dipole moment, the vector potential at the location of the electron is

\begin{equation}
\mbox{\boldmath$A$} = \frac{\mbox{\boldmath$\mu$}_p \times \mbox{\boldmath$r$}}{r^3},
\label{vec_pot_mu}
\end{equation}

where \(\mbox{\boldmath$r$}\) in this case being the displacement of the electron charge from the positron is in agreement with the convention of the present analysis. 

With the electron the charged particle, \(q=-e\), and taking the point \(\mbox{\boldmath$r$}_1\) to be the center of mass then obtains that \(\mbox{\boldmath$r$}-\mbox{\boldmath$r$}_1 = \mbox{\boldmath$r$}_e\) (since \(\mbox{\boldmath$r$} \equiv \mbox{\boldmath$r$}_e -\mbox{\boldmath$r$}_p\) and the displacement of the center of mass from the positron position is \(-\mbox{\boldmath$r$}_p\)), and so the field angular momentum according to Eq. (\ref{L_field_Trammel}) becomes

\begin{equation}
\mbox{\boldmath$L$}_{\text{field}, p} = -\frac{e}{cr^3}\mbox{\boldmath$r$}_e \times \left( \mbox{\boldmath$\mu$}_p \times \mbox{\boldmath$r$}\right) = -\frac{e}{2 c r^3}\mbox{\boldmath$r$} \times \left( \mbox{\boldmath$\mu$}_p \times  \mbox{\boldmath$r$}\right) = -\frac{e}{2 c r^3}\left[ r^2 \mbox{\boldmath$\mu$}_p - \mbox{\boldmath$r$} \left( \mbox{\boldmath$\mu$}_p \cdot  \mbox{\boldmath$r$}\right) \right],
\label{L_field_p}
\end{equation}

where the first equality is based on that \( \mbox{\boldmath$r$} = 2\mbox{\boldmath$r$}_e\).  Noting that the expression of Eq. (\ref{L_field_p}) for the total field angular momentum of the positron charge Coulomb field crossed with the intrinsic magnetic field of the electron is formally identical to the expression of Eq. (\ref{L_hidden_e_reduced}) for the hidden orbital angular momentum of the electron, allows the rate of change of the field angular momentum due to this contribution accurate to one part in \(1/\beta^2\) to be written immediately, based on Eq. (\ref{Ldot_hidden_e}), as

\begin{equation}
\dot{\mbox{\boldmath$L$}}_{\text{field}, p}  =    \frac{e^2}{2 m c^2 r^3}\left[\mbox{\boldmath$r$}\left(\mbox{\boldmath$s$}_p \cdot \mbox{\boldmath$v$}\right) + \mbox{\boldmath$v$} \left(\mbox{\boldmath$s$}_p \cdot \mbox{\boldmath$r$}\right) \right]. 
\label{Ldot_field_p}
\end{equation}

Strictly, since Eq. (\ref{L_field_Trammel}) assumed the current distribution was stationary, Eq. (\ref{Ldot_field_p}) is the rate of change of field angular momentum in the positron rest frame, which is rotating due to Thomas precession.  However, applying the standard transformation \cite{GoldsteinClassMech_133} for the rate of change of a vector from a rotating to a nonrotating frame, it is found that this changes the expression of Eq. (\ref{L_field_p}) only by about one part in \(1/\beta^2 = 1/\alpha^2\) here and so is negligible to the present analysis. 

The rate of change of the total field angular momentum is found by summing the contribution of Eq. (\ref{Ldot_field_p}) and the similar quantity for the field angular momentum of the positron Coulomb field crossed onto the electron intrinsic magnetic field, with the result 
 
\begin{equation}
\dot{\mbox{\boldmath$L$}}_{\text{field}}  =    \frac{e^2}{2 m c^2 r^3}\left[\mbox{\boldmath$r$}\left(\left(\mbox{\boldmath$s$}_e + \mbox{\boldmath$s$}_p \right)\cdot \mbox{\boldmath$v$}\right) + \mbox{\boldmath$v$} \left(\left(\mbox{\boldmath$s$}_e + \mbox{\boldmath$s$}_p \right)\cdot \mbox{\boldmath$r$}\right) \right]. 
\label{Ldot_field_total}
\end{equation}

\subsection{Rate of Change of the Total Orbital Angular Momentum}

The rate of change of the total orbital angular
momentum is evaluated as 

\begin{equation}
\dot{\mbox{\boldmath$L$}}_{\text{total}} =  \dot{\mbox{\boldmath$L$}} + \dot{\mbox{\boldmath$L$}}_{\text{hidden}} + \dot{\mbox{\boldmath$L$}}_{\text{field}}.
\label{Ldot_total_def}
\end{equation}

From Eqs. (\ref{Ldot_hidden_total}) and (\ref{Ldot_field_total}), 

\begin{equation}
\dot{\mbox{\boldmath$L$}}_{\text{hidden}} + \dot{\mbox{\boldmath$L$}}_{\text{field}} =  \frac{e^2}{ m c^2 r^2}\left[\mbox{\boldmath$n$}\left(\left(\mbox{\boldmath$s$}_e + \mbox{\boldmath$s$}_p \right)\cdot \mbox{\boldmath$v$}\right) + \mbox{\boldmath$v$} \left( \left(\mbox{\boldmath$s$}_e + \mbox{\boldmath$s$}_p\right)\cdot \mbox{\boldmath$n$}\right) \right]. 
\label{Ldoth_plus_Ldotf}
\end{equation}

Substitution of the results of Equations (\ref{TotalLdot}) and (\ref{Ldoth_plus_Ldotf}) into Eq. (\ref{Ldot_total_def}) obtains

\begin{equation}
\dot{\mbox{\boldmath$L$}}_{\text{total}} =   \frac{e^2}{ m^2 c^2 r^3} \left[ \mbox{\boldmath$s$}_p\times
3 \mbox{\boldmath$n$} \left(\mbox{\boldmath$s$}_e\cdot\mbox{\boldmath$n$}\right) + \mbox{\boldmath$s$}_e\times3\mbox{\boldmath$n$} \left(\mbox{\boldmath$n$} \cdot \mbox{\boldmath$s$}_p\right)  \right]  + \frac{e^2}{ m c^2 r^3}\left[\mbox{\boldmath$r$}\left(\left(\mbox{\boldmath$s$}_e + \mbox{\boldmath$s$}_p \right)\cdot \mbox{\boldmath$v$}\right) - \mbox{\boldmath$v$} \left( \left(\mbox{\boldmath$s$}_e + \mbox{\boldmath$s$}_p\right)\cdot \mbox{\boldmath$r$}\right) \right]. 
\label{torqueporbdtemm3}
\end{equation}

With the vector identity \(\mbox{\boldmath$a$} \times (\mbox{\boldmath$b$} \times \mbox{\boldmath$c$}) = (\mbox{\boldmath$a$} \cdot\mbox{\boldmath$c$})\mbox{\boldmath$b$} - (\mbox{\boldmath$a$} \cdot \mbox{\boldmath$b$})\mbox{\boldmath$c$} 
\),

\begin{equation}
\dot{\mbox{\boldmath$L$}}_{\text{total}} =   \frac{e^2}{ m^2 c^2 r^3} \left[ \mbox{\boldmath$s$}_p\times
3 \mbox{\boldmath$n$} \left(\mbox{\boldmath$s$}_e\cdot\mbox{\boldmath$n$}\right) + \mbox{\boldmath$s$}_e\times3\mbox{\boldmath$n$} \left(\mbox{\boldmath$n$} \cdot \mbox{\boldmath$s$}_p\right)  \right]  + \frac{e^2}{ m c^2 r^3}\left[\left(\mbox{\boldmath$s$}_e + \mbox{\boldmath$s$}_p \right)\times \left(\mbox{\boldmath$r$} \times \mbox{\boldmath$v$}\right) \right], 
\label{torqueporbdtemm3}
\end{equation}

or (note \(\mbox{\boldmath$r$} \times m \mbox{\boldmath$v$} = 2 \mbox{\boldmath$L$}\) here),

\begin{equation}
\dot{\mbox{\boldmath$L$}}_{\text{total}} = \frac{e^2}{m^2 c^2 r^3} \left[ \mbox{\boldmath$s$}_p\times
3 \mbox{\boldmath$n$} \left(\mbox{\boldmath$s$}_e\cdot\mbox{\boldmath$n$}\right) + \mbox{\boldmath$s$}_e\times3\mbox{\boldmath$n$} \left(\mbox{\boldmath$n$} \cdot \mbox{\boldmath$s$}_p\right)  \right]  + \frac{2e^2}{m^2 c^2 r^3}\left[\left(\mbox{\boldmath$s$}_e + \mbox{\boldmath$s$}_p \right)\times \mbox{\boldmath$L$} \right]. 
\label{LdotTotalFinal}
\end{equation}

\section{Motion of the Total Angular Momentum}

With total angular momentum \(\mbox{\boldmath$J$} \equiv  \mbox{\boldmath$s$}_e + \mbox{\boldmath$s$}_p +  \mbox{\boldmath$L$}_{\text{total}}\) and \(\dot{\mbox{\boldmath$J$}} =  \dot{\mbox{\boldmath$s$}}_e + \dot{\mbox{\boldmath$s$}}_p +  \dot{\mbox{\boldmath$L$}}_{\text{total}},\) it is obtained from Eqs (\ref{ThomasEqRedElectron}), (\ref{ThomasEqRedPositron}), and (\ref{LdotTotalFinal}) that

\begin{eqnarray}
\dot{\mbox{\boldmath$J$}} = -\frac{e^2}{m^2 c^2 r^3}\mbox{\boldmath$s$}_e \times \left[ 3\mbox{\boldmath$n$} \left(
\mbox{\boldmath$n$} \cdot \mbox{\boldmath$s$}_p   \right)  -
\mbox{\boldmath$s$}_p + \left(\frac{3}{2}\right) \mbox{\boldmath$L$} \right] 
 - \frac{e^2}{m^2 c^2 r^3}\mbox{\boldmath$s$}_p \times \left[ 3\mbox{\boldmath$n$} \left(
\mbox{\boldmath$n$} \cdot \mbox{\boldmath$s$}_e   \right)  -
\mbox{\boldmath$s$}_e  + \left(\frac{3}{2}\right) \mbox{\boldmath$L$} \right] + \nonumber \\
\frac{e^2 }{m^2 c^2 r^3} \left[ 
\mbox{\boldmath$s$}_p \times 3\mbox{\boldmath$n$}\left(\mbox{\boldmath$s$}_e\cdot\mbox{\boldmath$n$}\right) + 
\mbox{\boldmath$s$}_e \times 3 \mbox{\boldmath$n$} \left(\mbox{\boldmath$n$} \cdot \mbox{\boldmath$s$}_p\right)  \right]  + \frac{2e^2}{m^2 c^2 r^3}\left[\left(\mbox{\boldmath$s$}_e + \mbox{\boldmath$s$}_p \right)\times \mbox{\boldmath$L$} \right],
\label{JDot1}
\end{eqnarray}

which reduces to

\begin{equation}
\dot{\mbox{\boldmath$J$}} = \frac{e^2}{m^2 c^2 r^3}\left( \frac{1}{2} \right) \left[\left(\mbox{\boldmath$s$}_e + \mbox{\boldmath$s$}_p \right)\times \mbox{\boldmath$L$} \right]. 
\label{JdotFinal}
\end{equation}

This implies that the total angular momentum is a constant of the motion if \(\mbox{\boldmath$s$}_{e\perp} = -\mbox{\boldmath$s$}_{p\perp}\), where \(\mbox{\boldmath$s$}_{e\perp},\mbox{\boldmath$s$}_{p\perp}\) are the components of the electron and positron spin angular momenta perpendicular to the total orbital angular momentum, but not otherwise, if the spins and orbital angular momenta are not all aligned.  In general, the amount of nonconservation of angular momentum is exactly that attributable to the Thomas precession, as given by the sum of Eqs. (\ref{ThomasEqRedElectronTPContrib}) and (\ref{ThomasEqRedPositronTPContrib}).

The result of the present analysis, that total angular momentum is nonconserved due to Thomas precession, is surprising and perhaps distressing if correct.  Therefore, it's worth supposing this result simply incorrect and considering what type of changes would be required to arrive at the expected result of \(\dot{\mbox{\boldmath$J$}}\equiv 0\). This may help in finding the expected errors. 

One simple change that will result in total angular momentum conservation is to change the \(3/2\) factor on \(\mbox{\boldmath$L$}\) in the two spin equations of motion to \(2\).  But, the value of \(3/2\) resulted as a collection of a factor of unity with a factor of one-half, the latter of which is the one-half that is recognized as the ``Thomas factor'' that resolved the spin-orbit coupling anomaly of the anomalous Zeeman effect.  This result is very well established.  So, keeping the  term with the one-half unchanged, the first term in the sum, that derives from the magnetic field at one particle due to the orbital motion of the other, would have to become \(3/2\) instead of unity.  It's difficult to see how a ``3'' could creep into this term, however since the electron-positron displacement is simply \(\mbox{\boldmath$r$}=2\mbox{\boldmath$r$}_e=2\mbox{\boldmath$r$}_p\), and the \(\mbox{\boldmath$L$}\) here has resulted from the particle velocities crossed onto the local Coulomb field.  None of these involve any factors of 3.

Alternatively, if the spin equations of motion derived here are taken as correct, then the other contributions to the total angular momentum must be in error, if Eq. (\ref{JdotFinal}) is incorrect.  Since the field angular momentum does not vanish generally (if Eq. (\ref{Ldot_field_total}) is correct, or correct to within a factor), this would imply that the total mechanical momentum is not itself a constant of the motion (a {\em prima facie} interesting result, in itself, which is also true according to the present analysis).  However, obtaining the expected result of \(\dot{\mbox{\boldmath$J$}}\equiv 0\) here is not a matter of simply increasing or decreasing the field angular momentum or hidden orbital angular momentum by a factor or factors, or finding sign errors or missing factors on the individual torques acting on the orbit, alone.  Rather, it would require at least two separate errors that when corrected arrived at the expected result.  This is true because the form of the contribution of the Thomas precession, which as noted is the entire angular  momentum nonconservation of Eq. (\ref{JdotFinal}), involves the cross product of the spin vectors with the kinetic orbital angular momentum vector.  Therefore it can be written as a vector triple product when the kinetic orbital angular momentum is expanded as \(2\mbox{\boldmath$L$} = \mbox{\boldmath$r$} \times m\mbox{\boldmath$v$}\), and then using the vector identity \(\mbox{\boldmath$a$} \times (\mbox{\boldmath$b$} \times \mbox{\boldmath$c$}) = (\mbox{\boldmath$c$} \cdot \mbox{\boldmath$a$})\mbox{\boldmath$b$} - (\mbox{\boldmath$b$} \cdot \mbox{\boldmath$a$})\mbox{\boldmath$c$}\) obtains alternatively

\begin{equation}
\dot{\mbox{\boldmath$J$}} = \frac{e^2}{m c^2 r^3}\left( \frac{1}{4} \right) \left[\left(\mbox{\boldmath$s$}_e + \mbox{\boldmath$s$}_p \right)\times \left(\mbox{\boldmath$r$} \times \mbox{\boldmath$v$}\right) \right] = \frac{e^2}{m c^2 r^3}\left( \frac{1}{4} \right) \left[\left( \mbox{\boldmath$v$} \cdot \left(\mbox{\boldmath$s$}_e + \mbox{\boldmath$s$}_p \right)\right) \mbox{\boldmath$r$} - \left(   \left(\mbox{\boldmath$s$}_e + \mbox{\boldmath$s$}_p \right)\cdot \mbox{\boldmath$r$}\right) \mbox{\boldmath$v$}\right]. 
\label{JDotAlternative}
\end{equation}

The form of the right hand side of this equation stands in contrast to the form of the equations for the hidden orbital or field angular momenta, Eqs. (\ref{Ldot_hidden_total}) and (\ref{Ldot_field_total}), involving a difference rather than a sum of otherwise similar terms.  The minus sign here follows directly from an indisputable vector identity, while the plus sign in the equivalent position of the hidden orbital and field angular momenta follows from the also indisputable law of the differentiation of a product. If these signs are taken to be fixed then multiplying the hidden orbital or field angular momenta by any factors will not arrive at angular momentum conservation.  It will have to be combined with a change in the total torque on the orbit, or the other term in the equation of motion of the kinetic orbital angular momentum of Eq. (\ref{TotalLdot}), that accounts for the presense of the hidden momentum.  This is not impossible but a determined search by the author has turned up no such errors, and in fact it involved a very careful analysis simply to arrive at the situation reported here. 

By this time the reader is probably seeing the futility of this exercise, which is mentioned primarily to demonstrate that the possibility of an error or errors has been seriously considered and that errors have been sought for.  Obviously the error search has been fruitless, or the result unhappy as it is would not be so presented. Therefore the reader is entreated to attempt to find the obstensible error, any identification of which will be welcomed.  Nonconservation of the total angular momentum here is perhaps only a distraction from the equally interesting result that there is a complex dynamic involved in the electromagnetic two-body problem with spin that seems not to have been noticed previously. This dynamic would still be present even were the total angular momentum conserved, because the kinetic angular momenta and the motion of the spins would then have to be counter to the hidden and field angular momenta, which don't vanish independently. The total electric and magnetic dipole moments would thus not be constrained to be generally stationary, separately from the well-recognized dipole acceleration due to orbital motion of the separated charges, which clearly causes decay of the classical atom in the absence of spin.  That there are additional dipole acceleration components not previously considered invalidates, at least until their full consideration, the conclusion that the classical atom must of necessity and unconditionally decay radiatively on a short time scale.

Alternatively, considering that perhaps the result herein is correct, it can be noted that although the total angular momentum is nonstationary, the total magnetic moment of the quasiclassical positronium atom is not so constrained.  This was noted previously in the quasiclassical Hydrogen atom \cite{LushY9oc3}, where although the total angular momentum was not a constant of the motion, the total magnetic moment could nonetheless be stationary. The situation in the quasiclassical Positronium atom however is much richer, owing to the more rapid motions of the spins remarked about near the end of Section III. This motion may conceivably give rise to both magnetic and electric dipole radiation, since the particles acquire electric dipole moments due to their motions.  In quasiclassical Hydrogen in the simple circular orbit, for example, the oscillation of the acquired electric dipole moment can cancel the oscillation of the dipole due to charge separation, at an orbital radius that is between the nuclear and atomic scale.  Therefore such radiative effects and their reaction may be physically significant and potentially extend the applicability of classical physics further into the quantum domain.

\section{Concluding Remarks}

As described in the introduction, the 1927 analysis by L. H. Thomas found no angular momentum nonconservation in hydrogenic atoms due to the relativity precession.  A detailed analysis showing how unawareness of the existence of hidden momentum can mask the angular momentum nonconservation and that this led to Thomas overlooking the nonconservation is provided in \cite{LushY9oc4}.  The present analysis can thus be seen as in agreement with expectations based on the results of the analysis approach taken by Thomas, if performed according to modern textbooks recognizing the need for accounting for hidden momentum.  

It's also worth observing that although Thomas's 1927 analysis  of hydrogenic atoms found no angular momentum nonconservation due directly to the ``relativity precession,'' it did not either obtain strict angular momentum conservation, except in the trivial case of alignment of the spin and orbit angular momenta.  Rather, in Thomas's analysis, only the ``secular,'' {\em i.e.}, orbit averaged, total angular momentum was a constant of the motion for general relative spin and orbit orientations. This is in contrast to the present analysis.  Despite angular momentum nonconservation due to the Thomas precession in certain configurations, the present analysis obtains exact angular momentum constancy under the nontrivial conditions described, without need to average over an orbital period, but requires accounting for both the hidden and field orbital angular momenta, as well as the kinetic orbital angular momentum.

\end{widetext}


\appendix


\section{Circular Orbit Rutherford and Bohr Models of Positronium}

Suppose a classical point-charge electron and positron, without intrinsic magnetic moments, are co-circular orbiting opposite their common center of mass.  Equating the centripetal acceleration of electron circular motion with the acceleration due to the Coulomb attraction from the positron obtains

\begin{equation}
m \frac{{v_e}^2}{r_e} = \frac{e^2}{r^2} 
\label{Feqmaelec}
\end{equation}

where \(m\) and \(v_e\) are the electron mass and velocity
measured in the center-of-mass frame, \(r_e\) is the electron
distance from the center of mass, and \(r\) is the electron-positron
separation. The positron mass  is also \(m\).  Also, \(r = r_e + r_p\), where \(r_p\) is the positron distance from the center of mass, and from the definition of the center of mass and since the electron and positron masses are equal, \( r_e = r_p \), and \( r = 2r_e =2r_p \). Also,  \(\mbox{\boldmath $v$}_p =  -\mbox{\boldmath $v$}_e\). If \(\mbox{\boldmath $v$}\) is the electron velocity relative to the
positron, then \(\mbox{\boldmath $v$} =  \mbox{\boldmath $v$}_e - \mbox{\boldmath$v$}_p \), or \(\mbox{\boldmath $v$} = 2\mbox{\boldmath$v$}_e \).  Through substitutions in Eq. (\ref{Feqmaelec}) it is then obtained that

\begin{equation}
v = \frac{\sqrt{2} e}{\sqrt{m r}} 
\label{eprelvel}
\end{equation}

The electron velocity as measured in the laboratory frame (assuming the Positronium atom center of mass is stationary there) is

\begin{equation}
v_e = \frac{v}{2} = \frac{\sqrt{2}}{2} \frac{e}{\sqrt{m r}} =  \frac{1}{2} \frac{e}{\sqrt{m r_e}}.
\label{elecvelmag}
\end{equation}

The electron orbital angular momentum is

\begin{equation}
\mbox{\boldmath$L$}_e = \mbox{\boldmath$r$}_e \times m \mbox{\boldmath$v$}_e = L_e \hat{\mbox{\boldmath$L$}}= m r_e v_e \hat{\mbox{\boldmath$L$}}. \label{Leval}
\end{equation}

so

\begin{equation}
\mbox{\boldmath$L$}_e = \mbox{\boldmath$L$}_p =  e \sqrt{m r_e} \left(  \frac{1}{2}\right) \hat{\mbox{\boldmath$L$}}. 
\label{Leval6}
\end{equation}

and since \(\mbox{\boldmath$L$} =  \mbox{\boldmath$L$}_e +  \mbox{\boldmath$L$}_p \),

\begin{equation}
\mbox{\boldmath$L$} =  
e \sqrt{m_e r_e}  \hat{\mbox{\boldmath$L$}}.
\label{Ltoteval5}
\end{equation}

Alternatively, in terms of the electron-positron separation,

\begin{equation}
\mbox{\boldmath$L$} =  
e \sqrt{m r} \left(  \frac{1}{2}\right)^{1/2} \hat{\mbox{\boldmath$L$}}.
\label{LtotevalFinal}
\end{equation}

Setting \(L\) to \(\hbar\) obtains the Bohr radius of positronium as

\begin{equation}
r_{\text{B}} = \frac{2\hbar^2}{e^2 m} \approx  1.05 \times 10^{-8} \text{cm}.
\label{RBohrPos}
\end{equation}

(Note, this is the electron-positron separation, not strictly the orbital radius, but is the more convenient quantity for determining scale.)

The electron velocity at the Bohr radius is, from (\ref{elecvelmag})

\begin{equation}
v_e =  \frac{\sqrt{2}}{2} \frac{e}{\sqrt{m}} \left( \frac{e^2 m}{2\hbar^2} \right)^{1/2} =  \frac{1}{2} \frac{e^2}{\hbar} , 
\label{elecvelmagnitude6}
\end{equation}

and

\begin{equation}
\beta_e \equiv  v_e/c  =    \frac{\alpha}{2} ,   
\label{betaisalphaby2}
\end{equation}

where \(\alpha = e^2/(\hbar c) \approx 1/137\) is the fine structure constant.

Also

\begin{equation}
v_e \approx  1.10 \times 10^{8} \text{cm/sec}\approx 0.004 c,
\label{elecvelmagnitude6}
\end{equation}

which obtains

\begin{equation}
{\beta_e}^2 \equiv (v_e/c)^2 \approx 1.33 \times 10^{-5},
\label{elecbetasqd}
\end{equation}

and

\begin{equation}
\gamma - 1 \equiv (1-{\beta_e}^2)^{-1/2} - 1 \approx 6.66 \times 10^{-6}.
\label{elecbetasqd}
\end{equation}

\section{Electrogmagnetic Field of Quasiclassical Positronium}

The electromagnetic field in the vicinity of the quasiclassical Positronium atom consists of contributions due to the particle electric charges and due to their intrinsic magnetic moments.  
An electron or positron in the Positronium atom will feel an electric field due to the charge of the other particle. There is also an electric field contribution due to the motion of the intrinsic magnetic moment, but this will be seen to be of magnitude order \((v/c)^2\) compared to the Coulomb field in the present analysis. As will also be shown, in the atomic scale the magnitude of the hidden momentum of an electron in the Coulomb electric field due to a proton is of order \((v/c)^2\) compared to the electron kinetic momentum magnitude.  

The Li\'enard-Wiechert field expressions in three-vector notation are \cite{JacksonCEDoc3}

\begin{widetext}

\begin{equation}
\mbox{\boldmath$E$}(\mbox{\boldmath$r$},t) = q \left[ \frac{\mbox{\boldmath$n$}-\mbox{\boldmath$\beta$}}{\gamma^2 \left( 1 - \mbox{\boldmath$\beta$} \cdot \mbox{\boldmath$n$}\right)^3 R^2 }\right]_{\text{ret}} - \frac{q}{c} \left[ \frac{\mbox{\boldmath$n$} \times \left((\mbox{\boldmath$n$}-\mbox{\boldmath$\beta$})\times \dot{\mbox{\boldmath$\beta$}}\right)}{\left(1 - \mbox{\boldmath$\beta$} \cdot \mbox{\boldmath$n$}\right)^3 R }\right]_{\text{ret}}
\label{LW_E_field}
\end{equation}

\end{widetext}

and

\begin{equation}
\mbox{\boldmath$B$}(\mbox{\boldmath$r$},t) = \left[ \mbox{\boldmath$n$} \times \mbox{\boldmath$E$}  \right]_{\text{ret}}, 
\label{LW_B_field}
\end{equation}

where, if \(\mbox{\boldmath$r$}_0\) is the displacement from the particle to a field point, then \(R \equiv |\mbox{\boldmath$r$}_0 |\),  \(\mbox{\boldmath$n$} = \mbox{\boldmath$r$}_0/R\).  Also \(\mbox{\boldmath$\beta$}  \equiv \mbox{\boldmath$v$}/c\), where \(\mbox{\boldmath$v$}\) is the particle velocity. The subscript ``ret'' refers to that the quantity in the brackets is evaluated at the retarded time.  The Li\'enard-Wiechert field expressions are the exact fields for a possibly moving point charge, but in the present application it must be taken into account that the particles have intrinsic magnetic fields as well.  This will cause electric fields at least due to translational motion of the particles.  However for the time being this will be disregarded as the electric field form needed due to the charges is determined.

The first term on the right hand side of Eq. (\ref{LW_E_field}) is called the velocity field.  The second is called the acceleration field.  In the present application, for the electric field experienced by one of the mutually-circularly-orbiting charges due to the other, the magnitude of the acceleration field is less than \((v/c)^2\) times the magnitude of the velocity field. It can also be seen by inspection that the magnitude of the field difference from the Coulomb field of the particle (that is, the electric field in the rest frame of the particle) due to motion is small when \(\beta << 1\).    

\subsection{Electric Field due to a Circular-Orbiting Charge}

Consider a positron and electron classical point charges in a classical circular orbit around their common center of mass.  The positron and electron are assumed bound together through the Coulomb force into a perfectly circular Keplerian orbit, such that the particles are exactly opposite each other in the orbit at all times to a laboratory-frame observer, and the particle velocity vectors are at all times perpendicular to the line between them.   From the conditions of orbit circularity and direct opposition it is apparent that the particle separation \(r\) is constant, and so \( [r(t)]_{\text{ret}} \equiv r(t) \equiv r\).  From the perpendicularity of the velocity to the line between the particles, it is apparent that \(\mbox{\boldmath$\beta$} \cdot \mbox{\boldmath$n$} \equiv 0 \) at both the present and retarded times.

It should be borne in mind that the perfectly circular orbit is only approximation, because forces other than the Coulomb attraction will cause deviations from circularity.  It seems reasonable though that since the non-Coulomb forces are relatively small, the deviations they cause from perfect circularity will also be small, and can be assessed later as perturbations.

The positron velocity magnitude in the circular Coulombic orbit is (see Eq. (\ref{elecvelmag}) )

\begin{equation}
v_p =  \frac{e\sqrt{2}}{2\sqrt{m r}}.
\end{equation}

Choosing a coordinate frame to suit, suppose 

\begin{equation}
\mbox{\boldmath$r$}_p(t) = -\frac{r}{2} \mbox{\boldmath$n$}(t) = \frac{r}{2}\left( \sin \omega t \hat{\mbox{\boldmath$x$}}  - \cos \omega t \hat{\mbox{\boldmath$y$}} \right)    
\label{rp_of_t}
\end{equation}

where the definitions have been retained from above that \(r\) is the electron-positron separation, and \(\mbox{\boldmath$n$}\) is a unit vector in the direction of the electron from the location of the positron.  Then   

\begin{widetext}

\begin{equation}
\mbox{\boldmath$v$}_p(t) \equiv \dot{\mbox{\boldmath$r$}}_p(t) = \omega (r/2) \left( \cos \omega t \hat{\mbox{\boldmath$x$}}  + \sin \omega t \right) \equiv v_p\left( \cos \omega t \hat{\mbox{\boldmath$x$}}  + \sin \omega t \hat{\mbox{\boldmath$y$}} \right) \equiv v_p\hat{\mbox{\boldmath$v$}}(t), 
\label{vp_of_t_def}
\end{equation}

where

\begin{equation}
\omega = \frac{v_p}{r_p}  = \frac{e\sqrt{2}}{\sqrt{m r}} \left[ \left( \frac{1}{2}  \right) r   \right]^{-1} = \frac{v}{r} = \frac{e\sqrt{2}}{{m}^{1/2} {r}^{3/2}}.
\end{equation}

The velocity at the retarded time \( t - r/c \equiv t - D \) is then

\begin{equation}
\left[\mbox{\boldmath$v$}_p(t) \right]_{\text{ret}} =  v_p\left( \cos \left(\omega \left( t - D \right)\right) \hat{\mbox{\boldmath$x$}}  + \sin \left(\omega \left( t - D \right)\right) \hat{\mbox{\boldmath$y$}} \right). 
\label{vp_ret}
\end{equation}

Now

\begin{equation}
\cos \left(\omega \left( t - D \right)\right)= \cos \omega t  \cos \omega D + \sin \omega t  \sin \omega D \approx  \cos \omega t \left(1-\frac{\omega^2D^2}{2}\right) + \sin \omega t \left(\omega D - \frac{\omega^3D^3}{6}\right)
\end{equation}

and

\begin{equation}
\sin \left(\omega \left( t - D \right)\right) = \sin \omega t  \cos \omega D - \cos \omega t  \sin \omega D \approx  \sin \omega t\left(\omega D - \frac{\omega^3D^3}{6}\right) - \cos \omega t \left(1-\frac{\omega^2D^2}{2}\right).
\end{equation}

With \(\omega D \equiv (v/r)(r/c) = \beta\), and retaining terms to order \(\beta^2\),

\begin{equation}
\cos \left(\omega \left( t - D \right)\right) \approx (1-\beta^2) \cos \omega t   + \beta \sin \omega t = (1-\beta^2) \cos \omega t   + \beta \cos \omega (t - \pi/(2\omega))
\end{equation}

and

\begin{equation}
\sin \left(\omega \left( t - D \right)\right)  \approx (1-\beta^2)\sin \omega t - \beta \cos \omega t = (1-\beta^2) \sin \omega t   + \beta \sin \omega (t - \pi/(2\omega)),
\end{equation}

so

\begin{equation}
\left[\mbox{\boldmath$v$}_p(t) \right]_{\text{ret}} \approx \mbox{\boldmath$v$}_p(t) + \omega D \mbox{\boldmath$v$}_p(t - T/4) = \mbox{\boldmath$v$}_p(t) + \omega D \mbox{\boldmath$v$}_p(t - \pi/(2\omega)),
\label{vp_ret_approx1}
\end{equation}

where \(T = 2\pi/\omega\) and

\begin{equation}
\left[\mbox{\boldmath$n$}(t) \right]_{\text{ret}} \approx \mbox{\boldmath$n$}(t) + \omega D \mbox{\boldmath$n$}(t - \pi/(2\omega)).
\label{nhat_ret_approx1}
\end{equation}

Alternatively (with \(\sin(\theta - \pi/2) = - \cos \theta \) and \(cos(\theta - \pi/2) = \sin \theta\)),

\begin{equation}
\left[\mbox{\boldmath$n$}(t) \right]_{\text{ret}} \approx \mbox{\boldmath$n$}(t) + \omega D \hat{\mbox{\boldmath$v$}}(t)
\label{nhat_ret_approx}
\end{equation}

and

\begin{equation}
\left[\mbox{\boldmath$v$}_p(t) \right]_{\text{ret}} \approx  \omega (r/2)\left[\hat{\mbox{\boldmath$v$}}(t) - \omega D \mbox{\boldmath$n$}(t) \right],
\label{vp_ret}
\end{equation}`

or

\begin{equation}
\left[\mbox{\boldmath$\beta$}_p(t) \right]_{\text{ret}} \approx  \mbox{\boldmath$\beta$}_p(t) - \omega^2 D^2 \mbox{\boldmath$n$}(t). 
\label{betap_ret_approx}
\end{equation}`

For the circular orbit and evaluation of the field due to the positron at the position of the electron, \(\mbox{\boldmath$v$}_p(t) \cdot  \mbox{\boldmath$n$}(t) \equiv 0\), so the Li\'enard-Wiechert electric field becomes at the electron (for the time being the subscript on the velocity indicating the positron will be omitted)

\begin{equation}
\mbox{\boldmath$E$}(\mbox{\boldmath$r$},t) = e \left[ \frac{\mbox{\boldmath$n$}-\mbox{\boldmath$\beta$}}{\gamma^2 r^2 }\right]_{\text{ret}} - \frac{e}{c} \left[ \frac{\mbox{\boldmath$n$} \times \left((\mbox{\boldmath$n$}-\mbox{\boldmath$\beta$})\times \dot{\mbox{\boldmath$\beta$}}\right)}{r}\right]_{\text{ret}}.
\label{LW_E_field_red}
\end{equation}

Note, the vector \(\mbox{\boldmath$n$}\) here is from the source to the field point. This will be consistent with the definition above provided that the field is due to the positron and the field point is at the electron.  The first term on the right hand side is the velocity field term and using Eqs. (\ref{nhat_ret_approx}) and (\ref{betap_ret_approx}) evaluates to

\begin{equation}
\mbox{\boldmath$E$}_{\text{vel}}(\mbox{\boldmath$r$},t) = e \left[ \frac{\mbox{\boldmath$n$}-\mbox{\boldmath$\beta$}}{\gamma^2 r^2 }\right]_{\text{ret}}\approx e \left[ \frac{\mbox{\boldmath$n$}(t)-\mbox{\boldmath$\beta$}(t)}{\gamma^2 r^2 }\right] + e \omega D \left[ \frac{\hat{\mbox{\boldmath$v$}}(t) +\omega D \mbox{\boldmath$n$}(t)}{\gamma^2 r^2 }\right],
\label{LW_E_vel_field_red}
\end{equation}

or (to the order of the approximation), with the Coulomb electric field due to the positron charge given as \(\mbox{\boldmath$E$}_{\text{Coul}} = e\mbox{\boldmath$n$}/r^2\), and replacing \(\mbox{\boldmath$\beta$} \) with \( \mbox{\boldmath$v$}/c\),

\begin{equation}
\mbox{\boldmath$E$}_{\text{vel}}(\mbox{\boldmath$r$},t) - \frac{1}{\gamma^2} \mbox{\boldmath$E$}_{\text{Coul}} =   -  \frac{e \omega r \hat{\mbox{\boldmath$v$}}(t)}{c\gamma^2 r^2 } + e \omega D \left[  \frac{\hat{\mbox{\boldmath$v$}}(t) + \omega D \mbox{\boldmath$n$}(t)}{\gamma^2 r^2 }\right].
\label{LW_E_field_red2}
\end{equation}

With \(D \equiv r/c\), and \(\hat{\mbox{\boldmath$v$}}(t)\) given by Eq. (\ref{vp_of_t_def}),

\begin{equation}
\mbox{\boldmath$E$}_{\text{vel}}(\mbox{\boldmath$r$},t) - \frac{1}{\gamma^2} \mbox{\boldmath$E$}_{\text{Coul}} =    \frac{e \omega^2  \mbox{\boldmath$n$}(t)}{c^2 \gamma^2}. 
\label{LW_E_field_red2}
\end{equation}

Substituting \(\gamma^{-2} = 1-\beta^2\), and \(\beta=\omega D = \omega r / c\),

\begin{equation}
\mbox{\boldmath$E$}_{\text{vel}}(\mbox{\boldmath$r$},t) -  \mbox{\boldmath$E$}_{\text{Coul}} + \frac{\omega^2 r^2}{c^2 }\mbox{\boldmath$E$}_{\text{Coul}} =  \frac{e \omega^2  \mbox{\boldmath$n$}(t)}{c^2 \gamma^2}. 
\label{LW_E_field_red3}
\end{equation}

Using again that \(\mbox{\boldmath$E$}_{\text{Coul}} = e\mbox{\boldmath$n$}/R^2\), and that \(r \equiv v/\omega\),

\begin{equation}
\mbox{\boldmath$E$}_{\text{vel}}(\mbox{\boldmath$r$},t) -  \mbox{\boldmath$E$}_{\text{Coul}}  =  - \beta^2\left[1 - \frac{1}{\gamma^2}\right] \mbox{\boldmath$E$}_{\text{Coul}} =  - \beta^4 \mbox{\boldmath$E$}_{\text{Coul}}.
\label{LW_E_minus_Coul}
\end{equation}

Therefore representing the electric field at one particle due to the other as merely the instantaneous Coulomb field introduces an error only of order \(\beta^4\) compared to the exact velocity field.

The second term on the right hand side of Eq. (\ref{LW_E_field_red}) is the acceleration field term and using Eqs. (\ref{vp_ret_approx1}) and (\ref{nhat_ret_approx1}) evaluates to

\begin{equation}
\mbox{\boldmath$E$}_{\text{accel}}(\mbox{\boldmath$r$},t) =  - \frac{e}{c} \left[ \frac{\mbox{\boldmath$n$} \times \left((\mbox{\boldmath$n$}-\mbox{\boldmath$\beta$})\times \dot{\mbox{\boldmath$\beta$}}\right)}{r}\right]_{\text{ret}} = - \frac{e}{c} \left[ \frac{\mbox{\boldmath$n$} \times \left((\mbox{\boldmath$n$}-\mbox{\boldmath$\beta$})\times \dot{\mbox{\boldmath$\beta$}}\right)}{r}\right]_t + \frac{e(\omega D)^3}{c} \left[ \frac{\mbox{\boldmath$n$} \times \left((\mbox{\boldmath$n$}-\mbox{\boldmath$\beta$})\times \dot{\mbox{\boldmath$\beta$}}\right)}{r}\right]_{t'} 
\label{LW_E_field_accel_red}
\end{equation}

where here \(t'\equiv t-\pi/(2\omega)\).  For the nonrelativistic circular orbit the acceleration is perpendicular to the velocity and parallel to \(\mbox{\boldmath$n$}\) so

\begin{equation}
\mbox{\boldmath$n$} \times \left((\mbox{\boldmath$n$}-\mbox{\boldmath$\beta$})\times \dot{\mbox{\boldmath$\beta$}}\right)=  -\mbox{\boldmath$n$} \times \left(\mbox{\boldmath$\beta$}\times \dot{\mbox{\boldmath$\beta$}}\right) = -\beta \dot{\beta} \hat{\mbox{\boldmath$\beta$}}. 
\label{LW_E_field_red4}
\end{equation}

Disregarding initially the delay-related term, it is of interest to determine the relative magnitude of the velocity and acceleration field strengths in the system at hand.  Consider an electron and positron orbiting in the Bohr model ground state with a separation of \(r_{\text{B}} = 2\hbar^2/(m e^2)\), then 

\begin{equation}
\frac{|\mbox{\boldmath$E$}_{\text{accel}}|}{|\mbox{\boldmath$E$}_{\text{vel}}|} \approx \left( \frac{\beta \dot{\beta}}{cr} \right)\left( r^2 \right) = \frac{\beta \dot{\beta}r}{c} = \frac{\beta a r}{c^2} = \frac{v^3}{c^3} = \beta^3 
\label{LW_E_field_red5}
\end{equation}

and

\begin{equation}
\frac{|\mbox{\boldmath$B$}_{\text{accel}}|}{|\mbox{\boldmath$B$}_{\text{vel}}|} \approx \left( \frac{\beta \dot{\beta}}{cr} \right)\left( \frac{r^2}{\beta}\right) = \frac{\dot{\beta}r}{c} = \frac{a r}{c^2} = \frac{v^2}{c^2} = \beta^2 .
\label{LW_E_field_red6}
\end{equation}

The acceleration fields will be negligible to the analysis in both cases.  Corrections to the electric field below order \(\beta^2\) are not needed.  Since the magnetic field enters the dynamics only at the order \(\beta^2\) compared to the electric field, magnetic field corrections at order \(\beta^2\) are also clearly irrelevant here.

Next, to consider the magnitude of the delay-related terms, and recalling that \( \omega D \equiv \beta\), the magnitude of the delayed term of the acceleration electric field of Eq. (\ref{LW_E_field_accel_red}) can be seen to be a factor of \(\beta^3\) smaller than the non-delayed component.  Therefore the effect of delay on the acceleration fields of the point charges is irrelevant to the present analysis.

\end{widetext}

The electric acceleration field can thus be approximated to order \(\beta^2\) as

\begin{equation}
\mbox{\boldmath$E$}_{\text{accel}}(\mbox{\boldmath$r$},t) =    \frac{e}{c} \left[\frac{\beta \dot{\beta} \hat{\mbox{\boldmath$\beta$}} }{r}\right] = \frac{e}{c} \left[\frac{v^3  }{c^2 r^2 }\right] \hat{\mbox{\boldmath$\beta$}} = \left[\frac{e \beta^2}{r^2 }\right] \mbox{\boldmath$\beta$}
\label{LW_E_field_accel_approx_final}
\end{equation}

\subsection{Magnetic Field Due to Charge Motion}

From Eq. (\ref{LW_B_field}), the magnetic field ``velocity'' term is

\begin{equation}
\mbox{\boldmath$B$}_{\text{vel}}(\mbox{\boldmath$r$},t) \equiv \left[\mbox{\boldmath$n$} \times q \left[ \frac{\mbox{\boldmath$n$}-\mbox{\boldmath$\beta$}}{\gamma^2 \left( 1 - \mbox{\boldmath$\beta$} \cdot \mbox{\boldmath$n$}\right)^3 R^2 }\right]\right]_{\text{ret}} 
\label{LW_B_vel_field}
\end{equation}

or

\begin{equation}
\mbox{\boldmath$B$}_{\text{vel}}(\mbox{\boldmath$r$},t) =  \left[ \frac{-q(\mbox{\boldmath$n$} \times \mbox{\boldmath$\beta$})}{\gamma^2 \left( 1 - \mbox{\boldmath$\beta$} \cdot \mbox{\boldmath$n$}\right)^3 R^2 }\right]_{\text{ret}} .
\label{LW_B_vel_field}
\end{equation}

As for the case of the electric field described above, for the circular orbit \(\mbox{\boldmath$\beta$} \cdot \mbox{\boldmath$n$} \equiv 0\) at the retarded time as well as the present and so the magnetic field due to the positron charge motion is

\begin{equation}
\mbox{\boldmath$B$}_{\text{vel}}(\mbox{\boldmath$r$},t) =  \left[ \frac{-e(\mbox{\boldmath$n$} \times \mbox{\boldmath$\beta$})}{\gamma^2 r^2 }\right]_{\text{ret}} ,
\label{LW_B_vel_field}
\end{equation}

or

\begin{equation}
\mbox{\boldmath$B$}_{\text{vel}}(\mbox{\boldmath$r$},t) =  \frac{-e}{\gamma^2 r^2 } \left[ \mbox{\boldmath$n$} \times \mbox{\boldmath$\beta$}\right]_{\text{ret}} ,
\label{LW_B_vel_field}
\end{equation}

or, using Eqs. (\ref{nhat_ret_approx}) and (\ref{betap_ret_approx}),

\begin{equation}
\mbox{\boldmath$B$}_{\text{vel}}(\mbox{\boldmath$r$},t) \approx  \frac{-e}{\gamma^2 r^2 } \left[ (\mbox{\boldmath$n$} +\omega D \hat{\mbox{\boldmath$v$}} )\times (\mbox{\boldmath$\beta$}- \omega^2D^2 \mbox{\boldmath$n$})\right] .
\label{LW_B_vel_field}
\end{equation}

This approximation is accurate to order \(\beta^2\).  So, to order \(\beta^2\), 

\begin{equation}
\mbox{\boldmath$B$}_{\text{vel}} =  \frac{-e}{\gamma^2 r^2 } \left[ \mbox{\boldmath$n$}\times \mbox{\boldmath$\beta$} -\omega^3 D^3 \hat{\mbox{\boldmath$v$}} \times \mbox{\boldmath$n$}\right] .
\label{LW_B_vel_field}
\end{equation}

With \(\omega D \equiv \beta\) (see above), 

\begin{equation}
\mbox{\boldmath$B$}_{\text{vel}} =  \frac{-e}{\gamma^2 r^2 } \left[ (1+\beta^2)\mbox{\boldmath$n$}\times \mbox{\boldmath$\beta$}\right] 
\label{LW_B_vel_field}
\end{equation}

or

\begin{equation}
\mbox{\boldmath$B$}_{\text{vel}} =  \frac{-e}{r^2 } \left[ (1-\beta^4)\mbox{\boldmath$n$}\times \mbox{\boldmath$\beta$}\right]. 
\label{LW_B_vel_field}
\end{equation}

From Eqs. (\ref{LW_B_field}) and (\ref{LW_E_field_accel_approx_final}), the magnetic field ``acceleration'' term is

\begin{equation}
\mbox{\boldmath$B$}_{\text{accel}}(\mbox{\boldmath$r$},t) \approx \left[\mbox{\boldmath$n$} \times \left[\frac{e \beta^2}{r^2 }\right] \mbox{\boldmath$\beta$}\right]_{\text{ret}} .
\label{LW_B_vel_field}
\end{equation}

Recognizing that the effect of retardation is only an order \(\beta^2\) change from the field neglecting retardation, and that the magnetic acceleration field is overall only an order \(\beta^2\) contribution to the magnetic field due to charge motion, compared to the magnetic velocity field, an approximation for the magnetic field that is accurate to first order in \(\beta\) is

\begin{equation}
\mbox{\boldmath$B$}_{\text{vel}} =  \frac{-e}{R^2 } \left[ \mbox{\boldmath$n$}\times \mbox{\boldmath$\beta$}\right]  =  \frac{e}{cR^3 } \left[ \mbox{\boldmath$v$}\times \mbox{\boldmath$r$}\right]. 
\label{LW_B_vel_field}
\end{equation}

The magnetic field at the electron due to the orbital motion of the positron charge using this approximation is then

\begin{equation}
\mbox{\boldmath$B$}_v = \frac{q_p}{c r^3}\mbox{\boldmath$v$}_p \times \mbox{\boldmath$r$} = -\frac{2q_p}{c r^3}\mbox{\boldmath$v$}_p \times \mbox{\boldmath$r$}_p = \frac{2q_p}{m c r^3}\mbox{\boldmath$L$}_p, 
\label{Bdue2poscharge}
\end{equation}

since \(\mbox{\boldmath$r$} = -2 \mbox{\boldmath$r$}_p = 2 \mbox{\boldmath$r$}_e \), and \( \mbox{\boldmath$L$}_p  = \mbox{\boldmath$r$}_p \times m \mbox{\boldmath$v$}_p \).    Noting \( \mbox{\boldmath$L$} = 2 \mbox{\boldmath$L$}_p \) obtains that

\begin{equation}
\mbox{\boldmath$B$}_v = \frac{q_p}{m c r^3}\mbox{\boldmath$L$}.
\label{Bdue2poscharge}
\end{equation}

The torque generated by the magnetic field of the other particle charge translational motion will be shown to vanish, but there is a nonvanishing radial force component.  It is worthwhile to compare this force magnitude to the Coulomb force magnitude.  The force on the electron transiting the magnetic field due to the positron charge motion is

\begin{equation}
\left|\mbox{\boldmath$F$}\right| = \left|-e\mbox{\boldmath$\beta$}_e \times \mbox{\boldmath$B$}_v\right| = \frac{e^2 \beta }{m c r^3}\left|\mbox{\boldmath$L$}\right|.
\label{Bdue2poscharge}
\end{equation}

At the Bohr radius

\begin{equation}
\left|\mbox{\boldmath$F$}\right| = \frac{e^2 \beta \hbar}{m c {r_{\text{B}}}^3}
\label{Bdue2poscharge}
\end{equation}

The relative magnitude of this force to the Coulomb attractive force between the electron and positron at the Bohr separation is

\begin{equation}
\frac{\left|\mbox{\boldmath$F$}\right|}{\left|\mbox{\boldmath$F$}_{\text{Coul}}\right|} = \frac{e^2 \beta \hbar}{m c {r_{\text{B}}}^3} \left( \frac{e^2}{{r_{\text{B}}}^2}\right)^{-1}
\label{Bdue2poscharge}
\end{equation}

or

\begin{equation}
\frac{\left|\mbox{\boldmath$F$}\right|}{\left|\mbox{\boldmath$F$}_{\text{Coul}}\right|} = \frac{\beta \hbar}{m c {r_{\text{B}}}} = \frac{\beta \hbar}{m c } \left( \frac{2 \hbar^2}{e^2 m}\right)^{-1}
\label{Bdue2poscharge}
\end{equation}

or

\begin{equation}
\frac{\left|\mbox{\boldmath$F$}\right|}{\left|\mbox{\boldmath$F$}_{\text{Coul}}\right|} =  \frac{\beta}{2} \left( \frac{e^2}{\hbar c}\right) =  \beta^2, 
\label{Bdue2poscharge}
\end{equation}

where the result of Appendix A that the fine structure constant \(\alpha = e^2/(\hbar c) = 2 \beta\) has been used. The change in the electron-positron binding force due to this component of the magnetic field is therefore of order \(\beta^2\), which is of the same order as the spin-orbit coupling, and so this effect is not negligible to a calculation of the total coupling magnitude.  However because it is unrelated to the angular momentum analysis, due to that it is a radial force and so does not contribute a torque, it is of no concern to the present analysis.

\subsection{Magnetic Field Due to Intrinsic Magnetic Moments}

The total magnetic field at the electron is due to the intrinsic magnetic moment of the positron and to the motion of the positron charge.  This can be expressed as \(\mbox{\boldmath$B$} = \mbox{\boldmath$B$}_\mu + \mbox{\boldmath$B$}_v\).  The magnetic field component due to a magnetic dipole of moment \(\mbox{\boldmath$\mu$}\) is

\begin{equation}
\mbox{\boldmath$B$}_\mu = \frac{3\mbox{\boldmath$n$} \left(
\mbox{\boldmath$n$} \cdot \mbox{\boldmath$\mu$} \right)  - \mbox{\boldmath$\mu$}}{r^3},
\label{Bdue2mu_gen}
\end{equation}

where \(\mbox{\boldmath$n$} = \mbox{\boldmath$r$}/r\) is a unit vector in the direction from the dipole to the field point.

It will be illustrative to consider the relative magnitude of the magnetic fields due to orbital motion and intrinsic moment, for the quasiclassical Positronium atom at the Bohr ground state radius, and for the intrinsic moment perpendicular to the orbital plane.  Then \(\mbox{\boldmath$B$}_\mu = -\mbox{\boldmath$\mu$}/r^3\) and

\begin{equation}
B_\mu \equiv |\mbox{\boldmath$B$}_\mu| = \frac{\mu}{r^3} = \frac{g e s}{2mcr^3} = \frac{e \hbar}{2mcr^3}.
\label{Bdue2mu_gen}
\end{equation}

The ratio of the positron intrinsic magnetic field maagnitude to the positron charge velocity magnetic field magnitude at the electron is

\begin{equation}
\frac{B_\mu}{B_v} = \frac{e \hbar}{2mcr^3} \left(\frac{eL}{mcr^3} \right)^{-1} =\frac{\hbar}{2 L}.
\label{Bdue2mu_gen}
\end{equation}

At the Bohr radius, where \(L=\hbar\),

\begin{equation}
\frac{B_\mu}{B_v} = \frac{1}{2}.
\label{Bdue2mu_gen}
\end{equation}

Therefore the intrinsic magnetic field is also not negligible to the spin-orbit coupling magnitude.  Also, unlike the magnetic field due to the other-particle charge motion, it can generate a torque for non-perpendicular spin orientations relative to the orbital plane, and so has already been accounted for in the analysis of angular momentum motion.  As noted previously, the effect on the spin-orbit coupling magnitude of this force in the quasiclassical Hydrogen atom was addressed in \cite{KholmetskiiY10a, LushY10a, KholmetskiiY10b, LushY10b}.

\subsection{Electric field due to translational and orientational motion of intrinsically-magnetic particles}

The vector potential {\boldmath $A$} due to a magnetic moment
{\boldmath$\mu$} at a field point outside the source region is \cite{JacksonCEDoc4}

\begin{equation}
\mbox{\boldmath$A$} = \frac{\mbox{\boldmath$\mu$} \times
\mbox{\boldmath$r$}_0}{{r_0}^3},
\label{VecPotDueToMagMom}
\end{equation}

where \(\mbox{\boldmath$r$}_0\) here is the displacement from the dipole to the stationary field point, and \(r_0 = |\mbox{\boldmath$r$}_0|\).  Requiring the field point to be at the other particle obtains \(r_0 \equiv r\).

The electric field is obtained from the vector and scalar potentials as

\begin{equation}
\mbox{\boldmath $E$} = - \frac{1}{c} \frac{ \partial \mbox{\boldmath
$A$}}{\partial t}  - \nabla \Phi.
\label{elecfld}
\end{equation}

Although for a stationary magnetic dipole the scalar potential vanishes, it cannot be assumed to vanish in the present application where the magnetic dipole is translating.  From Eq. (\ref{VecPotDueToMagMom}),

\begin{equation}
\frac{ \partial \mbox{\boldmath
$A$}}{\partial t} =  -\frac{3}{r^4}  \left(  \mbox{\boldmath$\mu$}
\times \mbox{\boldmath$r$}_0 \right)  \frac{ \partial r}{\partial t} + \frac{1}{r^3} \frac{ \partial }{\partial t} \left(
\mbox{\boldmath$\mu$} \times \mbox{\boldmath$r$}_0 \right).
\label{elecfldmm2}
\end{equation}

In the present application only circular orbits are considered.  For the circular orbit, \( \partial r/\partial
t \equiv 0 \), obtaining that

\begin{equation}
\frac{ \partial \mbox{\boldmath
$A$}}{\partial t} =  \frac{1}{r^3} \left( \frac{
\partial\mbox{\boldmath$\mu$} }{\partial t}    \times
\mbox{\boldmath$r$}_0 - \frac{
\partial\mbox{\boldmath$r$}_0 }{\partial t}    \times
\mbox{\boldmath$\mu$} \right).
\label{Evcrossmu}
\end{equation}

Noting that the vector \(\mbox{\boldmath$r$}_0\) here is from the dipole to the arbitrary field point, and so  \(\mbox{\boldmath$v$} = - \partial \mbox{\boldmath$r$}_0 / \partial t\), the electric field component due to the motion of the positron magnetic moment relative to the electron, at the electron, may be rewritten as

\begin{equation}
\frac{ \partial \mbox{\boldmath$A$}}{\partial t} =  \frac{1}{r^3} \left( \dot{\mbox{\boldmath$\mu$} } \times
\mbox{\boldmath$r$} -  \mbox{\boldmath$v$}  \times \mbox{\boldmath$\mu$} \right).
\label{elecfldmmorb}
\end{equation}

The first term will be nonzero for a precessing spin moment while
the second will be nonzero due to the orbital motion of the electron
around the proton.  The relative magnitude of the two terms can be evaluated.  The first cross product inside the parentheses is upper bounded as 

\begin{equation}
\left| \dot{\mbox{\boldmath$\mu$}} \times
\mbox{\boldmath$r$}\right| \le \dot{\mu}r,
\label{elecfldmmco2}
\end{equation}

with \(\dot{\mu} \equiv  e \dot{s}/(m c)\), with \(\dot{s} \equiv |\dot{\mbox{\boldmath$s$}}|\).

An upper bound on \(\dot{s}\) is developed in Appendix D, for the quasiclassical Positronium atom model in the Bohr ground state, as

\begin{equation}
\left|\dot{\mbox{\boldmath$s$}}_e\right| \le \frac{7}{32} \frac{e^8 m}{\hbar^4 c^2}, 
\end{equation}

so, at the Bohr radius

\begin{equation}
\dot{\mu}r \le \frac{7}{32} \frac{e^8 m}{\hbar^4 c^2} \frac{e}{m c}\frac{2 \hbar^2}{m e^2}.
\end{equation}

The second term in the parentheses on the right hand side of Eq. (\ref{Evcrossmu}) can be upper bounded as

\begin{equation}
|\mbox{\boldmath$v$}\times \mbox{\boldmath$\mu$}| \le \frac{e}{{m_r}^{1/2}r^{1/2}}\frac{e \hbar}{m c} = \frac{e \sqrt{2}}{{m}^{1/2}r^{1/2}}\frac{e \hbar}{m c}.
\label{elecfldmmorb}
\end{equation}

At the Bohr radius

\begin{equation}
|\mbox{\boldmath$v$}\times \mbox{\boldmath$\mu$}| \le \frac{e \sqrt{2}}{{m}^{1/2}}\frac{e \hbar}{m c}\left(\frac{\hbar^2}{m_r e^2} \right)^{-1/2} = \frac{e^3}{m c}.
\label{elecfldmmorb}
\end{equation}

The ratio of the two term bounds is

\begin{equation}
\frac{7}{32} \frac{e^8 m}{\hbar^4 c^2} \frac{e}{m c}\frac{2 \hbar^2}{m e^2}\left(\frac{e^3}{m c}\right)^{-1} = \frac{7}{16} \frac{e^4}{\hbar^2 c^2} = \frac{7}{16} \alpha^2, 
\end{equation}

where \(\alpha = e^2/(\hbar c) \approx 1/137 \) is the fine structure constant.  As derived in the Appendix, \(v/c\) for Positronium in the Bohr ground state is \(\alpha/2\), so the force due to \(\dot{\mbox{\boldmath$\mu$}}\) is only about of magnitude \((v/c)^2\) compared to the \(\mbox{\boldmath$v$}\times \mbox{\boldmath$\mu$}\) force.  Since as shown in Appendix B5 below, the latter is already a \((v/c)^2\) magnitude compared to the Coulomb attraction force, the \(\dot{\mbox{\boldmath$\mu$}}\) related force can be neglected to order \((v/c)^2\) overall.

To evaluate the electric field term due to the scalar potential gradient, recall that a translating magnetic dipole acquires an electric dipole moment,  \(\mbox{\boldmath$d$}\),  as  \cite{Fisher71}

\begin{equation}
\mbox{\boldmath$d$} = \mbox{\boldmath$v$} \times \mbox{\boldmath$\mu$}/c,
\label{Edipoledtv}
\end{equation}

where \(\mbox{\boldmath$v$}\) here and until stated otherwise below is the laboratory frame translational velocity of the magnetic dipole (as opposed to the convention used otherwise throughout, that \(\mbox{\boldmath$v$}\) is the electron velocity relative to the positron).

The scalar potential due to an electric dipole is given by

\begin{equation}
\Phi(\mbox{\boldmath$r$}) = \frac{1}{r^3}\mbox{\boldmath$d$} \cdot \mbox{\boldmath$r$},
\label{potdtEd}
\end{equation}

and

\begin{equation}
-\nabla \Phi = \frac{1}{r^3}\left(3(\mbox{\boldmath$d$} \cdot \mbox{\boldmath$n$})\mbox{\boldmath$n$} - \mbox{\boldmath$d$}\right),
\label{Efielddtd}
\end{equation}

where \( \mbox{\boldmath$n$} = \mbox{\boldmath$r$}/r\) is a unit vector from the source to the field point.

For the electric dipole moment due to the translational motion of the magnetic dipole this becomes

\begin{equation}
-\nabla \Phi = \frac{1}{cr^3}\left(3((\mbox{\boldmath$v$} \times \mbox{\boldmath$\mu$}) \cdot \mbox{\boldmath$n$})\mbox{\boldmath$n$} - (\mbox{\boldmath$v$} \times \mbox{\boldmath$\mu$})\right).
\label{Efielddtd2}
\end{equation}

The total electric field due to translational motion of the magnetic dipole, accurate to within an error of \(\beta^2\) times the magnitude of the Coulomb electric field strength, is thus

\begin{equation}
\mbox{\boldmath $E$}_{\mbox{\boldmath$v$} \times \mbox{\boldmath$\mu$}} =  \frac{1}{c r^3}\left(3((\mbox{\boldmath$v$} \times \mbox{\boldmath$\mu$}) \cdot \mbox{\boldmath$n$})\mbox{\boldmath$n$} - 2(\mbox{\boldmath$v$} \times \mbox{\boldmath$\mu$})\right).
\label{elecfldmmorb}
\end{equation}

\subsection{Magnitude of Electric Field Due to  Motion of Intrinsic Magnetic Moments}

It is of interest to determine the relative magnitude of the electric field due to translational motion of the intrsinsically-magnetic particles to the magnitude of the Coulomb electric field.  It is needed for assurance that its contribution to the total momentum, through the hidden momentum that is proportional to the electric field, is negligible in the equation of motion of the momentum.  (It is not negligible through the force that it generates, but this is not the only way it enters into the equation of motion of the total angular momentum.)

Eq. (\ref{elecfldmmorb}) provided the total electric field at one circular-orbiting particle due to translational motion of the other particle magnetic dipole. The ratio this electric field magnitude to the Coulomb electric field magnitude is bounded as

\begin{equation}
\frac{\left|\mbox{\boldmath $E$}_{\mbox{\boldmath$v$} \times \mbox{\boldmath$\mu$}}\right|}{\left|\mbox{\boldmath $E$}_{\text{Coul}}\right|} \le \frac{1}{c e r}\left[9\left|((\mbox{\boldmath$v$} \times \mbox{\boldmath$\mu$}) \cdot \mbox{\boldmath$n$})\mbox{\boldmath$n$}\
\right|^2 + 4\left|\mbox{\boldmath$v$} \times \mbox{\boldmath$\mu$}\right|^2\right]^{1/2}
\label{elecfldmmorb_copy}
\end{equation}

or

\begin{equation}
\frac{\left|\mbox{\boldmath $E$}_{\mbox{\boldmath$v$} \times \mbox{\boldmath$\mu$}}\right|}{\left|\mbox{\boldmath $E$}_{\text{Coul}}\right|} \le \frac{1}{c e r}\left[13 v^2 \mu^2 \right]^{1/2} = \frac{\sqrt{13} v \mu}{c e r}
\label{elecfldmmorb_copy}
\end{equation}

or

\begin{equation}
\frac{\left|\mbox{\boldmath $E$}_{\mbox{\boldmath$v$} \times \mbox{\boldmath$\mu$}}\right|}{\left|\mbox{\boldmath $E$}_{\text{Coul}}\right|} \le  \frac{\sqrt{13} }{c e r} \frac{e\sqrt{2}}{m^{1/2}r^{1/2}} \frac{e \hbar}{2 m c} = \frac{\sqrt{13}}{\sqrt{2}}\frac{e \hbar}{c^2 m^{3/2}r^{3/2}}
\label{elecfldmmorb_copy}
\end{equation}

At the Bohr radius

\begin{equation}
\frac{\left|\mbox{\boldmath $E$}_{\mbox{\boldmath$v$} \times \mbox{\boldmath$\mu$}}\right|}{\left|\mbox{\boldmath $E$}_{\text{Coul}}\right|} \le  \frac{\sqrt{13}}{\sqrt{2}}\frac{e \hbar}{c^2 m^{3/2}} \left(\frac{2 \hbar^2}{e^2 m}\right)^{-3/2}
\label{elecfldmmorb_copy}
\end{equation}

or

\begin{equation}
\frac{\left|\mbox{\boldmath $E$}_{\mbox{\boldmath$v$} \times \mbox{\boldmath$\mu$}}\right|}{\left|\mbox{\boldmath $E$}_{\text{Coul}}\right|} \le  \frac{\sqrt{13}}{4}\frac{e^4}{c^2 \hbar^2} = \frac{\sqrt{13}}{4} \alpha^2
\label{elecfldmmorb_copy}
\end{equation}

This is of the order of \(\beta^2\).

\section{Time Rate of Change of the Electric Field}

Because the time derivative of the electric field appears in the time rate of change of the kinetic momentum (see Eq. (\ref{MassTimesAccel})), it is desirable to find a compact expression for it that is sufficiently accurate.  When it was assumed in Section IV that it was adequate to simply represent the electric field as the Coulomb field in the time-derivative-containing terms, it was concluded that these terms contributed only at order \(\beta^2\) and below compaered to the translational force term.  Therefore, since the needed accuracy of the time rate of change of the kinetic momentum is to relative order of \(\beta^2\), it is needed to confirm that there are no other contributions to the electric field time derivative that are significant to the order of the Coulomb field time derivative or higher.  This will be confirmed in this section. 

Based on the analyses in Appendix B, the electric field accurate to order of \(\beta^2\) is 

\begin{eqnarray}
\mbox{\boldmath$E$} = \frac{e\mbox{\boldmath$r$}}{r^3} +  \frac{1}{c r^3}\left(3((\mbox{\boldmath$v$} \times \mbox{\boldmath$\mu$}) \cdot \mbox{\boldmath$n$})\mbox{\boldmath$n$} - 2(\mbox{\boldmath$v$} \times \mbox{\boldmath$\mu$})\right).
\label{EFieldRestatement}
\end{eqnarray}

(The above omits the acceleration field, which is an order \(\beta^3\) correction, and deviation between the Coulomb field and the exact Lienard-Wiechert velocity electric field, which is an order \(\beta^4\) correction.)

The time rate of change of the approximate electric field is then

\begin{eqnarray}
\dot{\mbox{\boldmath$E$}} \approx  \frac{d}{dt}\left(\frac{e\mbox{\boldmath$r$}}{r^3} +  \frac{1}{c r^3}\left(3((\mbox{\boldmath$v$} \times \mbox{\boldmath$\mu$}) \cdot \mbox{\boldmath$n$})\mbox{\boldmath$n$} - 2(\mbox{\boldmath$v$} \times \mbox{\boldmath$\mu$})\right)\right).
\label{EFieldRestatement}
\end{eqnarray}

Assuming a circular orbit obtains that

\begin{widetext}

\begin{eqnarray}
\dot{\mbox{\boldmath$E$}} \approx \frac{e\mbox{\boldmath$v$}}{r^3} +  \frac{1}{c r^3}\left(3\mbox{\boldmath$n$}\left[((\mbox{\boldmath$a$} \times \mbox{\boldmath$\mu$}) \cdot \mbox{\boldmath$n$}) + ((\mbox{\boldmath$v$} \times \dot{\mbox{\boldmath$\mu$}}) \cdot \mbox{\boldmath$n$})  +  \left((\mbox{\boldmath$v$} \times \mbox{\boldmath$\mu$}) \cdot \frac{\mbox{\boldmath$v$}}{r}\right)  \right] + 3((\mbox{\boldmath$v$} \times \mbox{\boldmath$\mu$}) \cdot \mbox{\boldmath$n$})\frac{\mbox{\boldmath$v$}}{r} - 2(\mbox{\boldmath$a$} \times \mbox{\boldmath$\mu$})- 2(\mbox{\boldmath$v$} \times \dot{\mbox{\boldmath$\mu$}})\right).
\label{EFieldDotAll}
\end{eqnarray}

To determine which terms above are relevant to the analysis, consider first that the magnitude of the first term, that is the time derivative of the Coulomb field, can be evaluated as

\begin{equation}
\left|\frac{e\mbox{\boldmath$v$}}{r^3}\right| = \frac{e}{r^3} \frac{e\sqrt{2}}{m^{1/2}r^{1/2}}= \frac{e^2\sqrt{2}}{m^{1/2}r^{7/2}}.
\label{Edot1}
\end{equation}

This magnitude will serve as a reference for the magnitude bounds of the other terms in (\ref{EFieldDotAll}).

Next consider

\begin{equation}
\left|\frac{1}{c r^3}\left(3\mbox{\boldmath$n$}((\mbox{\boldmath$a$} \times \mbox{\boldmath$\mu$}) \cdot \mbox{\boldmath$n$})\right)\right| \le \frac{3\mu}{c r^3}\frac{v^2}{r} = \frac{3}{c r^4}\frac{e \hbar}{2 m c}\frac{2 e^2}{m r} = \frac{3e^3 \hbar}{m^2 c^2 r^5}.
\label{Edot2}
\end{equation}

The relative magnitude of this term compared to the Coulomb term is thus bounded as

\begin{equation}
\frac{\left|\frac{1}{c r^3}\left(3\mbox{\boldmath$n$}((\mbox{\boldmath$a$} \times \mbox{\boldmath$\mu$}) \cdot \mbox{\boldmath$n$})\right)\right|}{\left|\frac{e\mbox{\boldmath$v$}}{r^3}\right|} \le \left(\frac{3e^3 \hbar}{m^2 c^2 r^5}\right)\left(\frac{e^2\sqrt{2}}{m^{1/2}r^{7/2}} \right)^{-1} = \left(\frac{3e^3 \hbar}{m^2 c^2 r^5}\right)\left(\frac{m^{1/2}r^{7/2}}{e^2\sqrt{2}} \right) = \frac{3}{\sqrt{2}}\left(\frac{e \hbar}{m^{3/2} c^2 r^{3/2}}\right) .
\label{Edot2}
\end{equation}

At the Bohr radius

\begin{equation}
\frac{\left|\frac{1}{c r^3}\left(3\mbox{\boldmath$n$}((\mbox{\boldmath$a$} \times \mbox{\boldmath$\mu$}) \cdot \mbox{\boldmath$n$})\right)\right|}{\left|\frac{e\mbox{\boldmath$v$}}{r^3}\right|} \le \frac{3}{\sqrt{2}}\frac{e \hbar}{m^{3/2} c^2 } \left(\frac{2\hbar^2}{e^2m} \right)^{-3/2} = \frac{3}{\sqrt{2}}\frac{e \hbar}{m^{3/2} c^2 }\left(\frac{e^3m^{3/2}}{2^{3/2}\hbar^3} \right)= \frac{3}{4}\frac{e^4}{c^2\hbar^2} = \frac{3}{4}\alpha^2,
\label{Edot2}
\end{equation}

which is of order \(\beta^2\) here.

Next consider

\begin{equation}
\left|\frac{3\mbox{\boldmath$n$}}{c r^3}\left[ (\mbox{\boldmath$v$} \times \dot{\mbox{\boldmath$\mu$}}) \cdot \mbox{\boldmath$n$}  \right]  \right| \le \frac{3}{c r^3}v\dot{\mu} = \frac{3}{c r^3}\frac{\sqrt{2}e}{m^{1/2}r^{1/2}} \frac{e}{m c}\dot{s} = \frac{3}{c r^3}\frac{\sqrt{2}e}{m^{1/2}r^{1/2}} \frac{e}{m c}\frac{7}{32} \frac{e^8 m}{\hbar^4 c^2}  = \frac{21\sqrt{2}}{32}\frac{1}{m^{1/2}r^{7/2}}\frac{e^{10}}{\hbar^4 c^4}.
\label{Edot3}
\end{equation}

The relative magnitude of this term compared to the Coulomb term is thus bounded as

\begin{equation}
\frac{\left|\frac{3\mbox{\boldmath$n$}}{c r^3}\left[ (\mbox{\boldmath$v$} \times \dot{\mbox{\boldmath$\mu$}}) \cdot \mbox{\boldmath$n$}  \right]  \right|}{\left|\frac{e\mbox{\boldmath$v$}}{r^3}\right|} \le  \frac{21\sqrt{2}}{32}\frac{1}{m^{1/2}r^{7/2}}\frac{e^{10}}{\hbar^4 c^4}\left(\frac{e^2\sqrt{2}}{m^{1/2}r^{7/2}} \right)^{-1} = \frac{21}{32}\frac{e^{8}}{\hbar^4 c^4} = \frac{21}{32}\alpha^4  .
\label{Edot2}
\end{equation}

\end{widetext}

All other terms in Eq. (\ref{EFieldDotAll}) are of the same magnitude as one or the other of these two terms already considered.  Therefore, approximating the electric field as the Coulomb field introduces errors in the time rate of change of the electric field at only relative order \(\beta^2\) and so is fully adequate for the present purpose.

\section{Upper Bound on the Time Rate of Change of the Intrinsic Spin Vector}

The order of magnitude of the rates of change of the electron and positron spin vectors, \(\dot{s}_e\equiv|\dot{\mbox{\boldmath$s$}}_e|\) and \(\dot{s}_p\equiv|\dot{\mbox{\boldmath$s$}}_p|\),  is needed at various points in the analysis in order to determine the relative significance of various quantities.  From Eq. (\ref{ThomasEqRedElectron}),

\begin{equation}
|\dot{\mbox{\boldmath$s$}}_e| = \left|\frac{e^2}{m^2 c^2 r^3}\mbox{\boldmath$s$}_e \times \left[ 3\mbox{\boldmath$n$} \left(
\mbox{\boldmath$n$} \cdot \mbox{\boldmath$s$}_p\right)   -
\mbox{\boldmath$s$}_p + \left(\frac{3}{2}\right) \mbox{\boldmath$L$} \right]\right|.
\label{sdot_mag1}
\end{equation}

To develop an upper bound on \(\dot{s}\), consider first the quantity in the brackets on the right hand side of Eq. (\ref{sdot_mag1}). Also suppose that the orbital radius is the Bohr radius, where \(L=\hbar\).  Then  

\begin{equation}
\left| 3\mbox{\boldmath$n$} \left(
\mbox{\boldmath$n$} \cdot \mbox{\boldmath$s$}_p\right)   -
\mbox{\boldmath$s$}_p + \left(\frac{3}{2}\right) \mbox{\boldmath$L$}\right| \le \frac{7\hbar}{2}, 
\label{sdot_mag2}
\end{equation}

and

\begin{equation}
\left|\dot{\mbox{\boldmath$s$}}_e\right| \le \frac{e^2 \hbar}{2m^2 c^2} \left(\frac{2 \hbar^2}{m e^2} \right)^{-3} \frac{7\hbar}{2} = \frac{7}{32} \frac{e^8 m}{\hbar^4 c^2}. 
\label{sdot_bound_appendix}
\end{equation}

\section{Radiation Damping}

Because an electron in a circular orbit will radiate electromagnetic energy and momentum, as expected from conservation considerations a compensating force known as the radiation reaction force will cause the particles to decelerate. The radiation reaction is given by \cite{JacksonCEDoc5}

\begin{equation}
\mbox{\boldmath $F$}_{\text{rad}} =
\frac{2}{3}\frac{e^2}{c^3}\ddot{\mbox{\boldmath $v$}}.
\label{radresistforce}
\end{equation}

For the circular orbit, for the electron, the radiation reaction force magnitude is

\begin{eqnarray} 
F_{\text{rad}} = \left| \mbox{\boldmath $F$}_{\text{rad}} \right| =
\frac{2}{3}\frac{e^2}{c^3}\frac{{v_e}^3}{{r_e}^2} = \frac{2}{3}\frac{e^2}{c^3}\frac{1}{{r_e}^2}\left( \frac{e}{2\sqrt{m r_e}} \right)^3  \label{Fradcirce}
\end{eqnarray}

or

\begin{equation}
F_{\text{rad}} = \frac{2}{3}\frac{e^5}{c^3}{r}^{-7/2}\left(\frac{1}{2}\right)^{-7/2}\left( \frac{1}{2\sqrt{m}} \right)^3 \label{Fradcirce3}
\end{equation}

or

\begin{eqnarray}
F_{\text{rad}} =
\frac{2\sqrt{2}}{3}\frac{e^5}{c^3} r^{-7/2} m^{-3/2} .
\label{Fradcirce4}
\end{eqnarray}

To compare this with the Coulomb attraction force:

\begin{equation}
\frac{F_{\text{rad}}}{F_{\text{Coul}}}  =
\frac{2\sqrt{2}}{3}\frac{e^5}{c^3} r^{-7/2} m^{-3/2} \left(\frac{e^2}{r^2}\right)^{-1}
\label{Fradcirce4}
\end{equation}

or

\begin{equation}
\frac{F_{\text{rad}}}{F_{\text{Coul}}}  =
\frac{2\sqrt{2}}{3}\frac{e^3}{c^3} r^{-3/2} m^{-3/2} .
\label{Fradcirce4}
\end{equation}

At the Bohr radius

\begin{equation}
\frac{F_{\text{rad}}}{F_{\text{Coul}}}  =
\frac{2\sqrt{2}}{3}\frac{e^3}{c^3}  m^{-3/2}\left(\frac{\hbar^2}{e^2 m}\right)^{-3/2} 
\label{Fradcirce4}
\end{equation}

or

\begin{equation}
\frac{F_{\text{rad}}}{F_{\text{Coul}}}  =
\frac{2\sqrt{2}}{3}\frac{e^6}{c^3\hbar^3} = \frac{2\sqrt{2}}{3}\alpha^3.
\label{FradtoCoul}
\end{equation}

So (since \(\alpha = 2\beta\) here, by Eq. (\ref{betaisalphaby2})), the radiation reaction force is of order \(\beta^3\) smaller than the Coulomb attractive force, and so based on other results herein will be order \(\beta\) smaller than the non-Coulomb forces due to the motion of the dipole-carrying particles, and so is negligible to the present analysis.  This result is consistent with the finding that the radiation fields are negligible to the present analysis, since the angular momentum lost to radiation must be reflective of the angular momentum change due to the radiation damping forces.  Therefore the neglect of the radiation fields and radiation damping due to charge motion cannot change the central result of the present work, that the Thomas precession leads to the nonconservation of total angular momentum. 

Radiation damping will also be present associated with radiation due to acceleration of the particle magnetic dipole moments \cite{Itoh1991}, and including due to acceleration of the electric dipole moment that results when the dipole-carrying particles translate. However, these radiation damping forces can be shown \cite{LushInPrep} to be at most of similar magnitude to those due to charge motion, at the atomic scale. They will also be accompanied by angular momentum in the associated radiation fields, equal to the amount lost due to the damping forces, and so cannot be expected to overcome the nonconservation of total angular momentum due to Thomas precession.    

\section{Effect of Delay on Electric Force Induced by Motion of Intrinsically-Magnetic Particles}

It has already been shown that representing the exact, delayed, electric field at one particle due to the other particle charge as simply the instantaneous Coulomb field is a very close approximation. It follows directly that the effect of delay on the Coulomb attraction is very small. Since the instantaneous Coulomb field deviates from the retarded velocity electric field only at order \(\beta^4\) (see Eq. (\ref{LW_E_minus_Coul})), the ratio of the force error to the overall magnitude of the Coulomb attraction will also be of order \(\beta^4\), which makes it an order of \(\beta\) smaller than the radiation reaction force as determined relative to the Coulomb force by Eq. (\ref{FradtoCoul}).  Therefore it is of no interest to the present analysis.   

A second delay effect can be identified that has not yet been addressed, and that is the effect of propagation delay on the electric force induced by the motion of the magnetic dipoles. Although this latter delay effect will also be insignificant to the present analysis, the non-radial force it contributes will be seen to be an order of \(\beta\) larger than the non-radial Coulomb force component due to delay, at the scale under consideration here. That is, it will be about equal to the radiation reaction force, in the quasiclassical positronium atom in the Bohr gound state. 

For the case of non-relativistic circular orbital motion, the
propagation delay, \(D\), is approximately that associated with the
orbital radius as

\begin{equation}
D =  \frac{r}{c}.
\label{propdelay}
\end{equation}

The angular change of the velocity vector during the propagation
delay is

\begin{equation}
\phi_D =  \frac{vD}{r} =  \frac{v}{c} = \beta. \label{propdelayangle}
\end{equation}

To get an idea of the possible magnitude of the force resulting from
non-instantaneity, we will compute the force magnitude based on
(\ref{elecfldmmorb}) for the case of a circular orbit and spin perpendicular to the orbital plane. The
force acting on the positron due to the orbital motion of the electron intrinsic magnetic moment is then given by

\begin{equation}
\mbox{\boldmath $F$}_{\mbox{\boldmath$v$} \times \mbox{\boldmath$\mu$}} = e \mbox{\boldmath $E$}_{\mbox{\boldmath$v$} \times \mbox{\boldmath$\mu$}} = \frac{e}{cr^3} \left( \mbox{\boldmath$v$}  \times
\mbox{\boldmath$\mu$}_e \right).
\label{forceelecfldmmorb}
\end{equation}

The force magnitude is

\begin{eqnarray}
F_{\mbox{\boldmath$v$} \times \mbox{\boldmath$\mu$}} = \frac{e}{cr^3} \left( v  \mu \right) \approx
\frac{e}{cr^3} \left( \frac{e \sqrt{2}}{\sqrt{m r} }  \frac {e \hbar }{2m c} \right) \nonumber\\ = \frac{e^3 \hbar \sqrt{2}}{2 c^2 } r^{-7/2} m^{-3/2},
\label{forcemagelecfldmmorb}
\end{eqnarray}

where the approximation is to neglect the distinction between the
electron actual and reduced mass.  Accounting for propagation delay,
the component along the direction of the proton motion is
approximately

\begin{equation}
F_{\parallel} \approx \sin{\left(\frac{v}{c}\right)}\frac{e^3\sqrt{2}
\hbar}{2 c^2 } r^{-7/2} m^{-3/2} \approx \frac{v}{c}\frac{e^3
\sqrt{2}\hbar}{2 c^2 } r^{-7/2} m^{-3/2}, \label{forcemagelecfldmmorbpvel}
\end{equation}

or

\begin{equation}
F_{\parallel} \approx \frac{1}{c}\frac{e\sqrt{2}}{\sqrt{m r} }\frac{e^3\sqrt{2}
\hbar}{2 c^2 } r^{-7/2} m^{-3/2} = \frac{2e^4 \hbar}{2 c^3 } r^{-4}
m^{-2}.
\label{forcemagelecfldmmorbpvel2}
\end{equation}

For comparison, consider the relative magnitude of the force due to radiation reaction on the electron.  By equating the
non-radial force component due to delay and the radiation reaction force, which will be seen to have differing dependencies on the electron-positron
separation, the separation value at which the two
forces will be of equal magnitude can be determined.

Equating the radiation reaction on the electron, as given by (\ref{Fradcirce4}), with the force due to the moving electron intrinsic magnetic moment, that is,
evaluating

\begin{equation}
F_{\text{rad}} = F_{\parallel},
 \label{FradequalFpar}
\end{equation}

yields

\begin{equation}
\frac{2\sqrt{2}}{3}\frac{e^5}{c^3} {r}^{-7/2} {m_e}^{-3/2}
\approx \frac{e^4 \hbar}{c^3 } r^{-4} m_e^{-2}.
\label{Fradequalmmorb}
\end{equation}

So

\begin{equation}
r \approx \frac{9}{8}\frac{\hbar^2}{m e^2} 
\label{Fradequalmmorb4}
\end{equation}

is the electron-positron separation where the force on the positron due to the
electron intrinsic magnetic moment orbital motion is similar in
magnitude to the radiation reaction force on the electron. For
further comparison, (\ref{Fradequalmmorb4}) can be expressed in terms of the
Bohr radius as

\begin{equation}
r \approx \frac{9}{4} r_{\text{B}}.
\label{Fradbohrradius1}
\end{equation}

So, for the electron spin model and orientation under analysis, the separation where the non-radial force on the positron is equal to
the radiation reaction force on the electron is approximately
twice the Bohr separation.  Also, since the non-radial component of the electric force increases more rapidly with decreasing range than the radiation reaction force, it will be the larger for all smaller radii than (\ref{Fradbohrradius1}).

There is an equal and opposite force on the other particle that arises due to the effect of delay on the hidden momentum term in Eq. (\ref{ElecTranslationalEq}).

\bibliographystyle{plain}


\end{document}